\documentclass[12pt,twoside]{report}
\usepackage[utf8]{inputenc}
\usepackage{graphicx}
\graphicspath{ {images/} }
\usepackage{pdfpages}
\usepackage[a4paper,width=170mm,top=25mm,bottom=25mm,bindingoffset=6mm]{geometry}
\usepackage{caption}
\usepackage{listings}
\usepackage{subcaption}
\usepackage{fancyhdr}
\usepackage{amsmath}
\usepackage{float}
\usepackage[version=4]{mhchem}
\makeatletter
\newcommand*{\rom}[1]{\expandafter\@slowromancap\romannumeral #1@}
\makeatother

\usepackage[superscript]{cite}
\usepackage{wrapfig}

\begin{document}

\includepdf{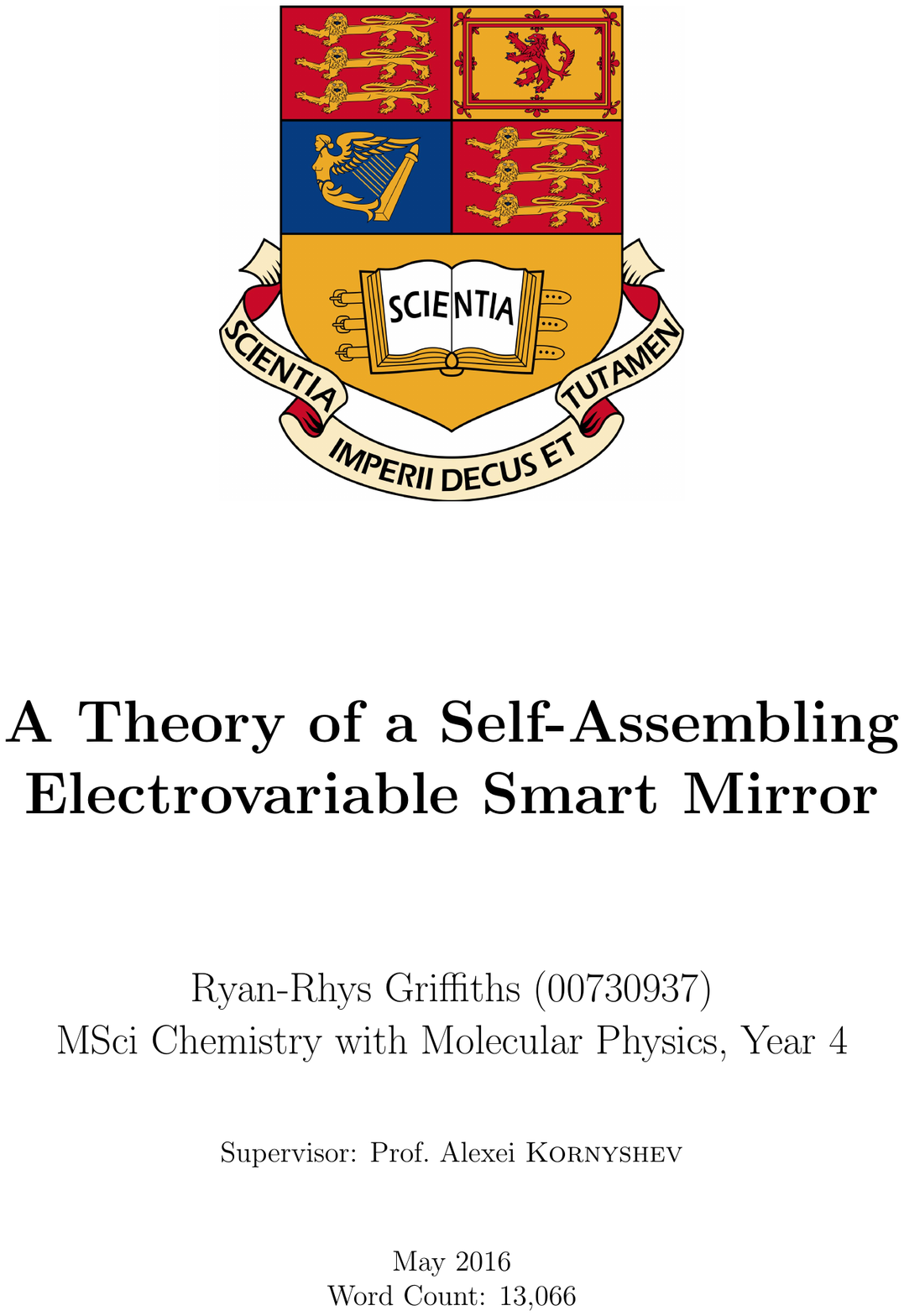}

\chapter*{Abstract}

A theory describing the forces governing the self-assembly of nanoparticles at the solid-liquid interface is developed. In the process, new theoretical results are derived to describe the effect that the field penetration of a point-like particle, into an electrode, has on the image potential energy, and pair interaction energy profiles at the electrode-electrolyte interface. The application of the theory is demonstrated for gold and ITO electrode systems, promising materials for novel colour-tuneable electrovariable smart mirrors and mirror-window devices respectively. Model estimates suggest that electrovariability is attainable in both systems and will act as a guide for future experiments. Lastly, the generalisability of the theory towards electrovariable, nanoplasmonic systems suggests that it may contribute towards the design of intelligent metamaterials with programmable properties.

\chapter*{Acknowledgements}

I would like to express my thanks to Professor Rudi Podgornik and Dr. Debabrata Sikdar for their advice on the theory of Van der Waals forces and nanoplasmonics respectively. Additionally, I would like to thank Dr. Yunuen Montelongo, Ye Ma and James Million for discussions on the experimental aspects of the electrovariable mirror project. Most of all, I would like to convey my sincerest gratitude to Professor Kornyshev for his altruistic guidance, patience and encouragement and for impressing on me the point, that while an invaluable tool, technology should never be used as a substitute for one's own judgement in the natural sciences.  

\tableofcontents

\listoffigures

\chapter{Introduction}












\section{A Brief Overview of Nanoplasmonics}

Research into the study of nanoparticle assemblies has progressed rapidly in recent years due to the novelty of the intrinsic optical\cite{2003 Kelly}, electronic\cite{1998 Feldheim}, magnetic\cite{2005 Duan} and catalytic\cite{2010 Crossley} properties exhibited in such systems. Moreover, the bulk properties of the assemblies, namely their surface-to-volume ratios and long range ordering, make them ideal candidates for implementation into practical devices\cite{2009 Velev}. This trio of desirable chemical, physical and engineering features has led to innovations ranging from influenza vaccines\cite{2013 Kanekiyo} and new-age transistors\cite{2016 Yang} to nanoscale thermometers and pH meters\cite{2010 Nie}.

\begin{figure}[h]
\begin{center}
\includegraphics[scale = 0.4]{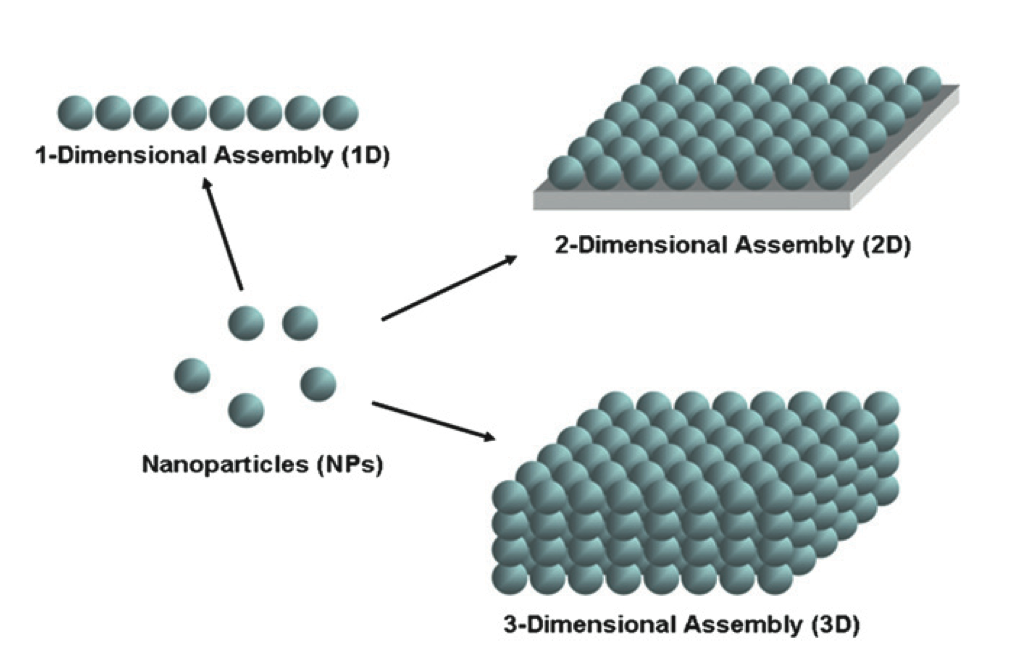}
\end{center}
\caption{Schematic of Nanoparticle Assemblies\cite{2010 Liu}.}
\end{figure}

\par Of the aforementioned properties of nanoparticle assemblies, it is their optical behaviour that has garnered the most widespread attention with respect to the fabrication of materials. Indeed, assemblies of metallic nanoparticles have enjoyed particular prominence in the area of nanoplasmonics, the field of research concerning the interaction of light with nanostructured metal surfaces\cite{2011 Stockman Today}. The optical phenomena arising from this interaction rely on the properties of surface plasmons, the oscillations of a metal's free electron gas. The behaviour of these surface plasmons, and hence the optical attributes of the material, may be controlled by making alterations to the surface structure of the metal\cite{2003 Barnes}. Due to the ease with which their structures can be regulated, solution-phase nanoparticle arrays represent some of the most flexible platforms towards the construction of novel optical devices\cite{Maier Book}. It is the use of such arrays, that makes nanoplasmonics an interdisciplinary science, integrating the chemistry-based knowledge of structure assembly with the theoretical knowledge of the physical properties of the assembled material. It is also true to say that nanoplasmonics is an interdisciplinary science with respect to its applications. In the field of chemistry, the complementary techniques of Localised Surface Plasmon Resonance and Surface-Enhances Raman Spectroscopy for molecular identification have found applications in biosensing\cite{2008 Anker}, trace analyte detection\cite{2013 Cecchini} and heavy metal sensing\cite{2014 Heavy Metal}, with signal-enhancement factors being so strong as to allow for single molecule detection in some cases.

\par It is however, the realisation of the science-fiction like qualities of nanoplasmonic metamaterials\cite{2012 Nanoplasmonic Metamaterials} that motivates the subject matter of this report. Metamaterials are so-called because they display properties that may not be replicated in nature. Some notable examples include materials with negative refractive indices\cite{2011 Metamaterials} and perhaps the best-known metamaterial of all, Sir John Pendry's proposed invisibility cloak\cite{2006 Invisible}. Fundamental to the function of such metamaterials is the localisation of optical energy in space\cite{2011 Stockman OSA}. Towards this end, the enablement of increasingly precise control over structural geometry, and hence energy localisation, through advances in fabrication methods, is catalysing the rate at which new metamaterials can be produced\cite{2012 Hafner}. In all of this, the idea of controlled assembly is crucial to the construction of nanoplasmonic metamaterials. Going from the idea of a static geometry to a variable one, the concept of an intelligent metamaterial is introduced, a system where the underlying nanostructure is fluid and amenable to be controlled by an external source, giving rise to programmable properties. A contribution towards the theoretical basis for such a metamaterial is presented in this report - the self-assembling electrovariable smart mirror.


\section{The Self-Assembling Electrovariable Smart Mirror}

\begin{figure}[h]
\begin{center}
    \centering
    \begin{subfigure}[!ht]{0.45\textwidth}
        \includegraphics[width=\textwidth]{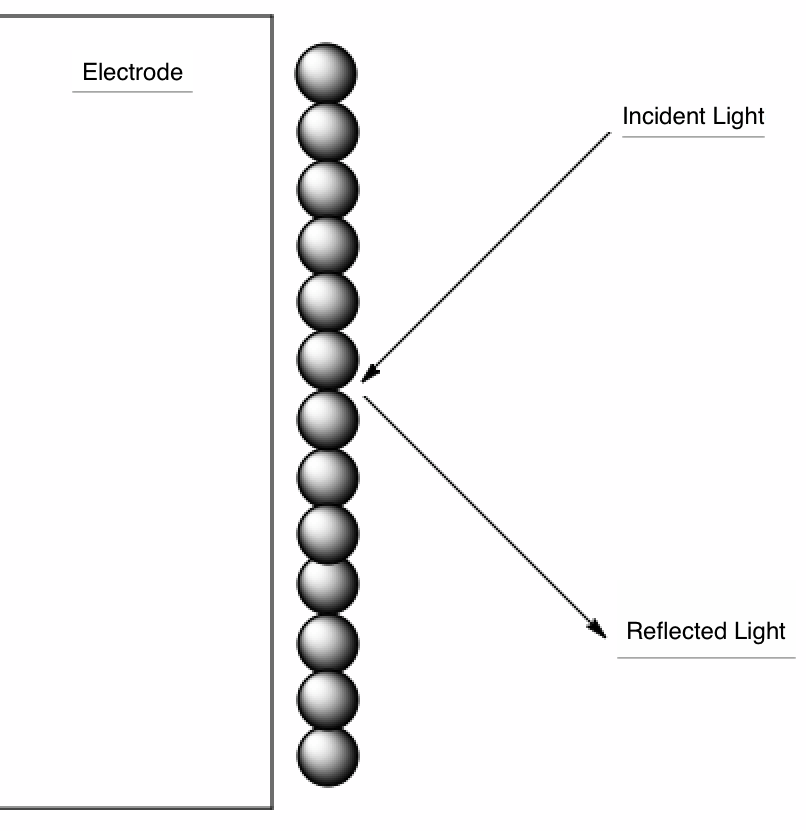}
        \caption{Mirror}
        \label{fig: Mirror}
    \end{subfigure}
    ~ 
    \begin{subfigure}[!ht]{0.525\textwidth}
        \includegraphics[width=\textwidth]{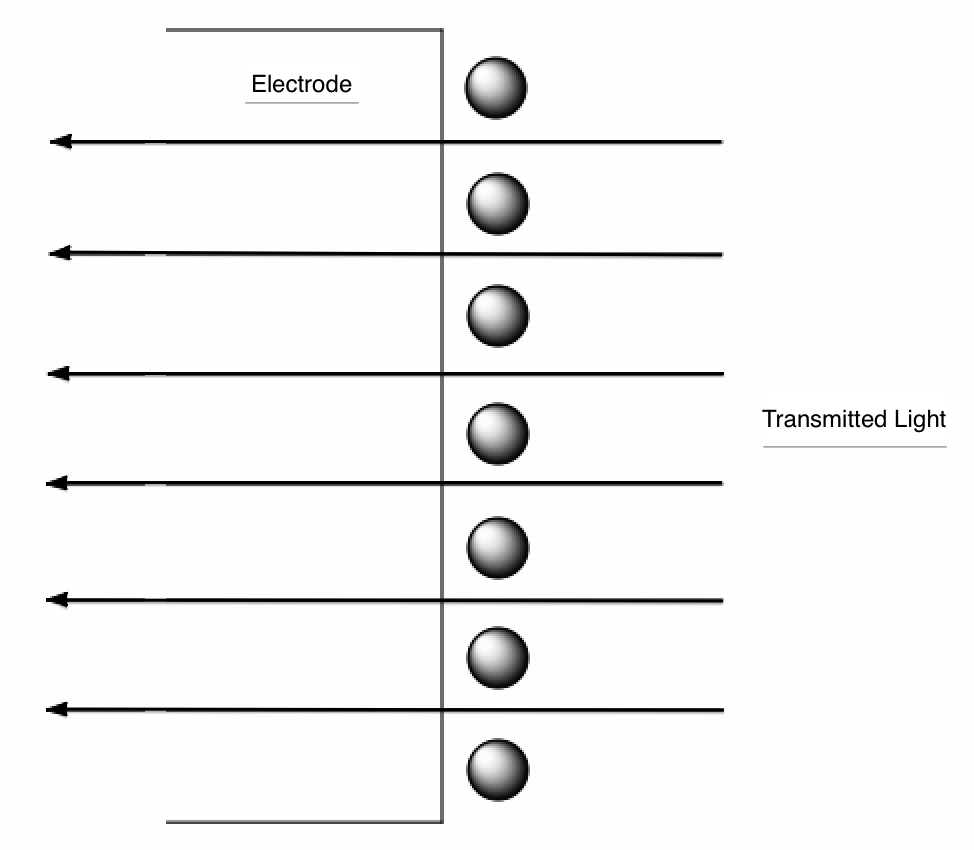}
        \caption{Window}
        \label{fig: Window}
    \end{subfigure}
    \caption{An Illustration of the Mirror-Window System.}
    \label{fig: An Illustration of the Mirror-Window System.}
\end{center}
\end{figure}

Although the concept of a liquid mirror is well established\cite{1988 Efrima}, it is only recently that metal films have been tuned for nanoplasmonic applications\cite{2003 Borra, 2005 Truong, 2006 Gingras}. A 50 nm thick film of metallic nanoparticles will behave unlike a metal film of the same thickness. The discrete structure of the nanoparticle assembly, in contrast to the film, imparts the opportunity to achieve almost full reflectance in the case of a dense monolayer of nanoparticles. Continuing with the theme of intelligent metamaterials, it will now be necessary to introduce the concept of the electrovariable mirror system\cite{2010 Flatte}. In this context, an electrovariable system is one in which it is possible, through the application of an external voltage, to reversibly control the adsorption and desorption of solution-situated nanoparticles to and from the surface of an electrode. This is one means of enabling directed self-assembly (DSA). Self-assembly (SA) is defined to be the spontaneous formation of a higher structure by a group of  nanoparticles. DSA is characterised by the use of an external source, voltage in this case, to modulate the thermodynamic forces governing the self-assembly process\cite{2010 Furst}.\par

A schematic detailing the operating mechanism of the electrovariable mirror is illustrated in figure 1.2. In this case, an applied voltage is being used to control the density of nanoparticles at the electrode surface. One may observe from figure 1.2, that under the voltage regime corresponding to figure 1.2 (a), the interface reflects light due to the presence of a dense monolayer and in so doing, acts as a mirror. In contrast, under the voltage regime corresponding to figure 1.2 (b), the surface nanoparticle density decreases and the interface transmits light, functioning as a window. In addition to the mirror-window system, a further application, not illustrated, is that of a colour-tuneable mirror. The operating principle in this case is that light is reflected or transmitted depending on its wavelength. For mirror-window applications, ITO is a particularly promising electrode material while for the colour-tuneable mirror system gold electrodes have been identified as promising \cite{2016 Edel}.

So far, the forces underpinning the assembly processes have not been discussed. One obstacle towards achieving the self-assembled monolayers illustrated in figure 1.2. is the tendency for nanoparticles to aggregate in the bulk solution due to attractive Van der Waals forces. In order to avoid this, the nanoparticles are functionalised with charge-bearing ligands that repel similarly charged ligands on other nanoparticles and impart stability towards aggregation. An encyclopedia of such ligands has been compiled by Sperling et al.\cite{2010 Sperling}. The presence of charge-bearing ligands on the nanoparticles however, also allows for the use of further experimental parameters for the fine-tuning of the surface coverage, the inter-nanoparticle spacing at the interface. These are:

\begin{enumerate}
\item Electrolyte Concentration - The presence of electrolyte results in screening of the Coulombic forces responsible for inter-nanoparticle repulsion. As such, it may be used to control the extent of surface coverage.
\item pH - the extent of ligand dissociation is governed by the solution pH. As such, in a similar manner to electrolyte concentration it may also be used to modulate inter-nanoparticle interactions.
\end{enumerate}

The aforementioned parameters have been demonstrated to be useful in controlling surface coverage experimentally at the interface of two immiscible electrolytic solutions (ITIES)\cite{2012 Turek}, a concept known as the plasmonic ruler. Considering this last point, the choice of interface is an important consideration in the search for a practical electrovariable mirror system.

\section{The Choice of Interface}

\begin{figure}[h]
\begin{center}
\includegraphics[scale = 0.22]{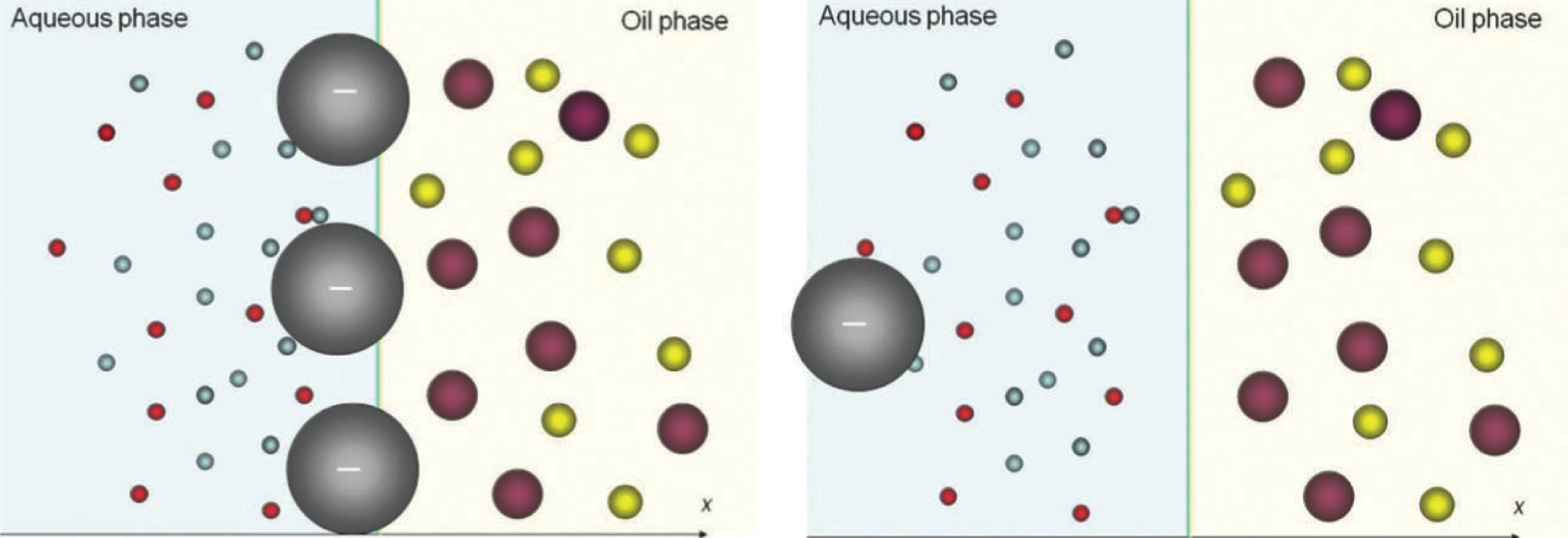}
\end{center}
\caption{Schematic of the LLI\cite{2016 Edel}.}
\label{fig: Schematic of the LLI.}
\end{figure}

Typically, this represents a choice between the liquid-liquid interface (LLI), and the solid-liquid interface (SLI), shown in figures 1.3 and 1.4 respectively. Of these two systems, it is the LLI that has attracted the most research to date. Indeed, in the context of electrovariability, Girault\cite{2004 Girault} has already demonstrated reversible adsorption-desorption at the LLI, albeit for nanoparticles with radii smaller than 2 nm. The size of the nanoparticles is crucial for mirror applications given that a sufficient amount of electronic density is necessary to achieve a high level of reflectance\cite{2016 Edel}, and in practice, the bar has been set at a radii of 20 nm or greater for most practical applications\cite{2016 Edel}. It is an unlikely prospect that Girault's work at the LLI will be replicated for larger nanoparticles. The depth of the capillary well  at the LLI has been shown theoretically to increase with the square of the nanoparticle radius\cite{2008 Flatte} and may be several electronvolts deep for the nanoparticle sizes required for mirror systems\cite{2016 Edel}. This creates significant barriers towards facile desorption from the LLI and by extension, electrovariability.\par

\begin{figure}[h]
\begin{center}
\includegraphics[scale = 0.242]{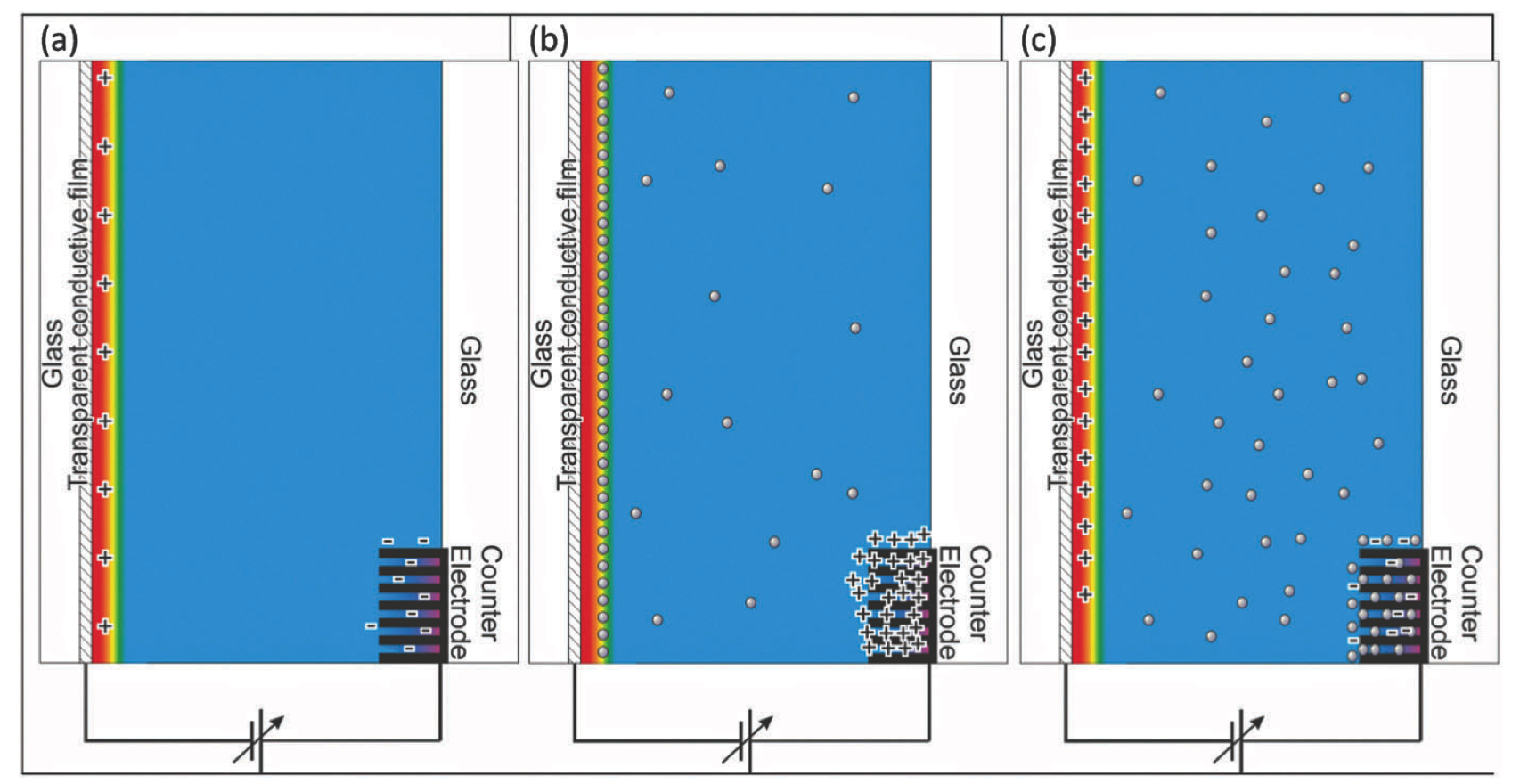}
\end{center}
\caption{Schematic of the SLI \cite{2016 Edel}.}
\label{fig: Schematic of the SLI.}
\end{figure}

    ~ 

In contrast, at the SLI, the attractive forces at play consist only of the van der Waals attraction and the image potential energy between nanoparticles and the electrode. It has been suggested\cite{2016 Edel}, that for a system consisting of gold nanoparticles and an ITO electrode, these attractive forces will be dwarfed by the electrostatic repulsion created between nanoparticles and the polarised electrode. As such, electrovariability could, in principle, be much easier to achieve at the SLI for large nanoparticles, making it an ideal platform for a prototype mirror. The optical behaviour of such systems has already been theorised under the assumption that control of the nanoparticle array geometry is achievable through electrovariability\cite{2012 Marinescu},\cite{2014 Walpole}. Although research continues to be carried out at the LLI\cite{2016 Girault}, the focus of this work will be strictly on the SLI, as this is currently considered to be the most promising system towards achieving electrovariability for the large nanoparticles needed for mirror applications. Given the absence of extensive work on the theory of self-assembly at the SLI, there is currently a pressing need for estimates of the magnitudes of the various forces governing the system behaviour in order to guide the development of experimental prototypes. \par

The theory of forces in nanoscale systems is known to be notoriously complex\cite{2008 Israelachvili}, and although many theoretical models exist for point-like molecules and macroscopic bodies, much remains to be considered for the nanoscale domain. It is this topic, the theory of the forces between nanoparticles, that will form the subject of this work.


\section{The Scope of the Present Work}

The overarching aim of the project is to develop a theoretical framework towards understanding how programmable nanoparticle geometries can be obtained. It should be stated at the outset that the resulting theory will not only be applicable to the electrovariable mirror, but will instead remain of general interest for research into the development of any device basing its function on a nanoscale metal-electrolyte system. Concretely, the individual objectives were set out as follows:

\begin{enumerate}
\item To model the total interaction potential energy between nanoparticles in the bulk electrolyte solution, choosing a level of approximation to be practical for the purposes of experiment.
\item To model the total interaction potential energy between nanoparticles and a planar electrode and as such, gauge the extent of the energy barriers acting as obstacles to electrovariability.
\item To model the third-body effect of the presence of the electrode on the interaction between nanoparticles in order to obtain an accurate representation of the energy profile close to the metal-electrolyte interface.
\item To illustrate the application of the theory to both gold and ITO electrode systems looking towards their likely future use in tuneable mirror and mirror-window applications.
\end{enumerate}

The first three of these objectives are discussed in order in chapters 3 through 5, while the fourth represents an additional consideration within the content of chapters 4 and 5. The validity of the model presented in this work will be assessed in the course of subsequent experiments undertaken as part of the Molecular Alarm grant project led by Prof. Alexei Kornyshev and Dr. Joshua Edel in the Department of Chemistry at Imperial College London\cite{2016 Edel, 2013 Edel}.




\chapter{Theoretical Methods}

In this section, the process of choosing an appropriate model for the computation of Van der Waals forces will be outlined. The methodologies involved in calculating the other interactions in the system, such as the Coulombic and image forces are less involved, and as such, will be discussed when they are arrived at in the main body of the results. Given the importance of this new venture into the theory of Van der Waals forces to the Kornyshev group, the principle theories will be reviewed in order to motivate the choice of the Hamaker hybrid formalism. It should be noted, that in all the discussions to follow in this work, it is only the interaction energy that is important. The terms force, potential and energy will be readily interchanged in this regard somewhat indifferently. Every result will be presented in the form of an energy. The overview of Van der Waals forces will take place under the following headings:

\begin{enumerate}
\item The Nature of Van der Waals Forces.
\item The Choice of Formalism.
\item Force Computation.
\item Additional Comments.
\end{enumerate}

\section{The Nature of Van der Waals Forces}


\begin{figure}[h]
\centering
\includegraphics[scale=0.5]{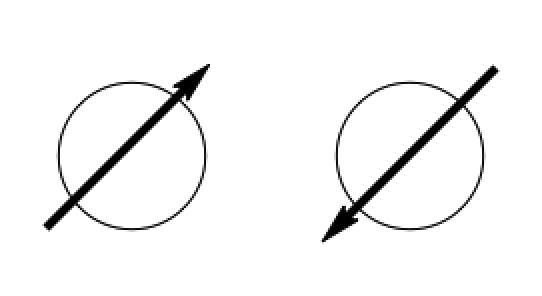}
\caption{An Analogy to Static Dipoles - The Absence of Fluctuations.}
\label{fig: An Analogy to Static Dipoles -The Absence of Fluctuations.}
\end{figure}





\noindent The Van der Waals force is one of electrodynamic origin, arising from the fluctuating charges in a neutral material. The static case of alignment of two dipoles is illustrated in figure 2.1. In this situation, the dipoles align in order to cancel each others charges. The Van der Waals force is characterised by spontaneous analogues of these static dipoles. In such a system, there will be disturbing effects present, such as thermal agitation as well as the intrinsic quantum mechanical uncertainty in the motion of the charges, that will drive the spontaneous dipoles in and out of alignment. It is the time-averaged effect of this jostling of charge that represents the Van der Waals force. The strength of the force between bodies is proportional to the level of correlation in the movement of their charge distributions. One may use the analogy of two dancers, where the quality of the performance is proportional to the extent to which the dancers are in step with each other. The strength of the force will also vary depending on the compositions of the materials. In particular, the force is known to be stronger in condensed phases relative to gaseous phases\cite{Parsegian}. There are a number of ways in which the force may be treated theoretically and these will now be considered.

\section{The Choice of Formalism}


The range of theoretical models available for the calculation of Van der Waals forces range from the simple language of pairwise summation to the complexities of quantum field theory. The historical development of the modern theory of Van der Waals forces will be briefly outlined, and the realms of applicability of the models discussed in the following order: 

\begin{enumerate}
\item Hamaker's Method - Pairwise Summation
\item The Lifshitz Model - Continuum Theory
\item The Hamaker Hybrid Formalism
\end{enumerate}

\subsection{Hamaker's Method - Pairwise Summation}


\begin{figure}[h]
\centering
\includegraphics[scale=0.5]{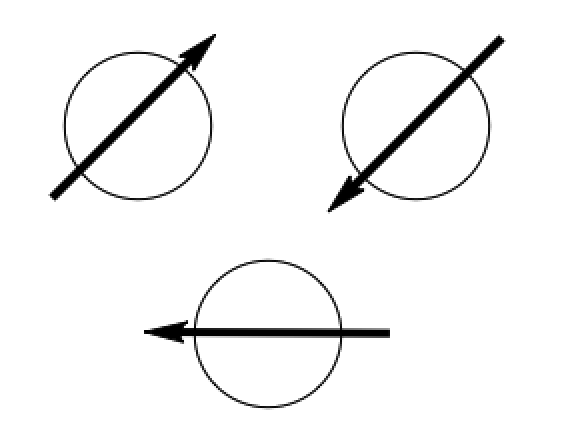}
\caption{An Illustration of the Third-Body Effect in Condensed Phases.}
\label{fig: An Illustration of the Third-Body Effect in Condensed Phases.}
\end{figure}




\noindent H. C. Hamaker, in 1937, was the first to develop a methodology for the computation of Van der Waals forces between macroscopic bodies\cite{1937 Hamaker}. He achieved this through invoking a pairwise summation of individual differential elements of each material. While this may constitute an accurate approach for dilute gases, where the molecules remain distant from each other, in condensed phases, the molecules comprising the material are more compressed, and third body effects start to become important. This effect is illustrated in figure 2.2. Clearly, there is no way in which all three spontaneous dipoles may align perfectly with each other. The Hamaker formalism treats the dipoles as if they do align perfectly, and so tends to overestimate the strength of interaction between condensed bodies. Nonetheless, the Hamaker formalism retains value through its simplicity, and indeed the geometric result for the distance dependence from the Hamaker model is still relevant in the hybridisation to the continuum theory, where relativistic effects are neglected.

\subsection{The Lifshitz Model - Continuum Theory}


The modern viewpoint on Van der Waals forces is exemplified by two key advances in their understanding made by Casimir and Lifshitz respectively. Casimir, in 1948, was the first to realise that the distance dependence of the Van der Waals interaction was affected by the relativistic velocity of light\cite{1948 Casimir}. The origin of this behaviour, known as the retardation effect, is illustrated in figure 2.3.

\begin{figure}[h]
\centering
\includegraphics[scale=0.5]{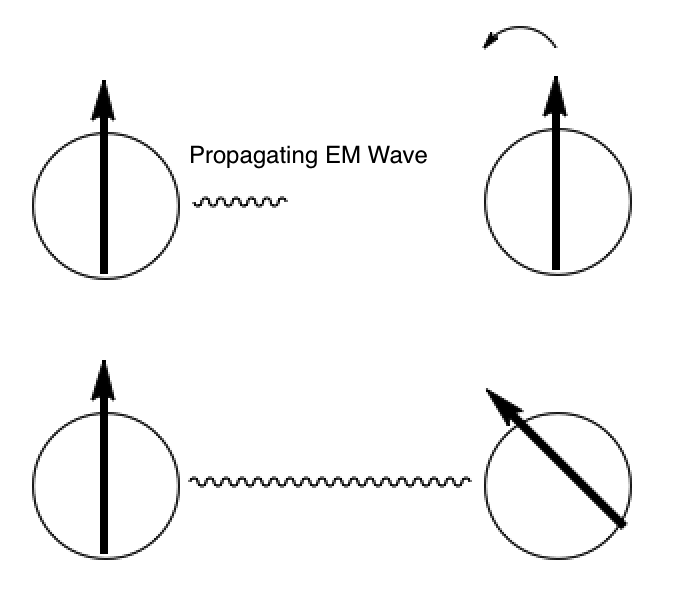}
\caption{Casimir Forces - Accounting for Retardation Effects.}
\label{fig: Casimir Forces - Accounting for Retardation Effects.}
\end{figure}

When intermolecular separations are large, the time it takes an electromagnetic (EM) wave to propagate towards another source is comparable to the time taken for the charge configuration of the other molecule to rearrange itself. Given that the strength of the Van der Waals force is determined by the correlated movement of charges, the breakdown of rapid signal communication between molecules means that the charge movements fall out of step with each other, resulting in a reduction in the magnitude of the force. The consequences for the distance dependence of the Van der Waals interaction, is that the classical $\frac{1}{r^6}$ dependence is replaced with a dependence of $\frac{1}{r^7}$, where $r$ is the intermolecular separation. Although the Hamaker hybrid formalism chosen for this work neglects these retardation effects, they represent important corrections in systems where the separation between bodies is large. In the context of nomenclature, it is worthy to note that the Van der Waals force is usually rebranded as the Casimir force when retardation effects are accounted for.\par

The contribution of Lifshitz in 1956, was to formulate a theory of macroscopic Van der Waals forces between condensed phases based on a continuum model\cite{1956 Lifshitz}. Making use of the fluctuation-dissipation principle, an idea that will be described in further detail in the section on connecting forces with spectra, Lifshitz's model enabled the calculation of Van der Waals forces from the bulk dielectric properties of the materials. In 1961, Lifshitz and Dzyaloshinskii extended the model to account for the presence of a medium between condensed phases. The Lifshitz theory is in essence, a description of Van der Waals forces at long ranges of distance. It begins to break down at the point where the separation between bodies approaches the length scale where discrete atomic structure is felt. While the Lifshitz theory has been demonstrated to be highly accurate, validating the experimental results of Derjaguin\cite{1956 Derjaguin}, it is also notoriously complex and its implementation requires a working knowledge of quantum field theory\cite{Israelachvili}. Fortunately, in recent years, much work has been undertaken in simplifying the complexity of the Lifshitz theory, producing a hybrid theory that manages to marry the simplicity of Hamaker's results to the accuracy of the Lifshitz continuum model.

\subsection{The Hamaker Hybrid Formalism}


The two key simplifying assumptions which see the Hamaker hybrid theory emerge from the Lifshitz theory are\cite{Parsegian}:

\begin{enumerate}
\item That the finite velocity of light may be neglected.
\item That the differences in the electromagnetic properties of the materials are small
\end{enumerate}

\noindent The resulting description of the energy between macroscopic bodies is the product of two terms, a geometric term taken from the Hamaker theory, and a Hamaker, or Lifshitz coefficient, $A$, that embodies the electromagnetic material properties and is computed within the full Lifshitz theory. The formula for the Van der Waals interaction energy of two spheres is given as

\begin{align} \label{equation 1}
G_{ss}(z; R_1, R_2) = \mspace{230mu} \notag\\ \notag\\ - \frac{A_{1m/2m}}{3}\left[\frac{R_1R_2}{z^2 - (R_1+R_2)^2} + \frac{R_1R_2}{z^2 - (R_1-R_2)^2} + \frac{1}{2} \ln \left(\frac{z^2 - (R_1 + R_2)^2}{z^2 - (R_1 - R_2)^2}\right) \right] \:.
\end{align}

\noindent $R_1$ and $R_2$ are the radii of spheres 1 and 2 respectively, $A_{1m/2m}$ is the Hamaker coefficient specific to materials 1 and 2 across medium $m$, and $z$ is the centre to centre separation between spheres. Correspondingly, for the case of a planar surface and a sphere, the interaction energy is given by

\begin{align} \label{equation 2}
G_{sp}(l; R) = - \frac{A_{1m/2m}}{6}\left[\frac{R}{l} + \frac{R}{2R + l} +  \ln \left( \frac{l}{2R + l}\right) \right] \:.
\end{align}


\noindent In this case, $l$ is the separation between the sphere and the surface. Equations 2.1 and 2.2 above will be reproduced in the relevant sections of the report for the sake of clarity. The topic of the size dependence of the Hamaker coefficient will be introduced in chapter 3, where it will be discussed in the context of its use in the literature on nanoparticle stability.

\section{Force Computation}


Regarding the task of computing Van der Waals forces, the following theoretical and practical aspects will be discussed.

\begin{enumerate}
\item Connecting Forces with Spectra
\item The Frequency-Dependent Dielectric Function
\item Imaginary Frequencies
\item The Hamaker Coefficient
\end{enumerate}

\noindent Borne out of the fact that the theory of Van der Waals forces is a new departure for the Kornyshev group, an annotated, step-by-step procedure for the computation of forces has been provided in section A.5 of the appendix.

\subsection{Connecting Forces with Spectra}



As mentioned previously, the fluctuation-dissipation theorem is fundamental to the computation of Van der Waals forces. The fluctuation-dissipation theorem presents a relation between the frequencies at which charges spontaneously fluctuate and the frequencies at which they absorb, or dissipate, electromagnetic radiation. The electromagnetic absorption spectrum encompasses the entire set of interactions between the atoms making up the material, and so, via the fluctuation-dissipation theorem, it will also characterise the spontaneous fluctuations of the Van der Waals force. The current convention in the literature is to represent these spectra in equation form, as fits to the spectral data. The reason for doing so, is for the purposes of yielding an understanding of the electromagnetic behaviour of the material from the parameters of the equation\cite{Parsegian}. Practically speaking, when one wants to compute accurate Van der Waals forces, this may not represent the most efficient approach. Within the Hamaker hybrid formalism, one could compute the Hamaker coefficient numerically from experimental spectra assuming sufficiently high-quality data. Nonetheless, given that the use of the equation representation, known as the frequency-dependent dielectric function, is so prevalent in the literature, it will be used throughout this work.

\subsection{The Frequency-Dependent Dielectric Function}


Frequency-dependent dielectric functions, designated by $\varepsilon(\omega)$, where $\varepsilon$ is the dielectric or permittivity function, and $\omega$ is the frequency, are generally featured in two forms in the literature:

\begin{enumerate}
\item Drude-Lorentz Form
\item Damped Oscillator Form
\end{enumerate}

\noindent The characteristics of these two representations as well as their conditions of use will be discussed here. It should be noted that the topic of the additional size dependence of the frequency-dependent dielectric function will be discussed in chapter 3 in conjunction with the literature on the size-dependent Hamaker coefficient.

\subsubsection{Drude-Lorentz Form}

%




The response of a metal's electrons to an incident EM wave involves displacement of the electron clouds about the nuclei. This displacement will depend on the frequency of the EM wave, and at resonant frequencies, the displacement of the clouds will be maximised. The cloud displacement induces the creation of secondary EM waves that act as obstacles towards the propagation of the incident wave through the metal, and it is this effect that imparts the metal's reflective quality. \par 

The Drude-Lorentz model captures the characteristics of this behaviour in three terms. The first, the asymptotic dielectric constant, $\varepsilon_{\infty}$, represents the response of the metal to fields or waves of sufficiently high frequencies as to not be felt by the core electrons. Typically, in the computation of Van der Waals forces, this parameter is set to a value of 1, in order that it is compatible with the formula for the Hamaker coefficient that will be introduced later in the chapter. Another practical reason for setting $\varepsilon_{\infty}$ equal to 1, is that there are less fitting parameters in the model\cite{Palik Quote}. The second term accounts for the free electrons of the metal which are electrons that remain unbound to nuclei. As such, in the treatment of these electrons, the restoring force associated with oscillations about the nuclei is negligible. The third term gives a sum of Lorentz resonances, accounting for the fact that real metals will have not one, but many regions of resonance, with each region corresponding to a separate Lorentz term in the sum. The form of the Drude-Lorentz representation is

\begin{align} \label{equation 3}
\varepsilon(\omega) = \varepsilon_{\infty} - \frac{\omega_p^2}{\omega^2 + i\omega_{\tau}\omega} - \sum_j \frac{f_j\omega_j^2}{ \omega^2 - \omega_j^2  + i\gamma_j\omega}\:,
\end{align}

\noindent where $\omega_p$ is the plasma frequency, representing the cutoff frequency beyond which waves of higher frequency will be transmitted, $\omega_{\tau}$ is the collision frequency of the free electrons, and $f_j$, $\omega_j$ and $\gamma_j$ are respectively, the oscillator strength, oscillator frequency and damping rate of the jth Lorentzian oscillator. In the more practical form of a function of imaginary, discrete Matsubara frequencies, $\xi$, the Drude-Lorentz model takes the form of

\begin{align} \label{equation 4}
\varepsilon(i\xi) = \varepsilon_{\infty} + \frac{\omega_p^2}{\xi^2 + \omega_{\tau}\xi} + \sum_j \frac{f_j\omega_j^2}{ - \xi^2 - \omega_j^2 - \gamma_j\xi}\:.
\end{align}

\noindent The use of the term imaginary frequencies as well as the concept of sampling over discrete Matsubara frequencies will be explained in a subsequent section.

\subsubsection{Damped Oscillator Form}

In contrast to the Drude-Lorentz form, the damped oscillator form normally functions as an interpolated fit to incomplete spectral data\cite{Parsegian}. The dielectric function is represented as

\begin{align} \label{equation 5}
\varepsilon(\omega) = 1  + \sum\limits_{k=1} ^n \frac{d_k}{1 - i\omega\tau_k} + \sum\limits_{j=1} ^n \frac{f_j}{\omega^2_j - \omega^2 - i\omega g_j}\:.
\end{align}

\noindent The parameters in this case are less physically meaningful than in the case of the Drude-Lorentz model, and act simply as fitting parameters to the data. Again, the damped oscillator model, as a function of imaginary, discrete Matsubara frequencies, is represented as

\begin{align} \label{equation 6}
\varepsilon(i\xi) = 1 + \sum\limits_{k}^n \: \frac{d_j}{1 + \xi\tau_j} + \sum\limits_{j=1}^n \: \frac{f_j}{\omega_j^2 + g_j\xi + \xi^2}\:.
\end{align}

\subsection{Imaginary Frequencies}

The primary motive for the use of imaginary frequencies in the evaluation of the dielectric function is to transform a sinusoidally oscillating function of real frequencies to an exponentially decaying function of imaginary frequencies following Euler's identity

\begin{align} \label{equation 7}
e^{i\theta} = cos\theta + isin\theta \:.
\end{align}

\noindent This is useful because, near resonance, the dielectric function of real frequencies begins to oscillate wildly, and so by having an exponential function instead, analytical difficulties may be avoided\cite{Parsegian}. The discrete Matsubara sampling frequencies,

\begin{align} \label{equation 8}
\xi_n = \frac{2\pi kT}{\hbar}n \:,
\end{align}

\noindent come from quantum theory and owe their origin to the properties of harmonic oscillators\cite{Parsegian}.

\subsection{The Hamaker Coefficient}

Within the Hamaker-hybrid formalism, the Hamaker coefficient is computed as


\begin{align} \label{equation 9}
A_{1m/2m} \approx \frac{3kT}{2} \sum\limits_{n=0} ^{\infty} {}^{'} \: \frac{\varepsilon_1 - \varepsilon_m}{\varepsilon_1 + \varepsilon_m} \frac{\varepsilon_2 - \varepsilon_m}{\varepsilon_2 + \varepsilon_m}\:,
\end{align}

\noindent where $kT$ is the thermal energy, $\varepsilon_1$ and $\varepsilon_2$ are the frequency-dependent dielectric functions of materials $1$ and $2$ respectively and $\varepsilon_m$ is the frequency-dependent dielectric function of the medium. The sum, is one over the discrete Matsubara frequencies. The prime in the sum indicates that the $n = 0$ term is multiplied by a factor of a half. In electrolytic solution, this zero-frequency term is screened, and as such, takes the form

\begin{align} \label{equation 10}
A_{\xi = 0} = \frac{3}{4}kT\left(\frac{\varepsilon_1(0) - \varepsilon_m(0)}{\varepsilon_1(0) + \varepsilon_m(0)}\right)\left(\frac{\varepsilon_2(0) - \varepsilon_m(0)}{\varepsilon_2(0) + \varepsilon_m(0)}\right)(1 + 2\kappa l)e^{-2\kappa l} \:,
\end{align}

\noindent where $\kappa$ is the inverse Debye length, $l$ is the separation between bodies and the $\varepsilon$'s are now the static dielectric constants of the materials. While important in hydrocarbon systems, the zero-frequency term is known to contribute less than 1\% to the magnitude of the Hamaker coefficient in metal systems\cite{Israelachvili}. As is demonstrated in the annotated program for force computation given in the appendix, the additional screening of this term means that it is discountable.

\section{Additional Comments}

One particular work in this area that should be mentioned is the tabulation of Hamaker coefficients for a wide range of systems by Bergstrom\cite{1997 Bergstrom}. At this point in time, this encyclopedic reference of Hamaker coefficients is in need of an update given the promise of hitherto unconsidered materials such as ITO for applications in nanoplasmonics. If such calculations could be performed using high-quality experimental data, it would be greatly beneficial for any field of research requiring Van der Waals force estimates.\par


\chapter{Nanoparticles in the Bulk}


\section{Motivation}

The challenges towards achieving precise control over nanoparticle geometry at the SLI are first experienced before the nanoparticles have reached the electrode. If left unopposed, the intrinsic Van der Waals forces between nanoparticles will lead to aggregation, the process by which nanoparticles fuse to form larger structures in the bulk solution. Once aggregated, the nanoparticles will lose their controllable plasmonic properties. As such, a means of stabilising the nanoparticles is required, and in practice, this is most frequently accomplished through the functionalisation of nanoparticles with charge-bearing ligands. These charged ligands will repel similarly-charged ligands on other nanoparticles, resulting in a Coulombic repulsion to counter the attraction of the Van der Waals force. The interplay of these two effects is illustrated graphically in figure 3.1.\par

\begin{figure}[h]
\centering
\includegraphics[scale=0.3]{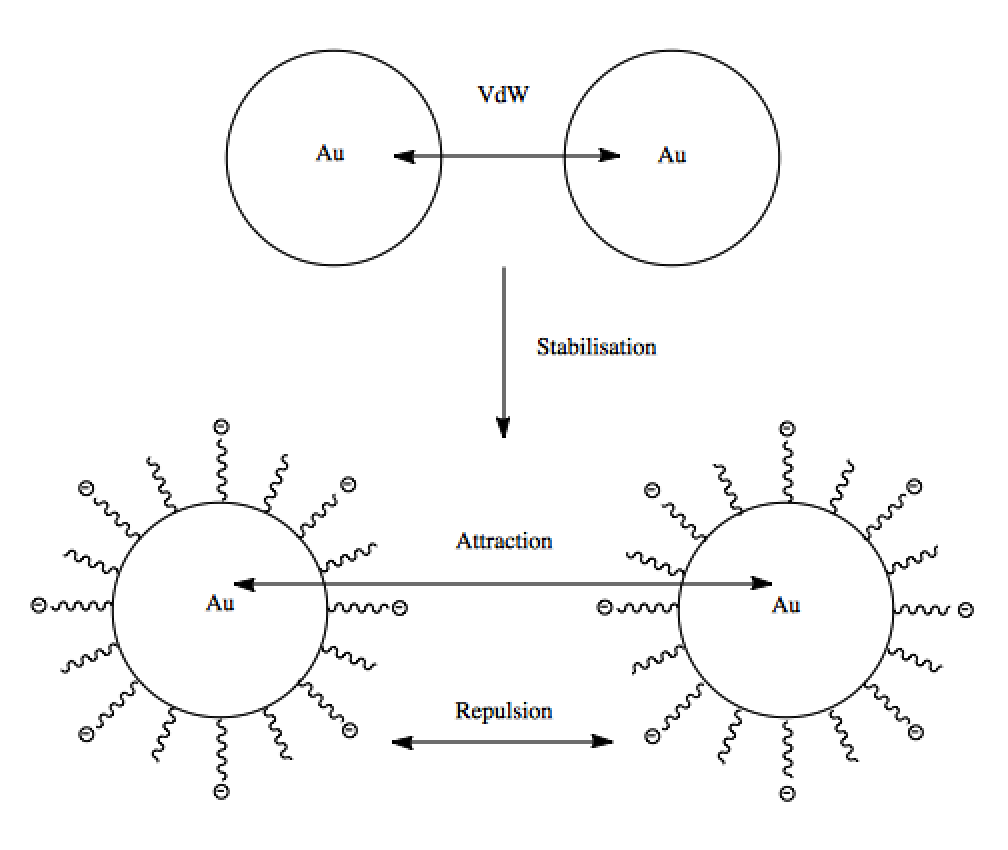}
\caption{Push and Pull - The Balance of Van der Waals (VdW) and Coulombic Forces.}
\label{fig:Push and Pull - The Balance of Van der Waals (VdW) and Coulombic Forces.}
\end{figure}

At the SLI however, control over nanoparticle stability through functionalisation alone will not lend itself to the creation of well-defined geometries. The nanoparticles in this instance must be sufficiently functionalised as to prevent bulk aggregation, yet not to the extent that they are repelled from the electrode. The need to optimise this balance of forces through adjustment of the number of ligands per nanoparticle would be a cumbersome task. Fortunately, further parameters for fine-tuning are available in the form of electrolyte concentration and solution pH. Indeed, such modulations have been demonstrated to provide superior control in the case of the plasmonic ruler at the liquid-liquid interface \cite{2012 Turek}. The effects of increased solution pH may be seen in figure 3.2, where ligand dissociation has been induced, leading to a larger Coulombic repulsion between particles. Conversely, an increase in the electrolyte concentration will have the effect of screening the Coulombic repulsion. This latter effect is illustrated in figure 3.3.

\begin{figure}[H]
\centering
\includegraphics[scale=0.3]{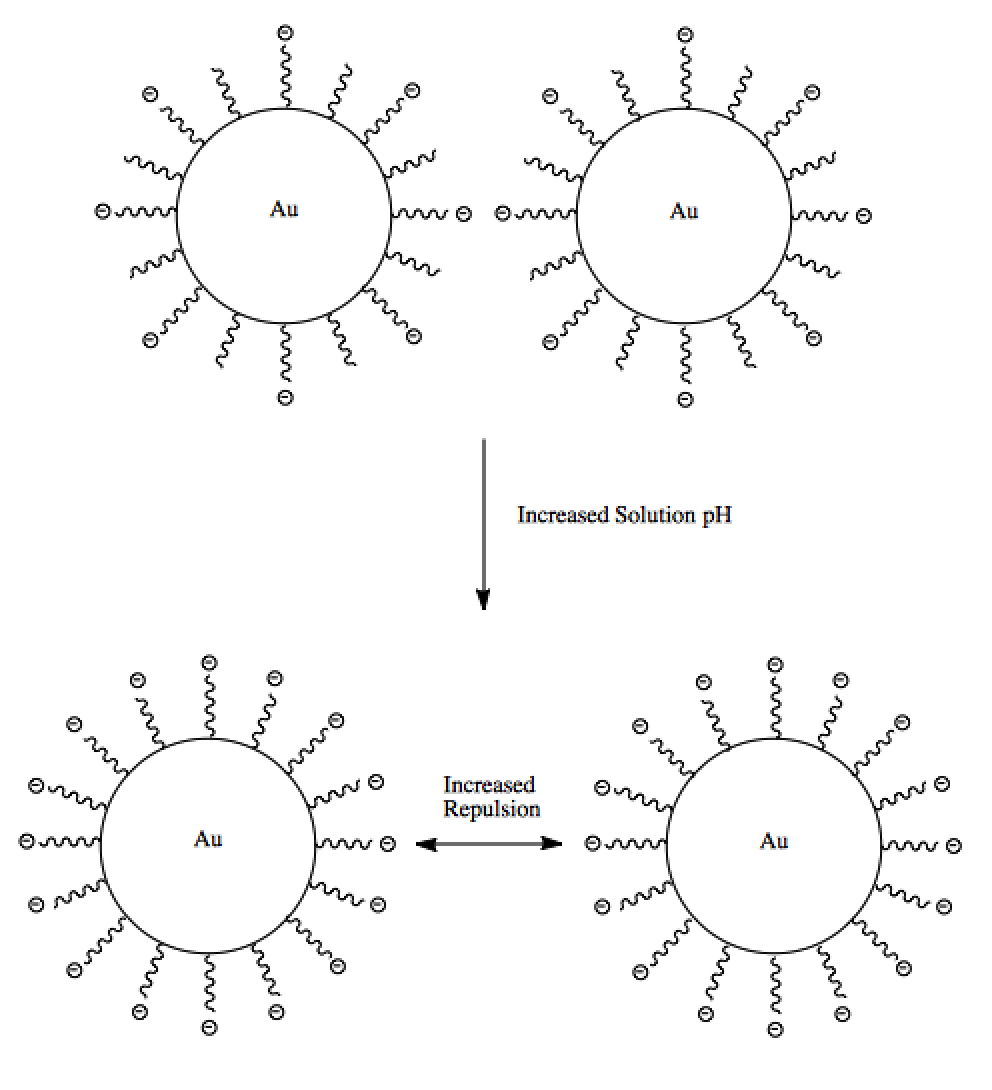}
\caption{Effect of Increased Solution pH.}
\label{fig:Effect of Increased Solution pH.}
\end{figure}

\begin{figure}[H]
\centering
\includegraphics[scale=0.3]{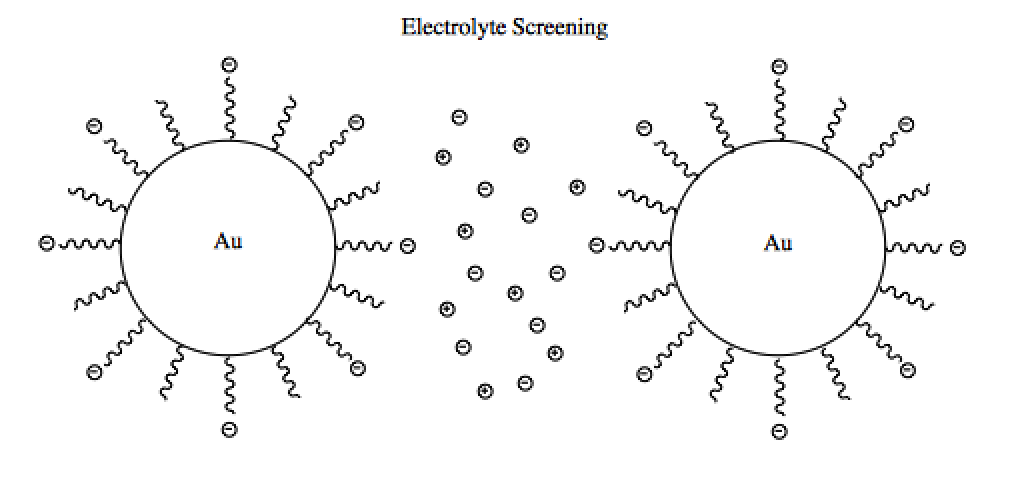}
\caption{Effect of Electrolyte Screening.}
\label{fig:Effect of Electrolyte Screening.}
\end{figure}

In this section, the distance dependence of the push and pull effects of the Coulomb and Van der Waals forces in the bulk will be modelled. The contributions made by pH and electrolyte concentration will also be accounted for in foresight of their presence in the SLI system, where their tuning may facilitate the nanoparticles' approach to the electrode. Needless to say, the resulting distance-dependent interaction potentials will be subject to the limitations of the theory describing them. This will be most evident in the case of the van der Waals force, for which accurate predictions for metals are difficult to achieve \cite{Israelachvili}. Historically \cite{1963 Enuston}, and indeed still in the most recent literature on nanoparticle stability \cite{2015 Wijenayaka, 2015 Gambinossi}, there have been a plethora of formalisms implemented for the computation of Van der Waals forces, involving mixed interpretations of both the pairwise Hamaker summation and the Lifshitz theory.\par

Often there is little justification given for the choice of theory, which in extreme cases may lead to a misrepresentation of the system being studied. In the work of Gambinossi et al. \cite{2015 Gambinossi} for example, an outdated theory of Van der Waals forces between colloids \cite{1961 Vold, 1973 Vincent} was applied to compute the forces between nanoparticles. Additionally, research into hitherto unexplored effects such as the size dependence of the dielectric function of nanoscale materials, and consequently the Hamaker coefficient \cite{2003 Pinchuk, 2012 Pinchuk}, is ongoing. Such considerations add extra dimensions of complexity towards the development of a quantitative theory of Van der Waals forces. Indeed, the literature is fraught with mismatched formalisms, with a stark example being the presentation of results treating nonlocality at short nanoparticle separations\cite{2012 Schatz} using the inaccurate Hamaker method\cite{2014 Pendry}.

As such, in light of the complexities involved in constructing an accurate, quantitative theory, the aim of this section will be to provide approximate estimates of the extent of the Van der Waals interaction between nanoparticles. In so doing, it is hoped that a realistic theory, with realistic error bounds, may be conveyed. The distance dependence of the Van der Waals and Coulombic forces will first be decoupled from each other before being combined in the following order:

\begin{enumerate}
\item The Van der Waals Force.
\item The Coulomb Force.
\item The Total Interaction Energy.
\end{enumerate}

\section{The Van der Waals Force}

The key model assumptions will be conveyed before the potential profile is presented. The discussion will take the following form:

\begin{enumerate}
\item Hamaker Hybrid Form for Two Nanoparticles
\item Nonlocality at Short Separations
\item Size Dependence of the Hamaker Coefficient
\item The Van der Waals Potential Profile
\end{enumerate}

\subsection{Hamaker Hybrid Form for Two Nanoparticles}




Reiterating the points made in chapter 2, the Hamaker hybrid form of the Lifshitz theory will be used for the computation of the Van der Waals force. Within this formalism, the general equation for the energy between nanoparticles,

\begin{align} \label{equation 1}
G_{ss}(z; R_1, R_2) = \mspace{230mu} \notag\\ \notag\\ - \frac{A_{1m/2m}}{3}\left[\frac{R_1R_2}{z^2 - (R_1+R_2)^2} + \frac{R_1R_2}{z^2 - (R_1-R_2)^2} + \frac{1}{2} \ln \left(\frac{z^2 - (R_1 + R_2)^2}{z^2 - (R_1 - R_2)^2}\right) \right] \:,
\end{align}

\noindent or the specific equation for nanoparticles of equal radius,

\begin{align} \label{equation 2}
G_{ss}(z; R) = - \frac{A_{1m/2m}}{3}\left[\frac{R^2}{z^2 - 4R^2} + \frac{R^2}{z^2} + \frac{1}{2} \ln \left(1 - \frac{4R^2}{z^2}\right) \right] \:,
\end{align} 

\noindent may be broken down into two factors. Firstly, the Hamaker coefficient, $A$, traditionally thought of as a size- and morphology-independent factor characterised by the dielectric properties of the materials\cite{2012 Pinchuk}, and secondly, a size-dependent factor accounting for the geometry of the system. The geometry is shown in figure 3.4.

\begin{figure}[h]
\centering
\includegraphics[scale=0.5]{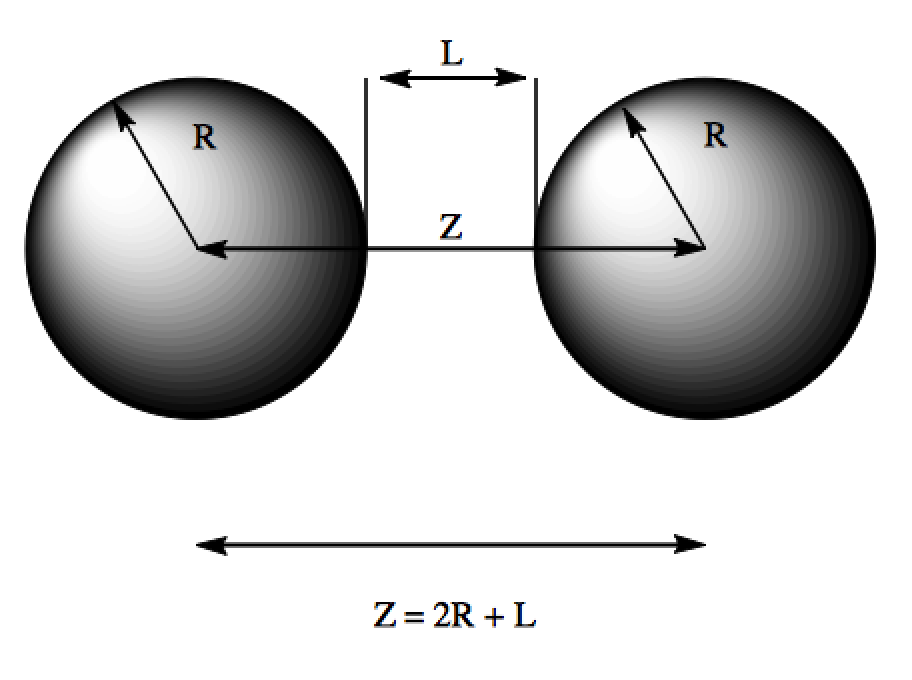}
\caption{System Geometry.}
\label{fig:System Geometry.}
\end{figure}

\noindent Whilst the geometric factor is well understood and subsequently common to both Hamaker and Lifshitz theories \cite{Israelachvili}, the Hamaker coefficient is the problem child of the majority of attempts to theorise Van der Waals forces. Contrary to the long-held belief, it has recently been suggested that the Hamaker coefficient is in fact neither size, nor geometry independent \cite{2012 Pinchuk, 2014 Pendry, 2012 Schatz}.

\subsection{Nonlocality}



Nonlocality, or spatial dispersion, is an effect that must be accounted for at short nanoparticle separations. Typically, at distances of less than 10 nm, classical electrodynamics fails to accurately describe the extent of the Van der Waals force \cite{2012 Schatz, 2014 Pendry}. The consequence being that instead of experiencing saturation, the Van der Waals force is predicted to grow infinitely large at small separations as shown in figure 3.5.


\begin{figure}[h]
\centering
\includegraphics[scale=0.75]{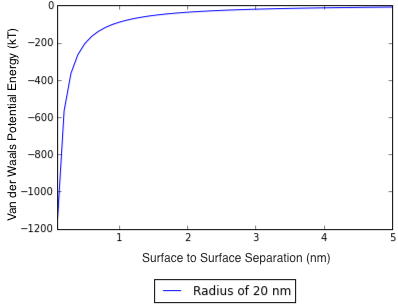}
\caption{Divergent Behaviour at Short Separation.}
\label{fig:Divergent Behaviour at Short Separation.}
\end{figure}

\noindent This behaviour is unphysical and comes about as a result of infinite compression of the charge fluctuations at wavelengths much smaller than that of the mean free path of the metal surface electrons \cite{1956 Lifshitz, Israelachvili}. A more correct representation is offered through the framework of quantum electrodynamics, whereby the surface charges are nonlocally smeared across a subnanometre boundary \cite{2014 Mortenson}. The term nonlocal in this instance refers to the fact that the dielectric response of the metal depends not only on the magnitude of the electric field experienced at the point of interest, but also on the value of the field at neighbouring points \cite{2012 Schatz}. The net result of incorporation of nonlocal effects is that a material's dielectric function becomes not only frequency dependent but wavevector dependent as well.\par

In the results to follow, it will be assumed that the nanoparticles will always remain at separations where nonlocal effects may be neglected. This assumption also happens to be consistent with that of steric interactions between ligands on different nanoparticles being unimportant due to the extent of the Coulombic repulsion. It is nonetheless important to always keep in mind the existence of nonlocal effects when drawing comparisons between theory and experimental systems where short nanoparticle separations remain feasible.

\subsection{Size Dependence of the Hamaker Coefficient}


It has been noted that for small metal nanoparticles, typically where the diameter is less than 10 nm \cite{1975 Kreibig, Kreibig, 2010 He}, the frequency-dependent dielectric function of the metal also becomes size dependent. A decrease in the mean free path of surface electrons at this scale gives rise to a broadening of the plasma resonance peak and consequently modifies the dielectric function. This has a knock-on effect for the Hamaker coefficient, which depends on the dielectric function of the materials in the system. One theory that accounts for this size dependence of the Hamaker coefficient for metal nanoparticles is offered by Pinchuk \cite{2012 Pinchuk, 2003 Pinchuk}. The size-dependent dielectric function is represented as

\begin{align} \label{equation 3}
\varepsilon(\omega, R) = 1 - \frac{\omega_p^2}{\omega^2 + i\gamma(R)\omega} \:,
\end{align}

\noindent where $\omega_p$ is the plasma frequency and $\gamma(R)$ is the size-dependent scattering rate of the material, dependent on the nanoparticle radius $R$. In turn, this scattering rate,

\begin{align} \label{equation 4}
\gamma(R) = \gamma + A\frac{v_F}{R} \:,
\end{align}

\noindent is dependent on $\gamma$, the size-independent scattering rate, $v_F$, the Fermi velocity of the metal, and $A$, a dimensionless constant. In the recent literature, this size dependence has been incorporated into the computation of Van der Waals forces between silver \cite{2012 Pinchuk} and gold \cite{2015 Wijenayaka, Wijenayaka} nanoparticles. An illustration of the size dependence of the Hamaker coefficient for gold, taken from the work of Wijenayaka et al. \cite{2015 Wijenayaka}, is given in figure 3.6.




\begin{figure}[h]
\centering
\includegraphics[scale=0.5]{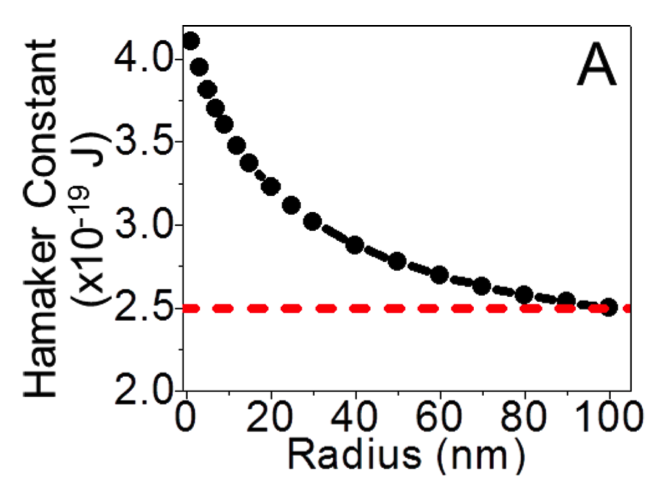}
\caption{An Illustration of the Size Dependence of the Hamaker Coefficient for Gold Nanoparticles in the Bulk.}
\label{fig: An Illustration of the Size Dependence of the Hamaker Coefficient for Gold Nanoparticles in the Bulk.}
\end{figure}

The plot has been constructed in a manner such that the calculated values of the Hamaker coefficient for increasing values of the nanoparticle radius are constrained to asymptotically approach that of bulk gold. As shall be discussed later, the reported values for bulk gold \cite{2015 Wijenayaka, 2005 Kim, 2008 Lundgren} are highly inaccurate as a result of the limitations of the experimental electromagnetic absorption spectra from which they are computed \cite{Parsegian}. When the y-axis values in the figure above are converted to more sensible units of kT, one may see that at 300 K, the asymptotic limit of bulk gold corresponds to a Hamaker coefficient of approximately 60 kT, and increases as the radius of the nanoparticle decreases. If the Hamaker coefficient were to increase with decreasing radius, importantly for nanoparticles of the sizes depicted in figure 3.6, one would expect, as the authors suggest, that smaller nanoparticles would be observed to aggregate more readily in the bulk. In experiment however, this is not the case, with larger particles being observed to aggregate more readily \cite{2016 Edel} in accordance with the increasing Van der Waals force associated with the geometric factor in equations 3.1 and 3.2.\par

One may indeed suspect, that at the nanoparticle sizes plotted in figure 3.6, with nanoparticle diameters vastly exceeding the 10 nm limit required for size dependence \cite{1975 Kreibig, Kreibig, 2010 He}, that the Hamaker coefficient should remain constant in this range. Further to this, Johnson and Christy found there to be no significant difference in the inferred values of the real and imaginary parts of the dielectric function for gold using thin films of thickness 34.3 nm and 45.6 nm respectively. This would add support to the point that there should be no size dependence of the dielectric function of gold, and by extension, the Hamaker coefficient of gold nanoparticles, for values of the diameter exceeding 10 nm. \par

Given that larger nanoparticles, with radii on the order of 20-40 nm, are needed to achieve the reflectivity necessary for mirror applications, size-dependent Hamaker coefficients will be neglected in this work. In conjunction with nonlocality however, it is clear that with increasingly small nanoparticles, and at increasingly short separations, the theory presented here is likely to break down due to these effects.

\subsection{The Van der Waals Potential Profile}

The ability to obtain accurate estimates of Van der Waals forces is encumbered by the fact that the Hamaker coefficient depends strongly on the frequency-dependent dielectric functions of the materials comprising the system. In particular, for noble metals, it is often the case that spectral data are incomplete\cite{Parsegian, 1981 Weiss}, and can show substantive variation based on the conditions under which the data were obtained\cite{2012 Olmon}. The Hamaker coefficient is computed over the full range of frequencies, and as such, the unavailability of consistent data over the entire spectral range means that interpolation is necessary between different experiments. As an illustrative example, figure 3.7(a), taken from the work of Olmon et al.\cite{2012 Olmon}, illustrates the fact that studies are often undertaken in different spectral ranges.

\begin{figure}[h]
    \centering
    \begin{subfigure}[h]{0.485\textwidth}
        \includegraphics[width=\textwidth]{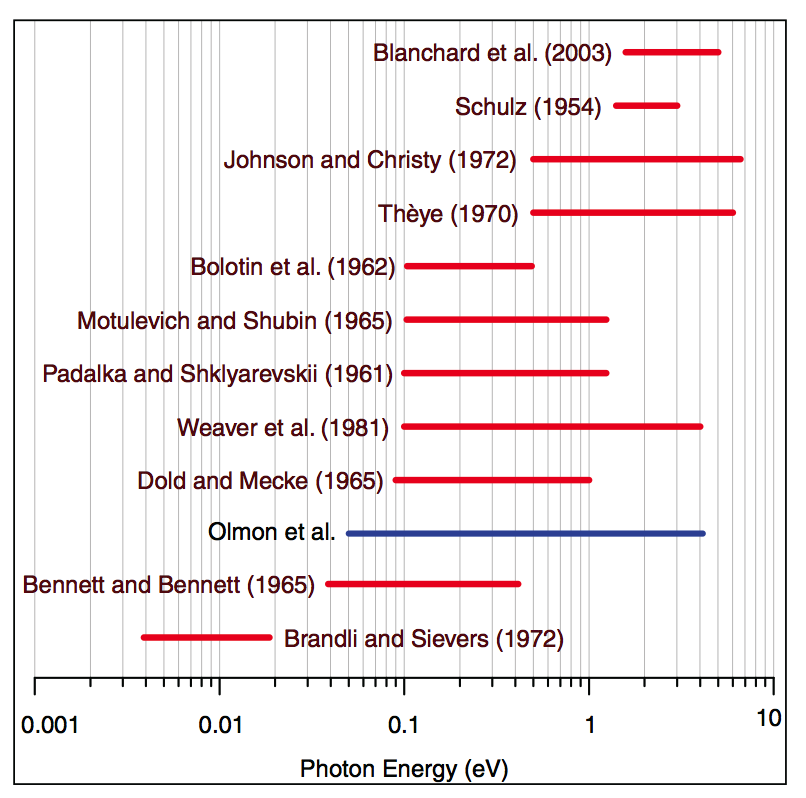}
        \caption{Spectral Ranges of the Available Dielectric Data for Gold.}
        \label{fig: Spectral Ranges of the Available Dielectric Data for Gold}
    \end{subfigure}
    ~ 
    \begin{subfigure}[h]{0.485\textwidth}
        \includegraphics[width=\textwidth]{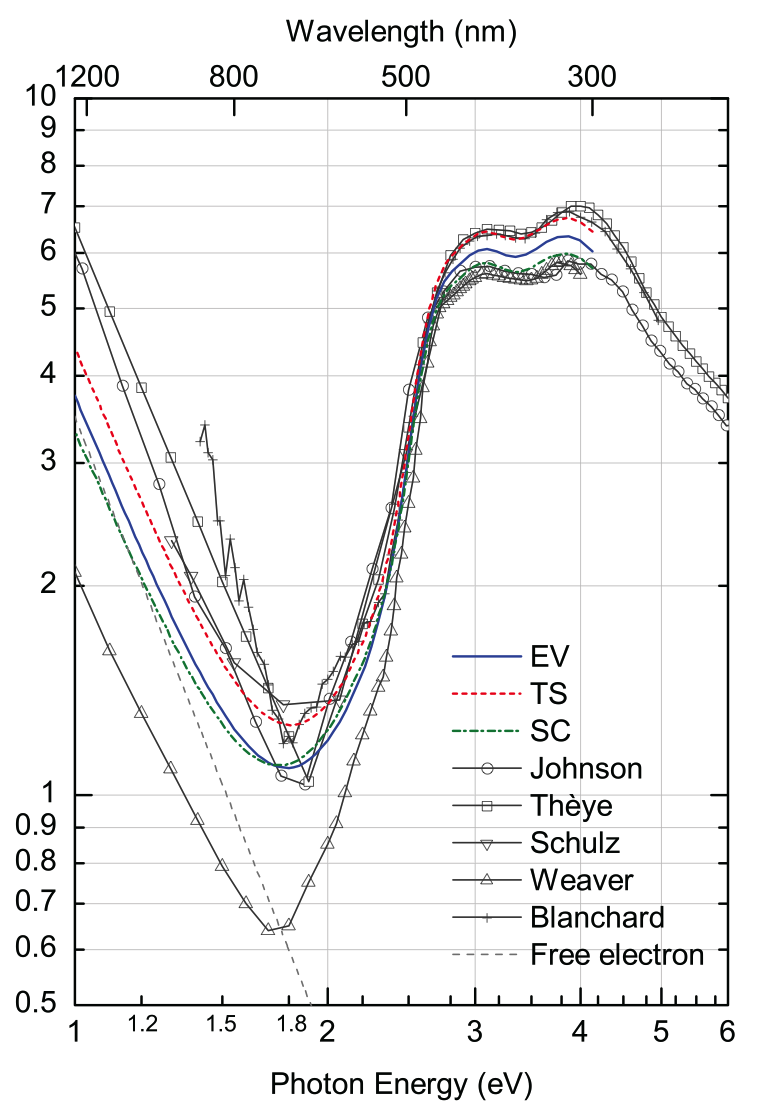}
        \caption{Variation in the Quality of Dielectric Data for Gold.}
        \label{fig: Variation in the Quality of Dielectric Data for Gold}
    \end{subfigure}
    \caption{Figures taken from the work of Olmon et al.\protect\cite{2012 Olmon}}
    \label{fig: Figures taken from the work of Olmon et al.}
\end{figure}

Additionally, it has been noted that the literature pertaining to the calculation of the dielectric function of gold in particular, has been plagued by systematic errors \cite{2012 Olmon}. As such, the unreliability of the data will impart an uncertainty to any Hamaker coefficient computed from it. One can see from the form of the equation describing the Hamaker coefficient,

\begin{align} \label{equation 5}
A_{1m/2m} \approx \frac{3kT}{2} \sum\limits_{n=0} ^{\infty} {}^{'} \: \frac{\varepsilon_{Gold} - \varepsilon_{Water}}{\varepsilon_{Gold} + \varepsilon_{Water}} \frac{\varepsilon_{Gold} - \varepsilon_{Water}}{\varepsilon_{Gold} + \varepsilon_{Water}}\:,
\end{align}

\noindent that it is strongly dependent on, what are in essence, the differences over sums of the experimental absorption spectra of the materials, the $\varepsilon(\omega)$s. Over the whole frequency range, any small discrepancy between the experimental conditions under which the spectra are obtained will be magnified, resulting in a large uncertainty in the calculated value of the Hamaker coefficient. This will ultimately give rise to inaccurate predictions of the Van der Waals force. An illustrative example of the variation in dielectric spectra inherent when experiments are carried out using different samples, and under different conditions is provided in figure 3.7(b), again taken from Olmon et al.\cite{2012 Olmon}. 


Barring these limitations, the bounds for estimates of the Hamaker coefficient of the system considered here, two gold nanoparticles separated by a medium of water are in the range of 90-300 zJ \cite{Parsegian, 1981 Weiss}, where zJ is the unit of the zeptojoule ($10^{-21}$ J). Translating these numbers into the units of kT, this corresponds to a range of approximately 22-74 kT. There would appear to be considerable confusion in the literature as to the origin of the inability to accurately predict Hamaker coefficients. Indeed, it has been fallaciously suggested \cite{2012 Pinchuk}, that the uncertainty arises out of a lack of consideration for the size dependence of the Hamaker coefficient. As has been discussed previously, the size effect has no bearing on the magnitude of the Hamaker coefficient at sizes of the nanoparticle diameter larger than 10 nm, and as such cannot be seen to be responsible for the margin of error. \par





Figures 3.8 and 3.9 correspond to two separate fits to the spectral data of gold by DESY\cite{1974 DESY} and Johnson and Christy\cite{1972 Johnson and Christy}. A third fit to the spectral data of Irani\cite{1971 Irani} was found to produce similar values of the Hamaker coefficient to Johnson and Christy and so is not shown for clarity. The spectral data of water is given by Parsegian \cite{Parsegian} and remains the same throughout. The parameters were computed by Parsegian and Weiss\cite{1981 Weiss}, who also provide tabulated values of the Hamaker coefficients for the gold-water-gold system in units of ergs. These values have been reproduced using the Hamaker hybrid procedure described by Parsegian \cite{Parsegian}, and incorporated with the geometrical factor featured in equation 3.2. The presence of electrolyte has been accounted for, and for further details on this subject, the reader is referred to section A.5 of the supporting information. The results are plotted as function of surface to surface separation for different values of the nanoparticle radius. The intention with these plots is to give some indication of the error inherent in the Van der Waals interaction profiles, that comes about as a result of the limitations on the accuracy of experimental spectral data. In a more practical fashion, the true value of the Van der Waals potential could be thought to lie somewhere between the bounds plotted. Alternatively, the Van der Waals potential could be thought of as being accurate only to an order of magnitude. \par

\begin{figure}[H]
    \centering
    \begin{subfigure}[h]{0.485\textwidth}
        \includegraphics[width=\textwidth]{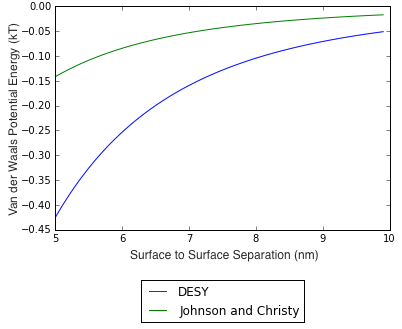}
        \caption{5 nm}
        \label{fig:5 nm}
    \end{subfigure}
    ~ 
    \begin{subfigure}[h]{0.485\textwidth}
        \includegraphics[width=\textwidth]{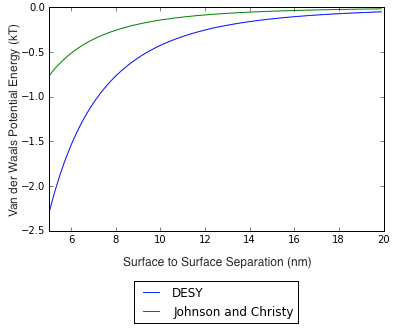}
        \caption{10 nm}
        \label{fig:10 nm}
    \end{subfigure}
    \caption{Van der Waals Interaction Profile between Two Nanoparticles of Radius (a) 5 nm, and (b) 10 nm, starting from 5 nm Surface to Surface Separation.}
 \label{fig: Van der Waals Interaction Profile between Two Nanoparticles of Radius (a) 5 nm, and (b) 10 nm, starting from 5 nm Surface to Surface Separation.}
\end{figure}

\begin{figure}[H]
    \centering
    \begin{subfigure}[h]{0.485\textwidth}
        \includegraphics[width=\textwidth]{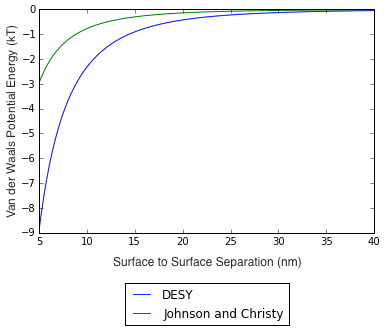}
        \caption{20 nm}
        \label{fig:20 nm}
    \end{subfigure}
    ~ 
    \begin{subfigure}[h]{0.485\textwidth}
        \includegraphics[width=\textwidth]{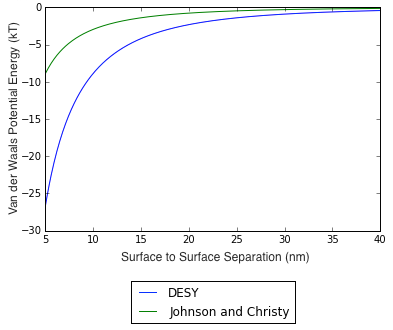}
        \caption{40 nm}
        \label{fig:40 nm}
    \end{subfigure}
    \caption{Van der Waals Interaction Profile between Two Nanoparticles of Radius (a) 20 nm, and (b) 40 nm, starting from 5 nm Surface to Surface Separation.}
    \label{fig: Van der Waals Interaction Profile between Two Nanoparticles of Radius (a) 20 nm, and (b) 40 nm, starting from 5 nm Surface to Surface Separation.}
\end{figure}



\clearpage

Distinct from the size-dependent Hamaker coefficient, the dependence of the Van der Waals profile on the size of the nanoparticles, using a size-independent Hamaker coefficient, is plotted in figure 3.10, using only the DESY data for clarity.


\begin{figure}[h]
\centering
\includegraphics[scale=0.75]{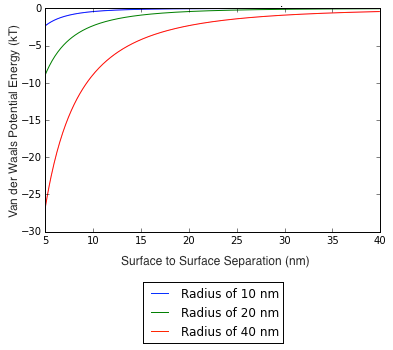}
\caption{Identical Gold nanoparticles - size comparison using DESY data.}
\label{fig:Identical Gold nanoparticles - size comparison using DESY data.}
\end{figure}

\clearpage

\section{The Coulomb Force}

The Coulombic interaction energy between two nanoparticles in electrolytic solution may be represented as

\begin{align} \label{equation 6}
U(L) = \left(\frac{R}{2}\right)Ze^{-\kappa L}
\end{align}

\noindent when the nanoparticles are of equal radius \cite{Israelachvili}, where $Z$ is given by

\begin{align} \label{equation 7}
Z = 64\pi\varepsilon_0\varepsilon\left(\frac{kT}{e}\right)^2tanh^2\left(\frac{ze\psi_0}{4kT}\right) \:.
\end{align}

\noindent In a similar fashion to the Hamaker coefficient, $Z$, the interaction constant, aside from the valency of the electrolyte, depends exclusively on the properties of the nanoparticle surfaces. Complementary to this, the inverse Debye length $\kappa$ depends only on the properties of the solution \cite{Israelachvili}. It was decided  to manipulate equation 3.6 above in order to express the interaction energy explicitly in terms of the surface charge density $\sigma$, as opposed to the surface potential $\Psi_0$, where

\begin{align} \label{equation 8}
\sigma = \frac{Be}{4\pi R^2} \:.
\end{align}

\noindent Here $e$ is the elementary charge of an electron, $R$ the radius of the nanoparticle and $B$ a constant accounting for the number of surface charges. The details of these manipulations are available in section A.1 of the supporting information. Experimentally, the surface charge density is easily obtained through the ratio of nanoparticles to charge-bearing ligands used to make up the solution. As such, a function of surface charge density would be more readily integrable with experimental work. The interaction energy was plotted, for two nanoparticles having a fixed radius of 20 nm, for varying values of:

\begin{enumerate}
\item The electrolyte concentration
\item The number of surface charges
\end{enumerate}

\noindent The electrolyte is a 1:1 electrolyte such as NaCL in all cases. In figure 3.11, depicting the effect of increasing electrolyte concentration, it may be seen that already at a concentration of 0.1 M, the repulsive interaction is practically entirely screened at separations of 5 nm or more, and will contribute nothing to the fight against the attractive Van der Waals force. On the other hand, an electrolyte concentration of 0.001 M results in very steep repulsive branches at 5 nm separations, and it is likely to be the case that at such concentrations, the nanoparticles, although stabilised, will also be repelled from the electrode at the SLI.

\begin{figure}[H]
\centering
\includegraphics[scale=0.35]{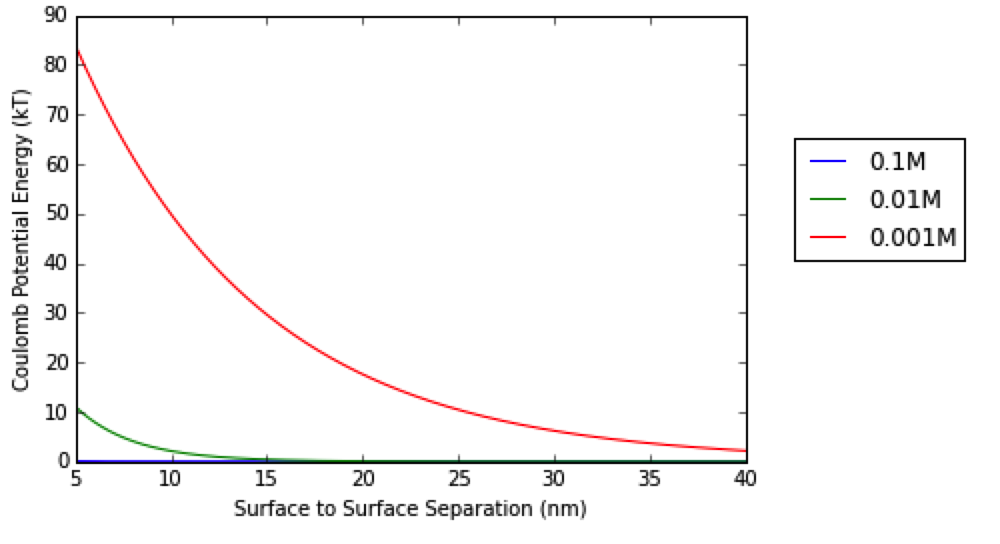}
\caption{Coulomb Potential Energy as a Function of Surface to Surface Separation for Varying Electrolyte Concentrations. Radius Fixed at 20 nm, 500 Charges per Nanoparticle.}
\label{fig: Coulomb Potential Energy as a Function of Surface to Surface Separation for Varying Electrolyte Concentrations. Radius Fixed at 20 nm, 500 Charges per Nanoparticle.}
\end{figure}

Secondly, the energy was plotted for different values of the number of surface charges per nanoparticle. The radius of the nanoparticle was again fixed at 20 nm, and the electrolyte concentration fixed at 0.01 M. The value of the latter, 0.01 M, was taken to be the most appropriate given the balance of attractive and repulsive forces required both to stabilise the nanoparticles, and allow them to approach the electrode. Drawing a comparison to the analogous figure for the Van der Waals force, figure 3.9(a), one may note that 1000 surface charges would seem to be the minimum value required to stabilise nanoparticles at 5 nm separations assuming the upper limit of the DESY Hamaker coefficient. We will see this more quantitatively in the following section, where the Van der Waals and Coulomb profiles will be combined.

\begin{figure}[H]
\centering
\includegraphics[scale=0.35]{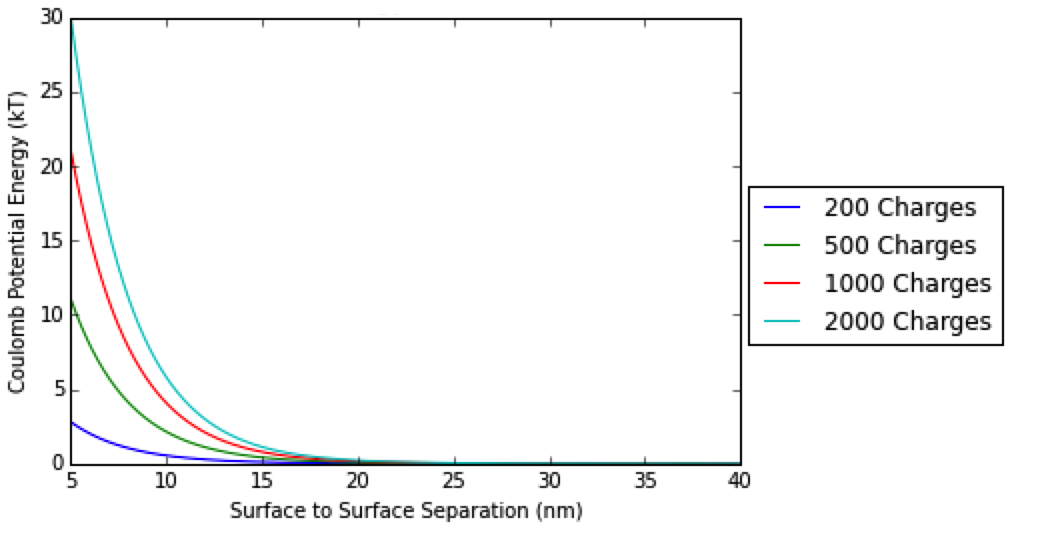}
\caption{Coulomb Potential Energy as a Function of Surface to Surface Separation for Varying Numbers of Surface Charges per Nanoparticle. Radius Fixed at 20 nm, Electrolyte Concentration of 0.01 M.}
\label{fig: Coulomb Potential Energy as a Function of Surface to Surface Separation for Varying Numbers of Surface Charges per Nanoparticle. Radius Fixed at 20 nm, Electrolyte Concentration of 0.01 M.}
\end{figure}

\section{The Total Interaction Energy}

In this section, the Van der Waals and Coulomb profiles will be combined in an attempt to estimate the optimal parameter values required to stabilise the nanoparticles in the bulk. Reiterating the points previously made, these parameters correspond to the largest number of charges i.e. ligand to nanoparticle ratio, and the largest electrolyte concentration, that will be effective in stabilising the nanoparticles, whilst simultaneously allowing them to approach the electrode. In all cases, the experimental error responsible for the uncertainty in the value of the Hamaker coefficient will be accounted for. Concretely, this will be accomplished by designating the Hamaker coefficient computed from the fit to the Johnson and Christy data as the lower bound on the strength of attraction, and the DESY Hamaker coefficient as the upper bound. From figures 3.13(a), and 3.13(b), showing the dependence of the interaction potential on the electrolyte concentration, it may be observed that electrolyte concentrations in the neighbourhood of 0.01 M would seem to represent the best compromise between stability and attraction to the interface. In both figures, it is clear that at concentrations of 0.1 M and higher, nanoparticles will become unstable, and aggregation will occur in the bulk. As such, the value of 0.01 M will be taken as the fixed electrolyte concentration when the variation of other parameters is investigated.



\begin{figure}[H]
    \centering
    \begin{subfigure}[h]{0.485\textwidth}
        \includegraphics[width=\textwidth]{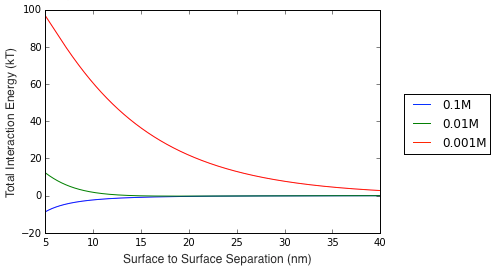}
        \caption{Total Interaction Energy Profile (DESY).}
        \label{fig: Total Interaction Energy Profile (DESY).}
    \end{subfigure}
    ~ 
    \begin{subfigure}[h]{0.487\textwidth}
        \includegraphics[width=\textwidth]{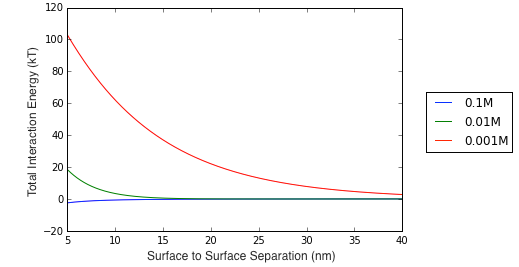}
        \caption{Total Interaction Energy Profile (JC).}
        \label{fig: Total Interaction Energy Profile (JC).}
    \end{subfigure}
    \caption{Total Interaction Energy as a Function of Surface to Surface Separation for (a) The DESY Hamaker Coefficient and (b) The Johnson and Christy (JC) Hamaker Coefficient. Radius Fixed at 20 nm, 1000 Charges per Nanoparticle.}
    \label{fig: Total Interaction Energy as a Function of Surface to Surface Separation for (a) The DESY Hamaker Coefficient and (b) The Johnson and Christy (JC) Hamaker Coefficient. Radius Fixed at 20 nm, 1000 Charges per Nanoparticle.}
\end{figure}


Secondly, when the number of surface charges is varied at constant electrolyte concentration, one may observe from figures 3.14(a) and 3.14(b), that agglomeration occurs for 200 surface charges in both the DESY, and Johnson and Christy cases. This corresponds to a ligand to nanoparticle ratio of 200:1. In contrast, for a ratio of 500:1, the nanoparticles would appear to be stable. As such, the optimum parameters would seem to involve electrolyte concentrations in the vicinity of 0.01M in conjunction with ligand to nanoparticle ratios of 500:1. The efficacy of these parameter values will be explored quantitatively in the following section when the electrode constituting the SLI is introduced into the picture.\par



\begin{figure}[H]
    \centering
    \begin{subfigure}[h]{0.485\textwidth}
        \includegraphics[width=\textwidth]{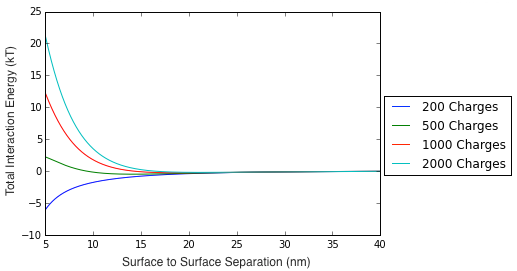}
        \caption{Total Interaction Energy Profile (DESY).}
        \label{fig: Total Interaction Energy Profile (DESY).}
    \end{subfigure}
    ~ 
    \begin{subfigure}[h]{0.487\textwidth}
        \includegraphics[width=\textwidth]{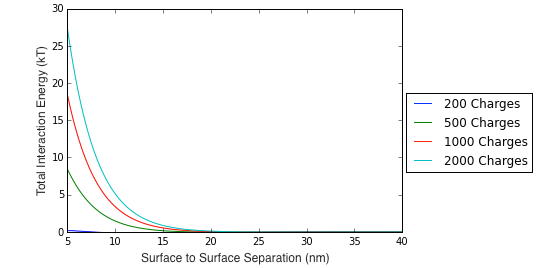}
        \caption{Total Interaction Energy Profile (JC).}
        \label{fig: Total Interaction Energy Profile (JC).}
    \end{subfigure}
    \caption{Total Interaction Energy as a Function of Surface to Surface Separation for (a) The DESY Hamaker Coefficient and (b) The Johnson and Christy (JC) Hamaker Coefficient. Radius Fixed at 20 nm, Electrolyte Concentration of 0.01 M.}
    \label{fig: Total Interaction Energy as a Function of Surface to Surface Separation for (a) The DESY Hamaker Coefficient and (b) The Johnson and Christy (JC) Hamaker Coefficient. Radius Fixed at 20 nm, Electrolyte Concentration of 0.01 M.}
\end{figure}

One further point to note is that the existence of a DLVO-type minimum can be observed in figures 3.14(a) and 3.14(b) when viewed sufficiently close up. Such a close-up view is illustrated in figure 3.15 for the case of 0.01 M electrolyte concentration, and a ligand to nanoparticle ratio of 1000:1. The depth of the minimum, -0.2 kT, will not be deep enough to induce nanoparticle flocculation.


\begin{figure}[h]
\centering
\includegraphics[scale=0.5]{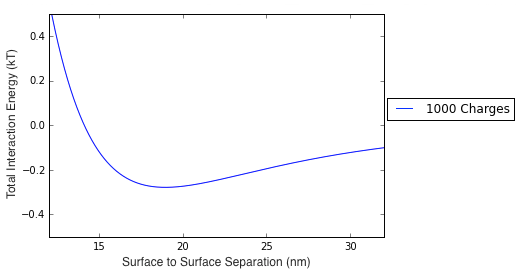}
\caption{Zoomed-in View of the Red Curve from Figure 3.14(a).}
\label{fig: Zoomed-in View of the Red Curve from Figure 3.14(a).}
\end{figure}

\chapter{A Single Nanoparticle at an Interface}

Now that the nanoparticles have been stabilised in the bulk, the next step towards achieving a functional electrovariable smart mirror is to bring the nanoparticles to the electrode. In principle, this may be accomplished simply through the polarisation of the electrode. One of the key challenges remains to hold the nanoparticles in place once they arrive at the SLI, and indeed the forces responsible for this are in need of investigation \cite{2016 Edel}. The purpose of this section will be to examine the forces at play between a single nanoparticle at the electrode, the system being illustrated in figure 4.1, where the electrode is depicted as being infinite in the lateral coordinate.

\begin{figure}[H]
\centering
\includegraphics[scale=0.45]{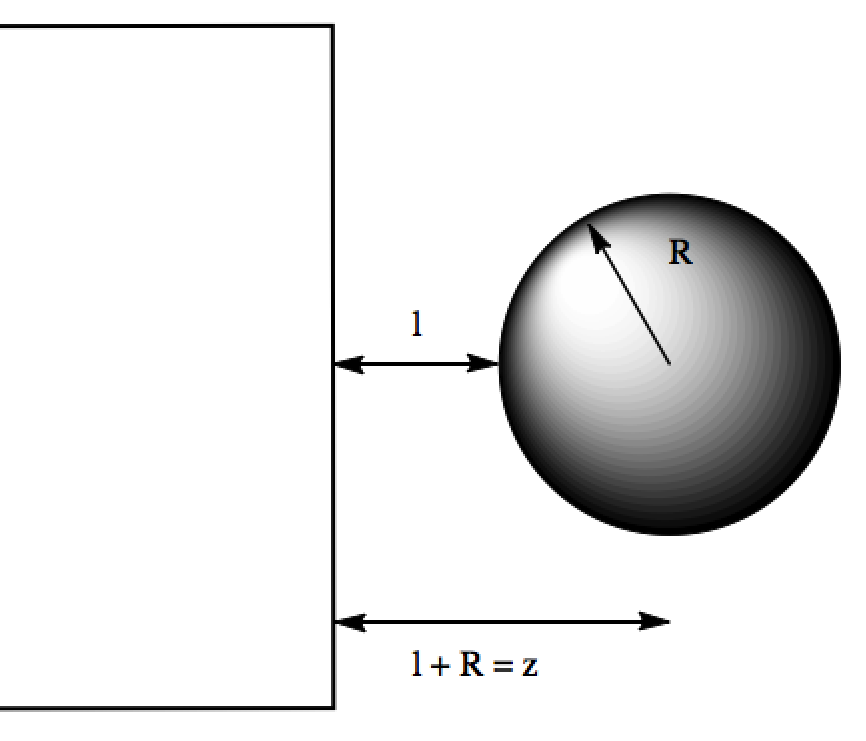}
\caption{System Geometry.}
\label{fig: System Geometry.}
\end{figure}





In a similar fashion to nanoparticles in the bulk solution, there will be a Van der Waals interaction between the electrode and the nanoparticle. In addition to the Van der Waals interaction however, there will also be an image force between the nanoparticle and the electrode. The extent of this latter interaction will be dependent on the material that is used for the electrode. With respect to mirror applications, both gold and Indium Tin Oxide (ITO) electrodes have been identified as promising materials with which to achieve electrovariability \cite{2016 Edel}. As such, when the forces governing the assembly process are examined in this section, they will be considered for both gold and ITO as the electrode material. The interactions will be examined with respect to:


\begin{enumerate}
\item The Van der Waals Potential Profile.
\item The Image Potential Profile.
\end{enumerate}

\section{The Van der Waals Potential Profile}

The computation of the Van der Waals force between nanoparticle and electrode will proceed analogously to that between nanoparticles. The interaction energy for the system,

\begin{align} \label{equation 1}
G_{sp}(l; R) = - \frac{A_{1m/2m}}{6}\left[\frac{R}{l} + \frac{R}{2R + l} +  \ln \left( \frac{l}{2R + l}\right) \right] \:,
\end{align}

\noindent comprises again, a Hamaker coefficient and a geometric factor. The potential for different electrode materials will be investigated in the order of:

\begin{enumerate}
\item Gold Electrodes.
\item ITO Electrodes.
\end{enumerate} 




\subsection{Gold Electrodes}

The range of values of the Hamaker coefficient for this system will remain the same as that between gold nanoparticles in the bulk. Again, the dielectric functions of both water and gold will be given by the damped harmonic oscillator model;

\begin{align} \label{equation 2}
\varepsilon(\omega) = 1  + \sum\limits_{k=1} ^n \frac{d_k}{1 - i\omega\tau_k} + \sum\limits_{j=1} ^n \frac{f_j}{\omega^2_j - \omega^2 - i\omega g_j} \:.
\end{align}

\noindent This representation may seem a poor choice in light of seemingly more accurate Drude-Lorentz type fits to more complete spectral data. It has been noted by Parsegian however, that it is often better to use the same form of approximation for the dielectric function of the materials comprising the system wherever possible \cite{Parsegian}. As a case in point, the most complete representation of the dielectric function of gold currently, is likely to be that offered by Pendry \cite{2014 Pendry},

\begin{align} \label{equation 3}
\varepsilon_{Gold}(\omega) = 1 - \frac{\omega_p^2}{\omega^2 + i\omega_{\tau}\omega} - \sum_{j=1}^{10} \frac{f_j\omega_j^2}{ \omega^2 - \omega_j^2  + i\gamma_j\omega}\:,
\end{align}

\noindent fitted to the joint data of both Palik \cite{Palik Quote} and Olmon\cite{2012 Olmon} across a broad spectral range. The use of such a dielectric representation for gold however, would not be compatible with the quality of the dielectric function available for water. The best offering for the latter is generally considered to be that of Dagastine et al. \cite{2000 Dagastine}, with the function being derived from an interpolated fit to incomplete spectral data. Looking at the expression for the Hamaker coefficient once again,

\begin{align} \label{equation 4}
A_{1m/2m} \approx \frac{3kT}{2} \sum\limits_{n=0} ^{\infty} {}^{'} \: \frac{\varepsilon_{Gold} - \varepsilon_{Water}}{\varepsilon_{Gold} + \varepsilon_{Water}} \frac{\varepsilon_{Gold} - \varepsilon_{Water}}{\varepsilon_{Gold} + \varepsilon_{Water}}\:,
\end{align}

\noindent it makes little sense to integrate a fit to high-quality spectral data for gold with an interpolated fit to incomplete spectral data for water. The asymmetry in the forms of approximation used will result in a larger error in the calculated value of the coefficient relative to coarser, yet symmetric approximations. As such, the damped oscillator model for gold and water can at least offer this symmetry. Until higher quality spectral data for materials become available, this approach would appear to represent the best means of attaining accurate estimates of Van der Waals forces.\par

One rather unimportant aspect to consider for Hamaker coefficient computation is the quality of the fit to spectral data. Roth and Lenhoff \cite{1996 Roth} offer an improved fit to the spectral data of Heller et al. \cite{1974 Heller} relative to that offered by Parsegian \cite{Parsegian}. This is illustrated graphically in figure 4.2.

\begin{figure}[H]
\centering
\includegraphics[scale=0.225]{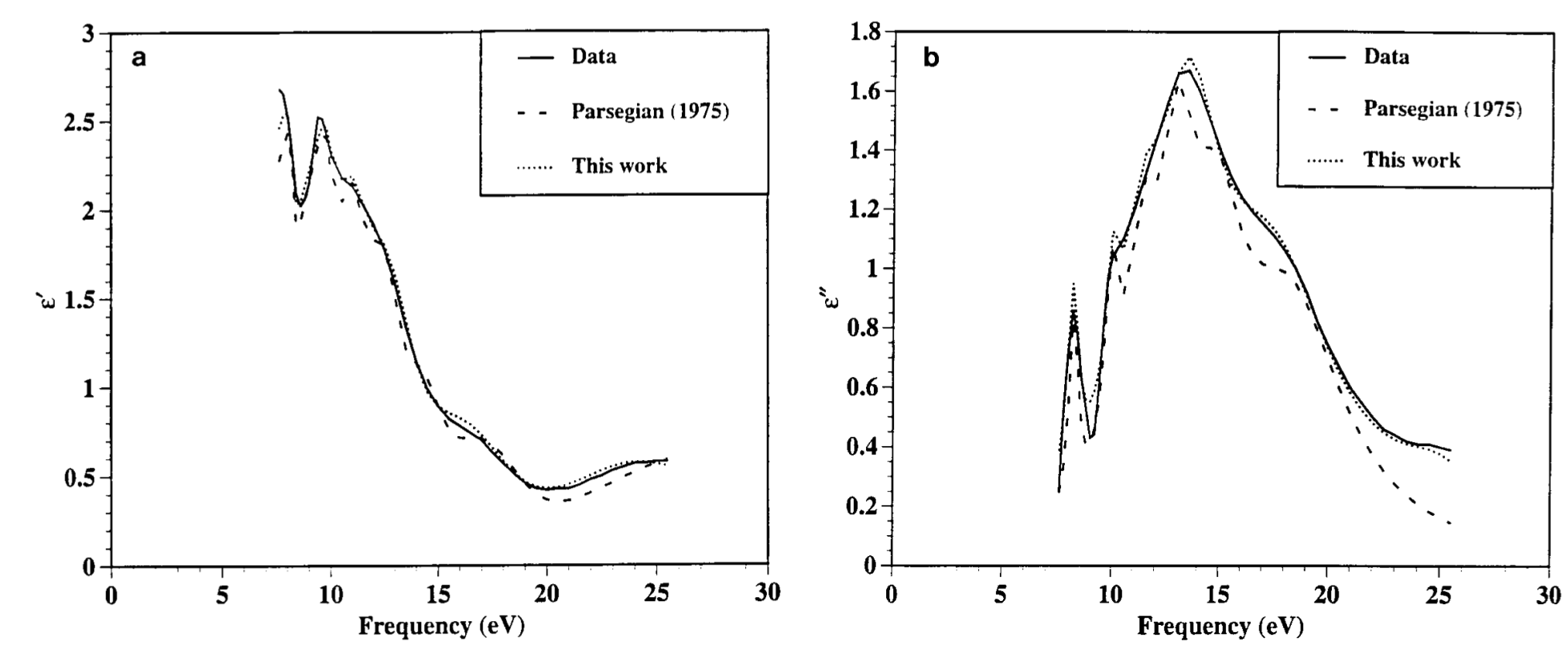}
\caption{Fits by Parsegian \cite{Parsegian}, and Roth and Lenhoff \cite{1996 Roth} to the spectral data of Heller et al. \cite{1974 Heller}. Figure taken from Roth and Lenhoff \cite{1996 Roth}.}
\label{fig: Fit to the Spectral Data of Water.}
\end{figure}

\noindent Quantitatively, the difference in the value of the calculated Hamaker coefficient incurred from the use of the alternative parameter set for water is on the order of 1 kT. In contrast, the uncertainty in the underlying spectral data for gold contributes a difference of ca. 50 kT for the gold-water-gold system. While it is indisputable that Roth and Lenhoff offer a better fit to the Heller data than Parsegian, this is an unimportant detail in the context of estimating the Van der Waals interaction. The process of overfitting data possessing a high degree of uncertainty should not yield a quantitative improvement in force estimation. The sets of parameters used in the damped oscillator models of gold and water are hence taken from Parsegian \cite{Parsegian}.\par

Figures 4.3 and 4.4 illustrate the Van der Waals potential  plotted for a range of nanoparticle sizes, as a function of separation from the interface. Once again, the area between the DESY, and Johnson and Christy curves represents the region of approximate energy magnitude based on the limitations of the spectral data for gold.


\begin{figure}[H]
    \centering
    \begin{subfigure}[h]{0.485\textwidth}
        \includegraphics[width=\textwidth]{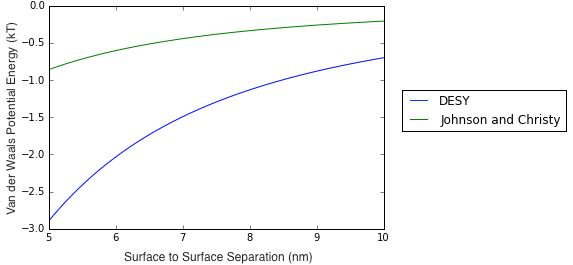}
        \caption{5 nm}
        \label{fig: 5 nm}
    \end{subfigure}
    ~ 
    \begin{subfigure}[h]{0.485\textwidth}
        \includegraphics[width=\textwidth]{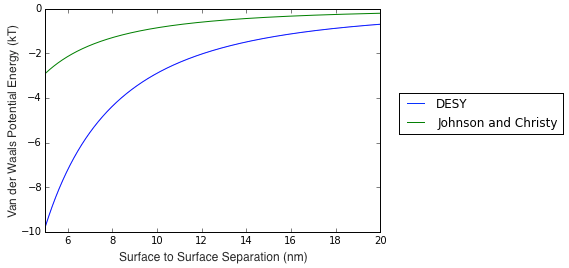}
        \caption{10 nm}
        \label{fig: 10 nm}
    \end{subfigure}
    \caption{Van der Waals Interaction Profile between a Gold Electrode and a Gold Nanoparticle of Radius (a) 5 nm, and (b) 10 nm, starting from 5 nm Surface to Surface Separation.}
    \label{fig: Van der Waals Interaction Profile between a Gold Electrode and a Gold Nanoparticle of Radius (a) 5 nm, and (b) 10 nm, starting from 5 nm Surface to Surface Separation.}
\end{figure}

\begin{figure}[H]
    \centering
    \begin{subfigure}[!ht]{0.485\textwidth}
        \includegraphics[width=\textwidth]{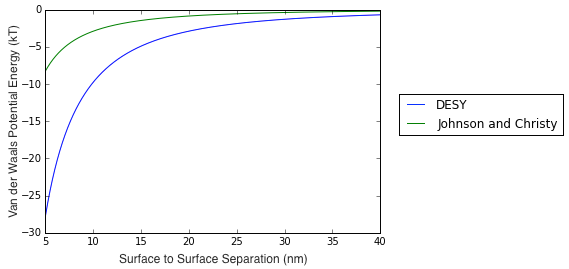}
        \caption{20 nm}
        \label{fig: 20 nm}
    \end{subfigure}
    ~ 
    \begin{subfigure}[!ht]{0.485\textwidth}
        \includegraphics[width=\textwidth]{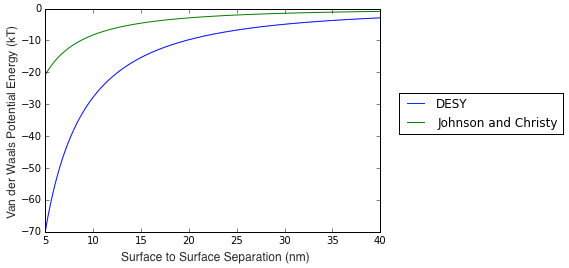}
        \caption{40 nm}
        \label{fig: 40 nm}
    \end{subfigure}
    \caption{Van der Waals Interaction Profile between a Gold Electrode and a Gold Nanoparticle of Radius (a) 20 nm, and (b) 40 nm, starting from 5 nm Surface to Surface Separation.}
    \label{fig: Van der Waals Interaction Profile between a Gold Electrode and a Gold Nanoparticle of Radius (a) 20 nm, and (b) 40 nm, starting from 5 nm Surface to Surface Separation.}
\end{figure}

\noindent As may be seen, when compared to the corresponding plots of the bulk interaction, figures 3.8 and 3.9, gold nanoparticles experience a stronger attraction to the electrode than to each other. Intuitively, this is to be expected from Hamaker's original formulation of pairwise additivity; there are more electrons in the electrode and hence there will be a stronger interaction.

\clearpage

\subsection{ITO Electrodes}

The Hamaker coefficient for the ITO-water-gold system was computed by representing the dielectric function of ITO in Drude-Lorentz form as

\begin{align} \label{equation 6}
\varepsilon_{ITO}(\omega) = 1 - \frac{\omega_p^2}{\omega^2 + i\omega_{\tau}\omega} - \sum_{j=1}^{2} \frac{f_j\omega_j^2}{ \omega^2 - \omega_j^2  + i\gamma_j\omega}\:.
\end{align}

\noindent Such a form, with two Drude-Lorentz terms, was given by Losurdo et al. \cite{2002 Rizzoli}. The parameters of this model were adapted so as to fit the more recent data of Konig \cite{2014 Konig}, as shown in figure 4.5. The parameter values are available in section A.6 of the appendix.


\begin{figure}[H]
\centering
\includegraphics[scale=0.3]{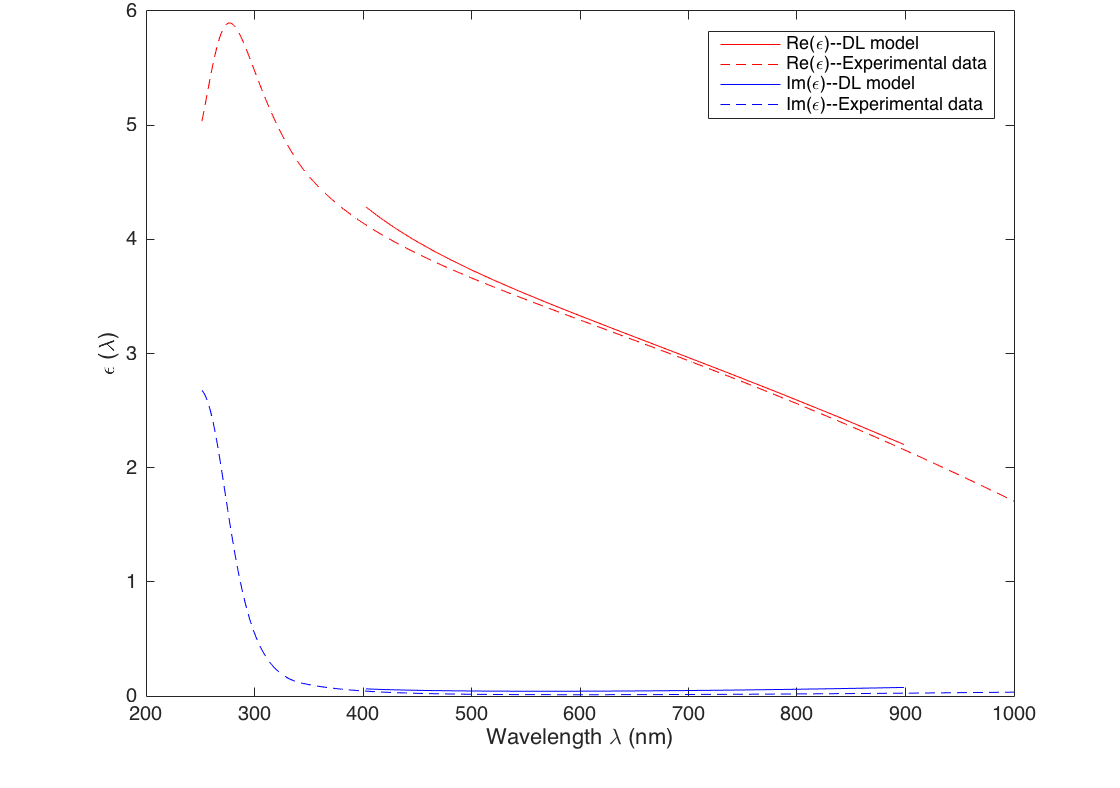}
\caption{Fit to the Spectral Data of Konig et al. \cite{2014 Konig}.}
\label{fig: Fit to the Spectral Data of Konig et al.}
\end{figure}

\noindent Ideally, as discussed in the previous section, it would be preferable to represent the dielectric functions of each material in the system in the same form. The added complexity of having three materials as opposed to two however, makes the task of finding one model to fit all three sets of experimental data more challenging. As a compromise, the damped oscillator models of water and gold were used in conjunction with the Drude-Lorentz model of ITO. Whilst such an approach cannot be expected to be quantitative to the nearest kT, it should be accurate to within an order of magnitude. Indeed, it is quite possible, as in the case with the representation of water, that the uncertainty in the experimental data of gold will again be the factor limiting the accuracy of the estimates. Therefore, the results have been treated as such, and, in figures 4.6 and 4.7, the DESY and Johnson and Christy curves once again mark the approximate range in which the interaction energy estimate lies.\par


\begin{figure}[H]
    \centering
    \begin{subfigure}[!ht]{0.485\textwidth}
        \includegraphics[width=\textwidth]{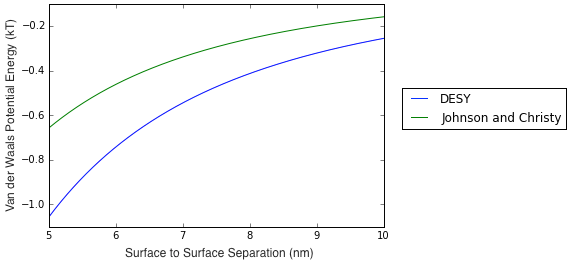}
        \caption{5 nm}
        \label{fig:5 nm}
    \end{subfigure}
    ~ 
    \begin{subfigure}[!ht]{0.485\textwidth}
        \includegraphics[width=\textwidth]{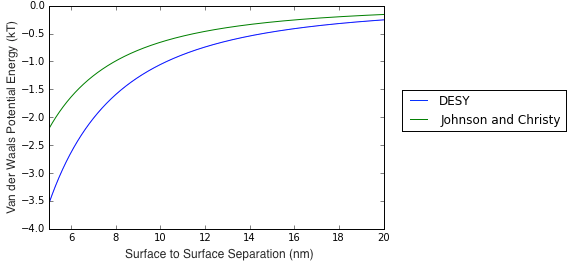}
        \caption{10 nm}
        \label{fig:10 nm}
    \end{subfigure}
    \caption{Van der Waals Interaction Profile between an ITO Electrode and a Gold Nanoparticle of Radius (a) 5 nm, and (b) 10 nm, starting from 5 nm Surface to Surface Separation.}
    \label{fig: Van der Waals Interaction Profile between an ITO Electrode and a Gold Nanoparticle of Radius (a) 5 nm, and (b) 10 nm, starting from 5 nm Surface to Surface Separation.}
\end{figure}

\begin{figure}[H]
    \centering
    \begin{subfigure}[!ht]{0.485\textwidth}
        \includegraphics[width=\textwidth]{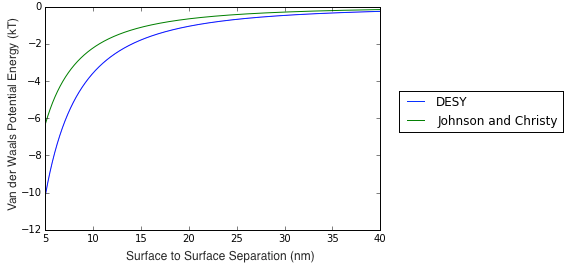}
        \caption{20 nm}
        \label{fig:20 nm}
    \end{subfigure}
    ~ 
    \begin{subfigure}[!ht]{0.485\textwidth}
        \includegraphics[width=\textwidth]{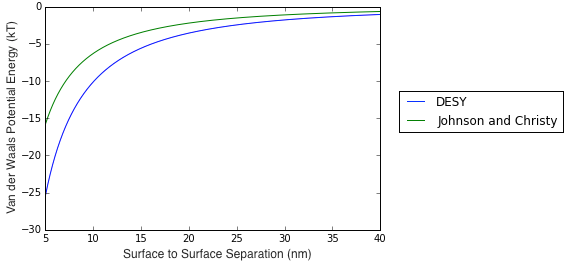}
        \caption{40 nm}
        \label{fig:40 nm}
    \end{subfigure}
    \caption{Van der Waals Interaction Profile between an ITO Electrode and a Gold Nanoparticle of Radius (a) 20 nm, and (b) 40 nm, starting from 5 nm Surface to Surface Separation.}
    \label{fig: Van der Waals Interaction Profile between an ITO Electrode and a Gold Nanoparticle of Radius (a) 20 nm, and (b) 40 nm, starting from 5 nm Surface to Surface Separation.}
\end{figure}

When compared to the gold electrode system, the switching-out of gold for ITO has led to a decrease in the magnitude of the Van der Waals interaction. A comparison is plotted for reference in figure 4.8.


\begin{figure}[H]
\centering
\includegraphics[scale=0.5]{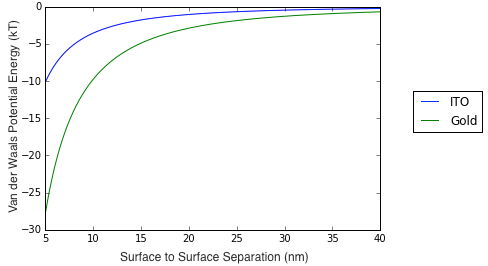}
\caption{ITO and Gold Comparison for NP radius = 20 nm. DESY results only for clarity.}
\label{fig:ITO and Gold Comparison}
\end{figure}

\noindent The Van der Waals interaction is not the only force with an attractive contribution at play at the SLI however. In order to fully appreciate the extent to which experimental parameters may be tuned to control nanoparticle array geometry, it will be necessary to consider the image contribution.

\section{The Image Potential Profile}

In an ideal metal, the field of an external point charge does not penetrate inside the body of the material. The physical origin of this behaviour is related to the plasma-like properties of the metal, the facility of the free electron gas to move its charges in such a manner as to cancel the effects of an external field. In other words, the dielectric permittivity goes to infinity. This malleability of the internal charge distribution of the metal gives rise to the classical law of attraction of the point charge to the metal surface, the impetus being for the point charge to have its field cancelled. In practice however, metals do not behave ideally. In real systems, electronic correlations give rise to finite field penetration into the body of the metal. The characteristic penetration depth is given by the Thomas-Fermi screening length.\par

This correction for real systems has been shown to lead to striking deviations from the classical image law at short distances from the interface between a dielectric and a plasma-like medium \cite{1977 Kornyshev}. The term image derives from the fact that the method of image charges is an oft-used process in electrostatics to solve problems where there is some constraint involving a boundary surface. By replacing the surface with an equivalent charge distribution, it is possible to simplify the problem. An example taken from Jackson's \textit{Classical Electrodynamics} \cite{Jackson} is provided in figure 4.9.

\begin{figure}[H]
\centering
\includegraphics[scale=0.3]{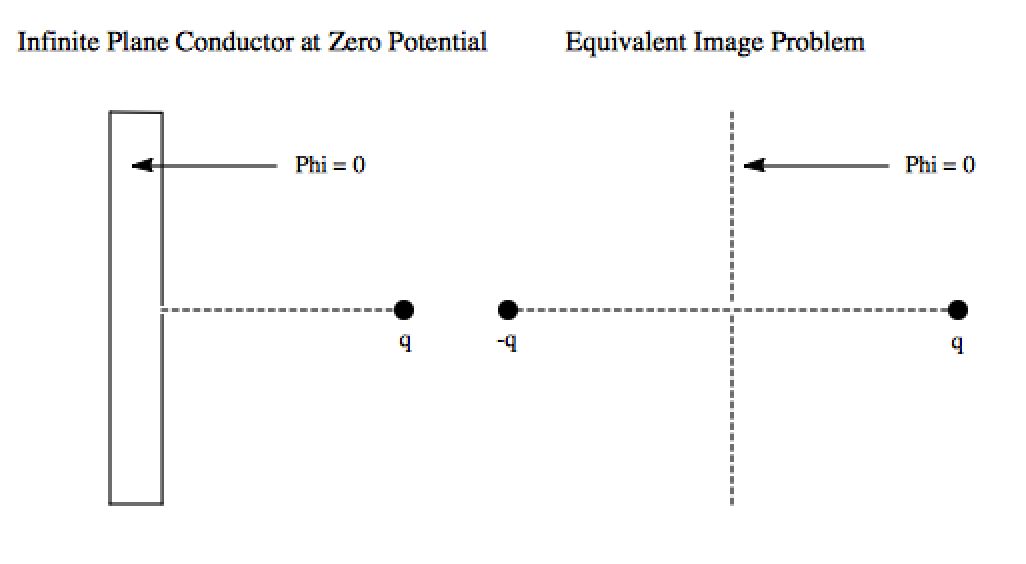}
\caption{Illustration of Method of Images.}
\label{fig:Illustration of Method of Images.}
\end{figure}


\noindent In this case, an infinite plane conductor has been replaced by an image charge of opposite sign to the external charge. One may note that the image problem on the right recreates the constraint of there being a potential of zero at the conductor boundary in the problem on the right. In this section, a new analytical result for the image potential created by a point charge in electrolytic solution near a plasma-like medium is presented. Whilst in itself, this is a valuable theoretical result, it only lays the foundations for the problem at hand, to estimate the image potential energy created by a nanoparticle. As such, the discussion will proceed as follows:

\begin{enumerate}
\item The Image Potential Energy of a Point Particle
\item The Image Potential Energy of a Nanoparticle
\end{enumerate}

\subsection{The Image Potential Energy of a Point Particle}

An illustration of the system is given in figure 4.10, where medium $\rom{1}$ represents the electrolyte and medium $\rom{2}$ represents the metal.

\begin{figure}[H]
\centering
\includegraphics[scale=0.3]{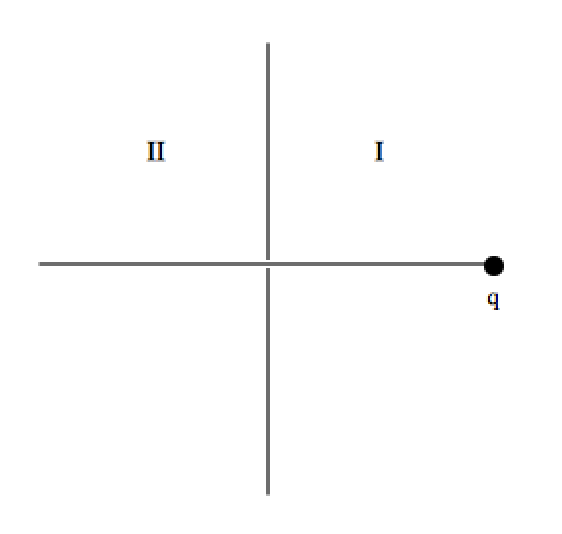}
\caption{Point Particle near the metal/electrolyte interface.}
\label{fig:Point Particle near the metal/electrolyte interface.}
\end{figure}

\noindent The analytical result obtained for the image potential energy,

\begin{align} \label{equation 7}
\frac{W(a)}{kT} = L_B\int_0^\infty dK \frac{\xi\sqrt{K^2 + \kappa_{\rom{1}}^2} - \sqrt{K^2 + \kappa_{\rom{2}}^2}}{\xi\sqrt{K^2 + \kappa_{\rom{1}}^2} +\sqrt{K^2 + \kappa_{\rom{2}}^2}}\frac{K}{\sqrt{K^2 + \kappa_{\rom{1}}^2}}\:e^{-2a\sqrt{K^2 + \kappa_{\rom{1}}^2}}\:,
\end{align}

\noindent as a function of the location of the point charge, $a$, is given in the dimensionless units of $kT$. $\kappa_{\rom{1}}$ and $\kappa_{\rom{2}}$ represent the Debye screening length in the electrolyte, and the Thomas-Fermi screening length in the metal respectively.

\begin{align}
\xi = \left(\frac{\varepsilon_{\rom{1}}}{\varepsilon_{\rom{2}}}\right)
\end{align}

\noindent is the ratio of the static dielectric constants of the metal and the solution, $\varepsilon_{\rom{1}}$ and $\varepsilon_{\rom{2}}$ respectively. The Bjerrum length,

\begin{align} \label{equation 8}
L_B = \frac{e^2}{\varepsilon kT}\:,
\end{align}

\noindent given in Gaussian units, represents the distance at which the electric potential energy between two point charges becomes comparable in magnitude to the thermal energy. It is dictated by the medium in which the point charges lie, and for water has a typical value of 7\AA. A more detailed diagram of the system in line with the notation of equation 4.6 is provided in figure 4.11.\par

\begin{figure}[!ht]
\centering
\includegraphics[scale=0.3]{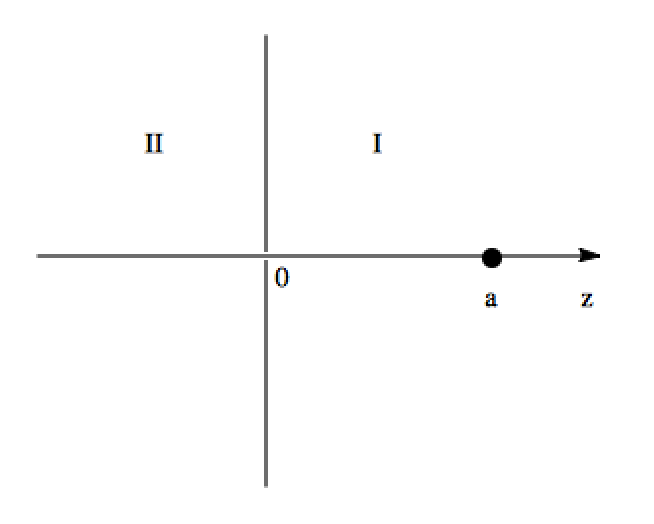}
\caption{Geometry of the System to accompany Equation 4.6.}
\label{fig:Geometry of the System to accompany Equation 4.6.}
\end{figure}




The image potential energy profile of the point charge will be plotted separately for the cases where gold and ITO occupy half-space $\rom{2}$ in figure 4.11. The origin of the difference in behaviour of the two systems will be determined by two parameters, the dielectric constant of the material and its Thomas-Fermi screening length. The specific values of these parameters for both materials will vary depending on the sample. As such, typical values for these parameters have been taken from the literature with the aim of highlighting the qualitative difference in behaviour between gold and ITO. These values are provided in table 4.1. The image potential energy profiles will be considered for both types of electrode in the following order:

\begin{table}[h]
\centering
\begin{tabular}{ |p{5.6cm}|p{2.8cm}|p{2.8cm}|  }
\hline
\multicolumn{3}{|c|}{Parameter Values} \\
\hline
Material     & Gold &ITO \\
\hline
Dielectric Constant -  $\varepsilon_{\rom{2}}$ & 7.0 \cite{2011 Hikita} & 3.62 \cite{2016 Moerland} \\
Thomas-Fermi Length - $\kappa_{\rom{2}}$ (\AA) & 0.5 \cite{2011 Hikita}   & 6.0 \cite{2007 Neumann, 2011 Melikyan} \\
\hline
\end{tabular}
\caption{Parameter Values for Gold and ITO.}
\label{table:1}
\end{table}

\begin{enumerate}
\item Gold Electrodes
\item ITO Electrodes
\end{enumerate}

\subsubsection{Gold Electrodes}

As may be seen in figure 4.12, the image potential energy profile for gold exhibits an adsorption minimum, of purely electrostatic origin, within 0.5 nm of the interface. The depth and location of this minimum will be important when the image potential energy of the nanoparticle is considered. Based on the estimates of electrolyte concentration required to stabilise nanoparticles in the bulk, it would appear that at electrolyte concentrations of 0.01 M, an adsorption minimum is already unavoidable. It will be important that the application of a voltage to the electrode can supply the energy required to desorb nanoparticles from this minimum and  hence achieve an electrovariable geometry. Additionally, the extent of the voltage that may be applied will be subject to the constraints imposed by the electrochemical window, the voltage range under which it is possible to operate without inducing a chemical reaction\cite{2016 Edel}. 


\begin{figure}[H]
\centering
\includegraphics[scale=0.7]{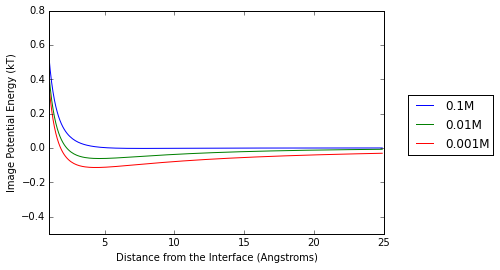}
\caption{Gold varying concentration.}
\label{fig:Gold varying concentration.}
\end{figure}

A comparison of the image potential energy profile of the gold electrode is compared with the classical image law of attraction to an ideal metal in figure 4.13. 


\begin{figure}[H]
\centering
\includegraphics[scale=0.7]{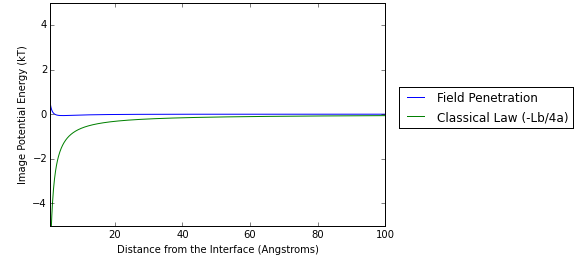}
\caption{A Comparison with the Classical Image Law.}
\label{fig: A Comparison with the Classical Image Law.}
\end{figure}

\noindent The effects of field penetration into the electrode may be observed to attenuate any attraction the point charge may have towards the interface. In addition, in the limit of close approach, a repulsive branch is created. At large distances, the image potential energy profile recovered is equivalent to the classical image law.

\subsubsection{ITO Electrodes}

The qualitative behaviour in the difference between gold and ITO may be observed in figure 4.14. The longer Thomas-Fermi screening length characteristic of ITO manifests itself in the absence of an adsorption minimum at short separations. This effect would seem to be invariant on the electrolyte concentration. Hence, it may already be inferred that there are fewer obstacles to electrovariability at the ITO interface relative to the gold interface. The decreased extent of Van der Waals attraction in addition to the non-existence of an attractive contribution to the image potential energy profile means that it should be easier to expel the nanoparticles with an applied voltage.


\begin{figure}[H]
\centering
\includegraphics[scale=0.7]{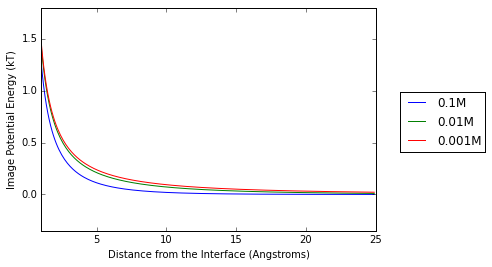}
\caption{ITO varying concentration.}
\label{fig:ITO varying concentration.}
\end{figure}

\noindent The image potential energy profiles of both Gold and ITO are compared in figure 4.15 for the electrolyte concentration of 0.01 M.


\begin{figure}[H]
\centering
\includegraphics[scale=0.7]{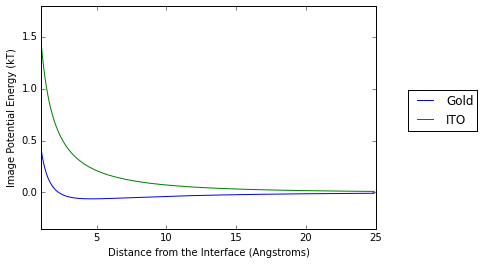}
\caption{Point Particle Result for ITO and Gold compared at 0.01M Concentration.}
\label{fig: Point Particle Result for ITO and Gold compared at 0.01M Concentration.}
\end{figure}

\subsection{The Image Potential Energy of a Nanoparticle}

While the image potential energy profile of a point particle may be used to elucidate some qualitative aspects of the differences between gold and ITO, it is not necessarily representative of the system behaviour when point particles and nanoparticles are interchanged. The smearing of charge over the surface of a sphere constitutes a different charge distribution entirely and one can easily imagine that at close separations, when the geometry of the sphere is readily detectable by the interface, that a different picture of the image potential energy will be observed. The approach presented here does not take into account the geometry of the sphere and is illustrated in figure 4.16.


\begin{figure}[H]
\centering
\includegraphics[scale=0.3]{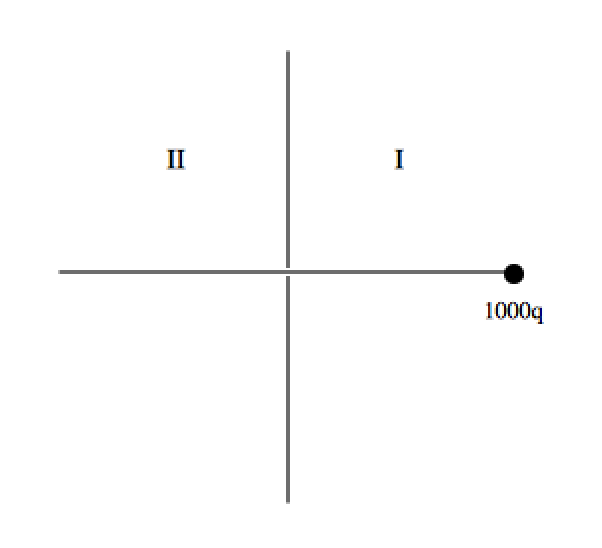}
\caption{Nanoparticle near an Interface.}
\label{fig:Nanoparticle near an Interface.}
\end{figure}


%

In this case, the sphere is treated as having its surface charges concentrated at a point. The attempt to model the nanoparticle as a sphere is discussed in the appendix. The coarse approach adopted here simply involves scaling the results of the previous section by the number of charges on the sphere's surface. The scaled results for each electrode will again be presented in the following order:

\begin{enumerate}
\item Gold Electrodes
\item ITO Electrodes
\end{enumerate}

\subsubsection{Gold Electrodes}

As may be seen in figure 4.17, the adsorption minimum created by the image potential for the scaled sphere is relatively deep but lies very close to the interface. It is doubtful that the nanoparticle will be able to approach this close given the size of the ligands. The influence of such steric considerations could have implications for electrovariability. If it is the case that the nanoparticle may avoid getting trapped in this minimum, it will be easier to push away with an applied voltage.


\begin{figure}[H]
\centering
\includegraphics[scale=0.7]{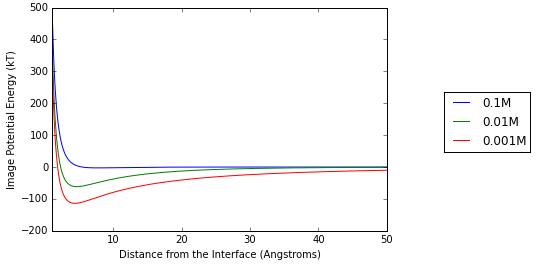}
\caption{Gold Electrode - Scaled Sphere in the Range of 1-50 Angstroms.}
\label{fig: Gold Electrode - Scaled Sphere in the Range of 1-50 Angstroms.}
\end{figure}

At larger distances the image potential energy would seem to be negligible for all values of the electrolyte concentration as may be seen in figure 4.18.

\begin{figure}[H]
\centering
\includegraphics[scale=0.5]{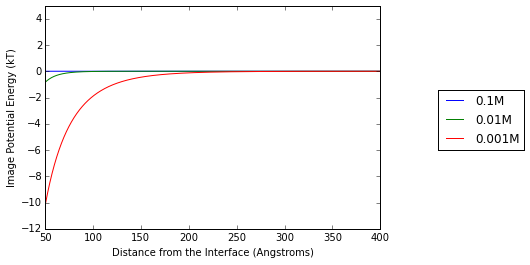}
\caption{Gold Electrode - Scaled Sphere in the Range of 50-400 Angstroms.}
\label{fig: Gold Electrode - Scaled Sphere in the Range of 50-400 Angstroms.}
\end{figure}

\subsubsection{ITO Electrodes}

At the ITO interface, the extent of repulsion at distances closer than 1 nm is tangible, as may be seen in figure 4.19. Such a feature is likely to remain unchanged in the case of a sphere model. As such, the point made previously regarding electrovariability being easier to achieve at the ITO interface is likely to remain valid.


\begin{figure}[H]
\centering
\includegraphics[scale=0.7]{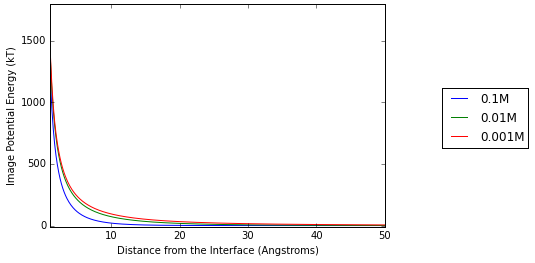}
\caption{ITO Electrode - Scaled Sphere in the Range of 1-50 Angstroms.}
\label{fig: ITO Electrode - Scaled Sphere in the Range of 1-50 Angstroms.}
\end{figure}

\noindent Likewise, comparably to the gold interface, the image potential stops becoming an important consideration at distances larger than 20 nm as may be seen in figure 4.20.

\begin{figure}[H]
\centering
\includegraphics[scale=0.7]{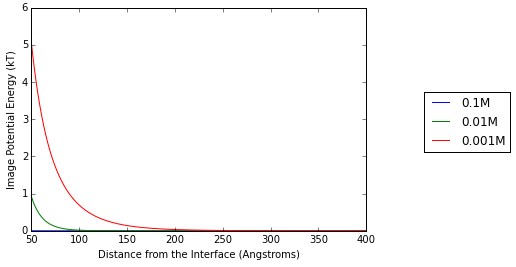}
\caption{ITO Electrode - Scaled Sphere in Range of 50-400 Angstroms.}
\label{fig: ITO Electrode - Scaled Sphere in Range of 50-400 Angstroms.}
\end{figure}

\noindent In terms of answering the question as to whether the combined effects of the Van der Waals, and image potential energies can represent a significant barrier to electrovariability, based on the estimates of this chapter, it would seem not. Even taking into consideration the constraints of the electrochemical window, it would appear that the magnitudes of the applied voltages possible are more than enough to bring about the reversible adsorption and desorption of nanoparticles from both gold and ITO interfaces. Though admittedly the scaling process to obtain the estimates for nanoparticles was a rather coarse one, the reader may see from the epilogue that progress towards accounting for the spherical geometry of the nanoparticle is underway.

\chapter{A Pair of Nanoparticles at an Interface}

Thus far, the two primary interactions in the system, the interaction between nanoparticles in the bulk as well as the interaction between nanoparticles and the electrode, have been treated separately. At the SLI however, it is clear that the interaction between nanoparticles will not remain the same as in the bulk solution. The presence of the metal electrode as a third body will be an important consideration, and dependent on the geometry of the system, may either enhance or suppress the forces between nanoparticles. This effect is illustrated graphically in figure 5.1.

\begin{figure}[H]
\centering
\includegraphics[scale=0.5]{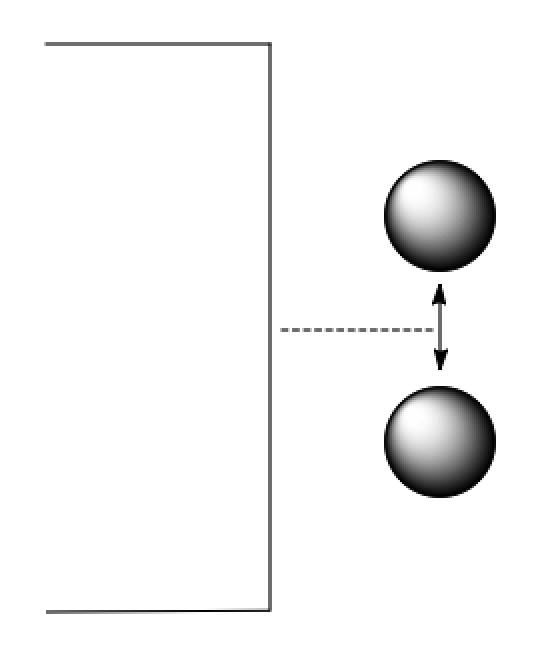}
\caption{Illustration of the Third Body Effect of the Electrode.}
\label{fig:Illustration of the Third Body Effect of the Electrode.}
\end{figure}

\noindent In this chapter, the contribution imparted by the electrode will be considered independently for the Van der Waals and image forces in the following order:



\begin{enumerate}
\item Van der Waals Interaction Energy
\item Image Potential Energy
\end{enumerate}

It must be said at the outset that there is a gulf in the current theoretical understanding of third body effects in nanoparticle systems with respect to these two forces. As such, the ability to construct a cohesive model of the total interaction potential will be hampered. Subject to these limitations, the goal in this section will be to develop this theory as far as is possible in the case of each constituent force. In the process, it is hoped that recommendations can be made about how to overcome this mismatch of theoretical understanding for the practical application of experiment.

\section{Van der Waals Interaction}


The influence of a planar third body on the Van der Waals forces between nanoparticles would not appear to be an effect thus far documented in the literature. As discussed previously, this can most likely be attributed to the complications involved in constructing a theory at the nanoscale, where the size and separation of the particles is comparable and hence the use of simplifying asymptotic formulae is impossible \cite{2008 Israelachvili}. In contrast, work on the incorporation of planar third body effects on the Van der Waals forces between point particles has been studied extensively \cite{1996 Silbey, Mahanty, 2005 Marcovitch}. Consequently, the third-body effect will be examined for point particles in this section with the aim of extricating qualitative aspects of the behaviour that may yield insight into the corresponding nanoparticle system.\par

The discussion will begin with an examination of the theoretical behaviour elucidated by Silbey et al.\cite{1996 Silbey}. A graphical representation of the system is provided in figure 5.2, taken from the paper itself \cite{1996 Silbey}.



\begin{figure}[H]
\centering
\includegraphics[scale=0.3]{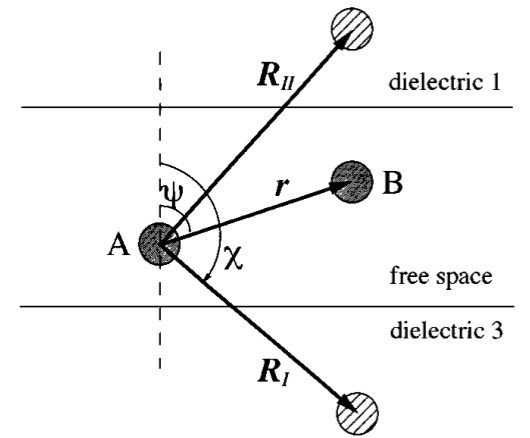}
\caption{Diagram of the Three-Body System considered by Silbey et al. \cite{1996 Silbey}.}
\label{fig:Diagram of the Three-Body System.}
\end{figure}

\noindent In the electrovariable mirror system, the only case of interest concerns that in which one of the two dielectrics in the diagram is retained and the other is replaced with water. Given that Silbey et al. consider only the case where one of the dielectrics is replaced with vacuum, and not water, the behaviour of the dielectric-water system will have to be inferred from that of the dielectric-vacuum system. Using a Thomas-Fermi like description of the planar metal, Silbey et al. show that at short interparticle separations, relative to the distance from the interface, the result of the third-body effect of the metal is negligible. Additionally, this feature seems to be independent of the point particle geometries. The two geometries considered are depicted in figures 5.3 and 5.4, and may be thought of as ninety degree rotations of figure 5.2.\par

\begin{figure}[H]
\centering
\includegraphics[scale=0.3]{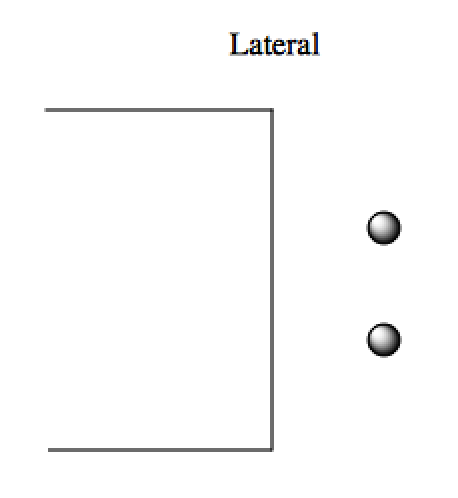}
\caption{Lateral Geometry.}
\label{fig:Lateral Geometry.}
\end{figure}

\begin{figure}[H]
\centering
\includegraphics[scale=0.3]{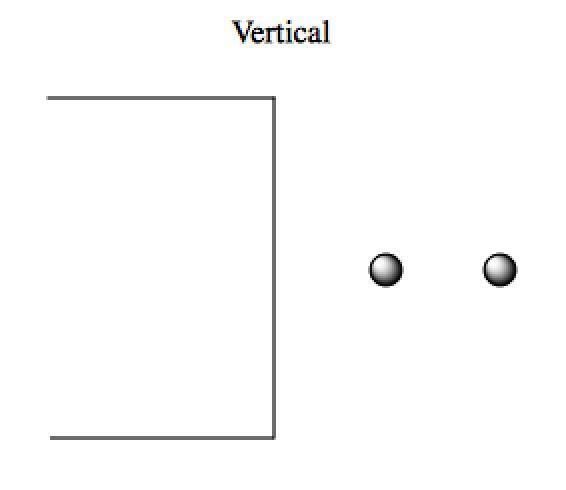}
\caption{Vertical Geometry.}
\label{fig:Vertical Geometry.}
\end{figure}

One interesting aspect of the system is the ability of the metal to either suppress or enhance the Van der Waals force between point particles dependent on whether they are distributed in either a lateral, or a vertical geometry. This behaviour is shown quantitatively in figure 5.5, taken from Silbey et al., where the axes are scaled such that the distance between the closest point particle and the metal is unity.

\begin{figure}[H]
\centering
\includegraphics[scale=0.3]{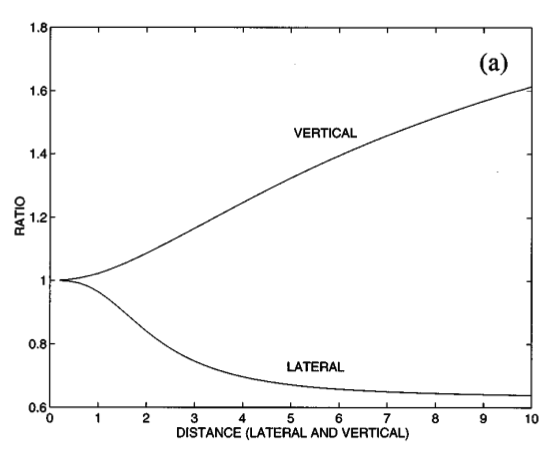}
\caption{An Illustration of the Ratio of the Van der Waals Force between Molecules when compared to the Interaction in Free Space.. \cite{1996 Silbey}.}
\label{fig: An Illustration of the Ratio of the Van der Waals Force between Molecules when compared to the Interaction in Free Space.}
\end{figure}


\noindent One may observe that when the distance from the metal is comparable to the interparticle separation, the interaction potential in free space is recovered. On going from vacuum to water, it could be expected that this ratio will be unchanged given that the universal presence of water should screen all interactions equally. Inferring information about the nanoparticle system from these results represents a more complex task and is something that should be analysed rigorously. Bearing in mind the application at hand however, ascertaining whether the Van der Waals force can offer a serious obstacle towards electrovariability, it should be possible to gauge that this effect should not be significant.\par

In the worst-case scenario, the nanoparticles will become trapped in an adsorption minimum too deep to overcome using an applied voltage. Treating the nanoparticles as being comprised of many point particles and even assuming Hamaker-style perfect additivity of the point particle interaction, the resultant force is vastly unlikely to exceed that of the bulk solution by greater than a factor of two for the vertical geometry. In the more feasible lateral geometry, the attraction may even be attenuated judging from the point particle system. When compared to the voltage that may be applied within the electrochemical window, this worst-case scenario contribution to the adsorption minimum should not present any significant barrier. \\\\

\noindent In summary, seeing as:

\begin{enumerate}
\item Third-body effects in point particle systems only take effect at distances from the interface larger than that between particles.
\item Even at close distances the upper bound for Van der Waals attraction in the vertical geometry should not be larger than a factor of two.
\end{enumerate}

\noindent it may be speculatively concluded that third-body effects for the Van der Waals force can be neglected. Ultimately, techniques will need to be developed to model Van der Waals interactions at the nanoscale to be able to determine the quantitative validity of this statement.

\section{Image Potential Energy}

In this section, a new quantitative theory is presented to model the third-body effect of the electrode on the image potential energy between point particles. The derivation of this theory is available in section A.2 of the appendix in addition to a rigorous demonstration, in section A.4, of its reduction to known results\cite{1979 Vorotyntsev} in the asymptotic limits. While the application of the theory towards modelling the third-body effect of gold and ITO electrodes only will be presented here, there exist further results, in section A.3 of the appendix, which incorporate a hydrocarbon film onto the surface of the electrode. Such an electrode design features in a number of nanoplasmonics applications and so the results presented there will have a wide range of practical applicability. The current discussion will take the following form:

\begin{enumerate}
\item Point Particle
\item Nanoparticle
\end{enumerate}

\subsection{Point Particle}

The effect of the field penetration of point charges into the electrode has been shown in chapter 4 to result in a modification of the classical image law of attraction to the surface of an ideal metal. Further to this, the electrostatic potential between point particles will be modulated by the same effect. This effect is illustrated in figure 5.6..\par

\begin{figure}[H]
\centering
\includegraphics[scale=0.3]{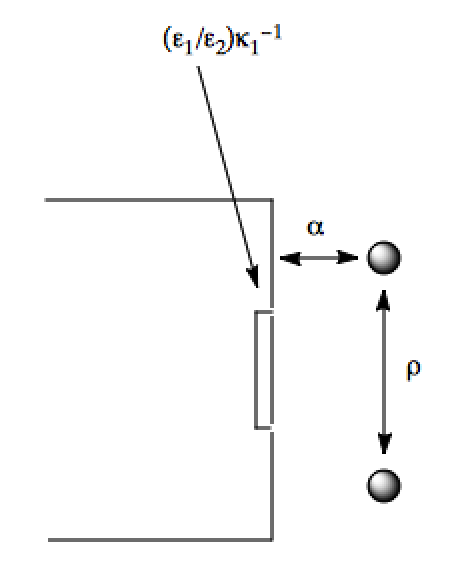}
\caption{Diagrammatic of Field Penetration into the Electrode.}
\label{fig:Diagrammatic of Field Penetration into the Electrode.}
\end{figure}

\begin{figure}[H]
\centering
\includegraphics[scale=0.3]{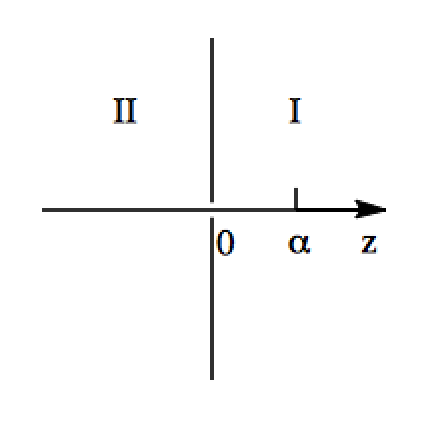}
\caption{Coordinate System.}
\label{fig:Coordinate System.}
\end{figure}

In the coordinate system provided in figure 5.7, the medium $\rom{1}$ represents the electrolyte solution, while the medium $\rom{2}$ represents the electrode. The method of images is used to recreate the boundary conditions imposed by the interface. In this instance, as opposed to an image charge, the artefact of the charge distribution transformation is a  pancake of charge with characteristic radius given by

\begin{align}\label{equation 9}
\left(\frac{\varepsilon_1}{\varepsilon_2}\right)\kappa_1^{-1}\:.
\end{align}

\noindent The significance of this characteristic radius in the asymptotic analysis of the modified pair interaction energy,

\begin{align}\label{equation 9}
\frac{q\Phi_{\rom{1}}(\rho, z)}{kT} = L_B\frac{e^{-\kappa_{\rom{1}}\sqrt{(z - \alpha)^2 + \rho^2}}}{\sqrt{(z - \alpha)^2 + \rho^2}} \mspace{400mu} \notag\\ \notag\\ + L_B\int_0^\infty dK J_0(K\rho)\frac{\xi\sqrt{K^2 + \kappa_{\rom{1}}^2} - \sqrt{K^2 + \kappa_{\rom{2}}^2}}{\xi\sqrt{K^2 + \kappa_{\rom{1}}^2} +\sqrt{K^2 + \kappa_{\rom{2}}^2}}\frac{K}{\sqrt{K^2 + \kappa_{\rom{1}}^2}} \: e^{-\sqrt{K^2 + \kappa_{\rom{1}}^2}\:(z + \alpha)} \:,
\end{align}

\noindent is explained in section A.4 of the appendix. The modified pair interaction energy is so-called because it represents the third-body presence of the electrode. In the results to follow, the modified pair interaction is plotted as a function of $\rho$, the distance between point particles in cylindrical coordinates, only. For simplicity, the particles will be considered only in a lateral geometry, corresponding in both figure 5.7 and equation 5.2 to $z = \alpha$, where $\alpha$ and $z$ represent the distances from the electrode at which the first and second point charges respectively, lie. Other parameters dictating the magnitude of the modified pair interaction energy include $\kappa_{\rom{1}}$, the inverse Thomas-Fermi length of the electrode, and $\kappa_{\rom{2}}$ the inverse Debye length of the electrolyte. The former parameter characterises the depth of field penetration into the electrode and will take on different values depending on whether the electrode is composed of gold or ITO.

\subsection{Nanoparticle}


In an analogous fashion to chapter 4, the point particle result will be scaled linearly by the number of charges on the surface of the nanoparticle in order to get an idea of the impact that this effect will have in the experimental system. The results of the pair interaction in the presence of the metal will be compared against the interaction potential of the bulk solution. For point particles, the interaction potential in the bulk is given by \cite{Israelachvili}

\begin{align}\label{equation 9}
\frac{L_B}{l}\frac{e^{-\kappa l}}{1 + \kappa\sigma} \:,
\end{align}

\noindent where $\kappa$ is the Debye length of the solution, $L_B$ is the Bjerrum length, $l$ is the interparticle separation and $\sigma$ is the particle diameter, which will be set to zero in this case. This pair interaction potential will be scaled in a similar fashion to the modified pair interaction potential by the number of charges on the surface of the nanoparticle. Comparisons are shown between the interaction potential in the bulk and that in the presence of the electrode for both gold and ITO in figures 5.8 and 5.9 respectively.

\begin{figure}[H]
\centering
\includegraphics[scale=0.3]{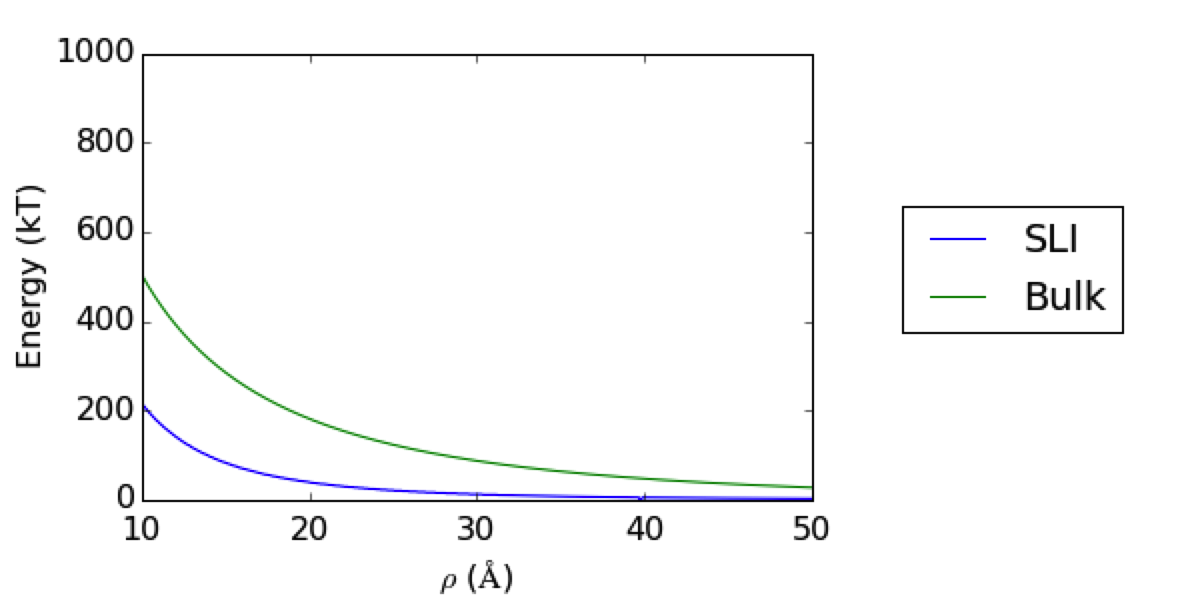}
\caption{Gold - Comparison against the Potential in the Bulk at 0.01 M Electrolyte Concentration.}
\label{fig: Gold - Comparison against the Potential in the Bulk at 0.01 M Electrolyte Concentration.}
\end{figure}

\begin{figure}[H]
\centering
\includegraphics[scale=0.3]{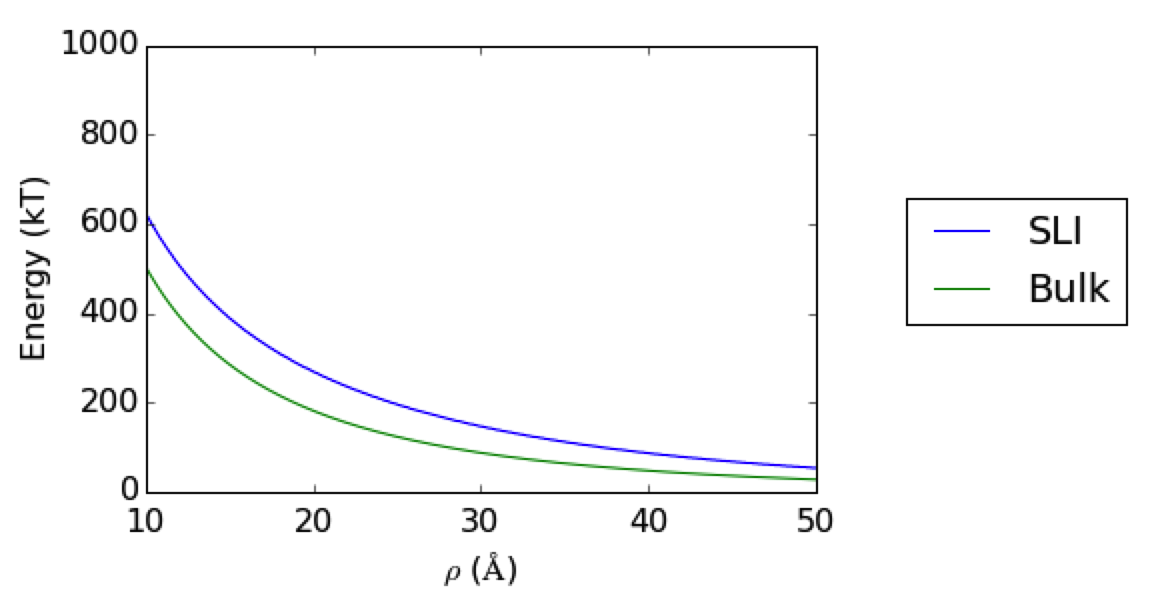}
\caption{ITO - Comparison against the Potential in the Bulk at 0.01 M Electrolyte Concentration.}
\label{fig: ITO - Comparison against the Potential in the Bulk at 0.01 M Electrolyte Concentration.}
\end{figure}

One may see that opposite effects are observed in the range corresponding to 5 nm from the SLI depending on the electrode material. In the case of gold, the pair interaction is suppressed in this range compared to the bulk. In contrast for ITO the interaction is enhanced. At distances larger than 5 nm however, one can see that the modified pair interaction potential converges quickly to that of the bulk. Given that large nanoparticles are unlikely to approach each other within this range, like the third-body modification to the Van der Waals force, it may also be inferred that the third-body contribution to the electrostatic potential between particles can be neglected. Once again, as in chapter 4, in the context of electrovariability, the effect of the third-body interaction is unlikely to present a barrier towards the reversible adsorption-desorption of nanoparticles from the interface. Nonetheless, as is so often the case in theoretical work, new results, although unimportant for the system considered, may find value for other applications where the sensitivity to such effects is greater.

\chapter{Conclusion}
The groundwork for a theory of the interactions of nanoparticle arrays at the SLI has been presented. The work is summarised firstly in terms of its applicability to the electrovariable mirror project, and secondly in terms of the information it has yielded regarding the current knowledge of nanoscale forces. The outlook and future work pertaining to the electrovariable project is discussed and, in addition, recommendations of a more general nature are provided for further research in the field concerning the theoretical modelling of  forces at the nanoscale.
\section{Summary}

\begin{enumerate}
\item Based on a consideration of the Van der Waals and electrostatic forces between functionalised nanoparticles, the experimental parameters required in order to achieve nanoparticle stability in the bulk solution have been elucidated. Typically, the concentration of 1:1 electrolyte needed lies in the neighbourhood of 0.01 M, and is independent of the nanoparticle size.
\item The forces comprising the energy profile between a single nanoparticle and an electrode have been analysed for both gold and ITO-based systems. In the process, a new theoretical result for the image potential energy of a point charge in electrolytic solution, accounting for field penetration into the electrode, has been derived. The results have been illustrated for a variety of electrode systems in line with their contribution to the modelling of the electrovariable mirror system. It should be emphasised however, that the equations derived are of a general nature and as such, are of interest for any theory attempting to describe the electrostatic forces at a metal-electrolyte interface.
\item The effect that the presence of the electrode may have on the forces between point particles has been studied. Taking the results as a guide, it was decided that this third-body effect is unimportant in terms of its ability to present an obstacle towards electrovariability.
\item As such, a proof of the concept of electrovariability has been supported. The estimates, while not yet rigorous, are hoped to be able to guide forthcoming experimental work.
\end{enumerate}

\clearpage

\section{Outlook}


\begin{enumerate}
\item The applicability of the estimates presented here will be tested in experiments undertaken by the Edel group, with the aim of demonstrating electrovariability at the SLI.
\item The refinement of the existing model will involve a more pointed consideration of the image potential energy profile for nanoparticles by treating them as spheres as opposed to scaled point charges. The reader is referred to the epilogue for more information on the state of progress towards this further, new theoretical result.
\item In terms of general recommendations for the field concerning the modelling of nanoscale interactions, it has been identified that the Van der Waals interaction remains the least well understood of the  forces acting in the system. Conflated information in the literature regarding the primary cause for uncertainty in the theoretical estimates of the Van der Waals force has given rise to a proliferation of misleading information. As such, it is highly recommended that the classic texts of Parsegian \cite{Parsegian} and Israelachvili \cite{Israelachvili} become obligatory reading for researchers working in the area. The information contained here is invaluable and there would appear to be a pressing need for it to be disseminated.
\item Towards the end of promulgating the instructive content of Parsegian's text, an annotated program detailing the step-by-step process of Van der Waals force computation within the Hamaker Hybrid Formalism has been included in section A.5 of the appendix. It is hoped that this may act as a guide for future researchers in the Kornyshev group when presented with the task of interpreting the swathe of recent literature on Van der Waals forces detailing hitherto unaccounted for, but largely unimportant effects.
\item The interplay between theory and experiment is rather pronounced when it comes to the field of Van der Waals forces. This is borne out of the fact that estimates of the Van der Waals potential energy rely heavily on the quality of the spectral data of the materials considered, which in turn will depend strongly on the specific sample being used. It is not sufficient to take an estimate from the same type of material, as is the current trend. Parameters such as film thickness and the extent of doping in the sample will produce markedly different spectra, and hence markedly different estimates of the Van der Waals interaction.\par

An ideal setup for obtaining accurate estimates would feature the recording of material-specific absorption spectra in advance of both theoretical modelling and experiment. An alternative suggestion for progress on this front would see some form of predictive software developed to simulate absorption spectra from material parameters. Given the extent of emerging applications of nanoscale systems, it is important that the Van der Waals force can be predicted with higher accuracy than is currently attainable in order to better understand system behaviours.
\item Lastly, in relation to the Van der Waals force, while continuum theories such as the Lifshitz theory \cite{1956 Lifshitz, 1961 Lifshitz} work well at macroscopic dimensions, and quantum mechanical theories \cite{Ginzburg, Bardeen} work well at the atomic level, the nanoscale currently presents something of a blank space in-between. When due consideration is given to the potential of this lengthscale for scientific applications, one can see that the ability to determine the extent of a fundamental force in this regime would no doubt lend itself favourably towards the design of novel devices. As such, a theory that can describe Van der Waals forces at the nanoscale is in high demand, and will offer rich rewards for those who possess the drive to develop it.

\end{enumerate}

\appendix
\chapter{Supporting Information}

\section{Derivation of the Coulomb Potential between Two Nanoparticles in Electrolyte}




\noindent The expression for the Coulomb potential energy between two nanoparticles in electrolytic solution is given in Israelachvili's \textit{Intermolecular and Surface Forces} as

\begin{align} \label{equation 1}
U(L) = \left(\frac{R_1R_2}{R_1 + R_2}\right)Ze^{-\kappa L}\:,
\end{align}

\noindent where $\kappa$ is the inverse Debye length, $L$ is the surface to surface separation and $R_1$, and $R_2$ are the radii of nanoparticles 1 and 2 respectively. The factor of,

\begin{align} \label{equation 2}
Z = 64\pi\varepsilon_0\varepsilon\left(\frac{kT}{e}\right)^2tanh^2\left(\frac{ze\psi_0}{4kT}\right) \:,
\end{align}

\noindent accounts for the dependence of the energy on the surface potential, $\psi_0$, of the nanoparticles. $z$ in this case, is the electrolyte valency. The goal of the manipulations that follow is to make $\sigma$, the surface charge density, the subject of the equation for the interaction potential, in place of $\psi_0$. This would allow for a more direct parameter input of the number of charges on the surface of the nanoparticle, in contrast to expressing this dependence through the surface potential. Starting from the Grahame equation given in Israelachvili's \textit{Intermolecular and Surface Forces}\cite{Israelachvili},

\begin{align} \label{equation 3}
\sigma = \sqrt{\frac{8c_0\varepsilon \varepsilon_0kT}{e^2}} sinh\left(\frac{e\psi_0}{2kT}\right) \:,
\end{align}

\noindent where $\sigma$ is the surface charge density and $c_0$ is the electrolyte concentration in the bulk, we define 

\begin{align} \label{equation 4}
\sigma^* = \sqrt{\frac{8c_0\varepsilon \varepsilon_0k_BT}{e^2}} \:.
\end{align}

\clearpage

\noindent To simplify matters the argument of the trigonometric functions can be defined as

\begin{align} \label{equation 5}
\frac{e\psi_0}{2kT} = x \:.
\end{align}

\noindent Thus

\begin{align} \label{equation 6}
\sigma = \sigma^* sinh(x) \: ; sinh(x) = \frac{\sigma}{\sigma^*}
\end{align}

\noindent and

\begin{align} \label{equation 8}
tanh^2\left(\frac{x}{2}\right) = \left(\frac{e^\frac{x}{2} - e^\frac{-x}{2}}{e^\frac{x}{2} + e^\frac{-x}{2}}\right)^2 = \frac{e^x + e^-x - 2}{e^x + e^-x + 2} = \frac{cosh(x) - 1}{cosh(x) + 1} \:.
\end{align}

\noindent From the relations

\begin{align} \label{equation 9}
cosh^2(x) - sinh^2(x) = 1 \: ; cosh^2(x) = 1 + sinh^2(x) \: ; cosh(x) = \sqrt{1+sinh^2(x)} \:,
\end{align}

\noindent it is found

\begin{align} \label{equation 12}
tanh^2\left(\frac{x}{2}\right) = \frac{\sqrt{1 + \left(\frac{\sigma}{\sigma^*}\right)^2} - 1}{\sqrt{1 + \left(\frac{\sigma}{\sigma^*}\right)^2} + 1} = \frac{1 + \left(\frac{\sigma}{\sigma^*}\right)^2 - 2\sqrt{1 + \left(\frac{\sigma}{\sigma^*}\right)^2 } + 1}{\left(\frac{\sigma}{\sigma^*}\right)^2} = 2\frac{1 - \sqrt{1 + \left(\frac{\sigma}{\sigma^*}\right)^2 } + \frac{1}{2}\left(\frac{\sigma}{\sigma^*}\right)^2 }{\left(\frac{\sigma}{\sigma^*}\right)^2} \:.
\end{align}

\noindent When this expression is simplified,

\begin{align} \label{equation 13}
tanh^2\left(\frac{x}{2}\right) = 2\left(\frac{\sigma^*}{\sigma}\right)^2\left(1 - \sqrt{1 + \left(\frac{\sigma}{\sigma^*}\right)^2 } + \frac{1}{2}\left(\frac{\sigma}{\sigma^*}\right)^2 \right) \:,
\end{align}

\noindent is obtained. Restoring x and simplifying the resultant expression, the interaction energy is found to be

\begin{align} \label{equation 14}
\frac{U(L)}{kT} = \frac{64c_0}{L_B^2\pi \sigma^2} \left(1 - \sqrt{1 + \left(\frac{\sigma}{\sigma^*}\right)^2 } + \frac{1}{2}\left(\frac{\sigma}{\sigma^*}\right)^2\right) \left(\frac{R_1R_2}{R_1 + R_2}\right)e^{-\kappa L} \:.
\end{align}

\clearpage

\section{Derivation of the Image Potential Energy and Pair Interaction Energy - Point Charge}



\begin{figure}[!ht]
\centering
\includegraphics[scale=0.5]{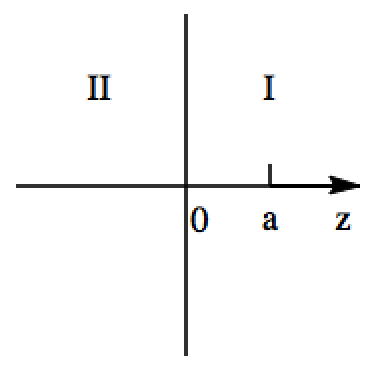}
\caption{A point charge near a dielectric-electrolyte interface, where half-space \rom{1} represents a dielectric with dielectric constant $\varepsilon$ and half-space \rom{2} represents the electrolyte medium. The location of the point charge is given by $a$.}
\end{figure}

\noindent The solutions for the potential distributions in mediums \rom{1} and \rom{2} are given by $\Phi_{\rom{1}}(\rho, z)$ and $\Phi_{\rom{2}}(\rho, z)$ respectively:\\\\

\begin{align}
\Phi_{\rom{1}}(\rho, z) = \frac{q}{\varepsilon_{\rom{1}}}\int_0^\infty dK J_0(K\rho) \frac{K}{\sqrt{K^2 + \kappa_{\rom{1}}^2}}\: e^{-\sqrt{K^2 + \kappa_{\rom{1}}^2}\:\left|z - a\right|} + \frac{q}{\varepsilon_{\rom{1}}}\int_0^\infty dK J_0(K\rho) \: e^{-\sqrt{K^2 + \kappa_{\rom{1}}^2}\:z}\: U(K)
\end{align}

\begin{align}
\Phi_{\rom{2}}(\rho, z) = \frac{q}{\varepsilon_{\rom{2}}}\int_0^\infty dK J_0(K\rho) \: e^{\sqrt{K^2 + \kappa_{\rom{2}}^2}\:z} \: V(K)\:.
\end{align}

\noindent The unknown functions $U(K)$ and $V(K)$ are obtained by solving under the boundary conditions corresponding to

\begin{align}
\Phi_{\rom{1}}(\rho, z)\Big|_{z \:=\: +0} = \Phi_{\rom{2}}(\rho, z)\Big|_{z \:=\: -0}
\end{align}

\noindent and

\begin{align}
\varepsilon_{\rom{1}}\left(\frac{\partial \Phi_{\rom{1}}(\rho, z) }{\partial z}\right)\Big|_{z \:=\: +0} = \varepsilon_{\rom{2}}\left(\frac{\partial \Phi_{\rom{2}}(\rho, z) }{\partial z}\right)\Big|_{z \:=\: -0} \:.
\end{align}

\noindent The boundary conditions are valid for all values of $\rho$, yielding the following expressions for the functions $U(K)$ and $V(K)$:

\begin{align}
U(K) = \frac{\xi\sqrt{K^2 + \kappa_{\rom{1}}^2} - \sqrt{K^2 + \kappa_{\rom{2}}^2}}{\xi\sqrt{K^2 + \kappa_{\rom{1}}^2} +\sqrt{K^2 + \kappa_{\rom{2}}^2}}\frac{K}{\sqrt{K^2 + \kappa_{\rom{1}}^2}}\:e^{-a\sqrt{K^2 + \kappa_{\rom{1}}^2}} \:,
\end{align}

\begin{align}
V(K) = \frac{2K}{\xi\sqrt{K^2 + \kappa_{\rom{1}}^2} + \sqrt{ K^2 + \kappa_{\rom{2}}^2}}\:e^{-a\sqrt{K^2 + \kappa_{\rom{1}}^2}}\:.
\end{align}

\noindent Where 

\begin{align}
\xi = \left(\frac{\varepsilon_{\rom{1}}}{\varepsilon_{\rom{2}}}\right) \:.
\end{align}

\noindent The image potential energy is calculated according to,

\begin{align}
W(a) =  \frac{q}{2}\lim_{\substack {z\to a \\ \rho\to 0}} \left\{\Phi_{\rom{1}}(\rho, z) - \frac{q}{\varepsilon_{\rom{1}}}\int_0^\infty dK J_0(K\rho) \frac{K}{\sqrt{K^2 + \kappa_{\rom{1}}^2}}\: e^{-\sqrt{K^2 + \kappa_{\rom{1}}^2}\:\left|z - a\right|} \right\}  \notag\\ \notag\\  = \frac{q^2}{2\varepsilon_{\rom{1}}}\int_0^\infty dK\:U(K)\:e^{-a\sqrt{K^2 + \kappa_{\rom{1}}^2}}\:.
\end{align}

\noindent Yielding

\begin{align}
W(a) = \frac{q^2}{2\varepsilon_{\rom{1}}}\int_0^\infty dK \frac{\xi\sqrt{K^2 + \kappa_{\rom{1}}^2} - \sqrt{K^2 + \kappa_{\rom{2}}^2}}{\xi\sqrt{K^2 + \kappa_{\rom{1}}^2} +\sqrt{K^2 + \kappa_{\rom{2}}^2}}\frac{K}{\sqrt{K^2 + \kappa_{\rom{1}}^2}}\:e^{-2a\sqrt{K^2 + \kappa_{\rom{1}}^2}}\:.
\end{align}

\noindent The pair-interaction energy in the electrolyte medium is given by $q\Phi_{\rom{1}}(\rho, z)$ yielding,

\begin{align}
q\Phi_{\rom{1}}(\rho, z) = \frac{q^2}{\varepsilon_{\rom{1}}}\int_0^\infty dK J_0(K\rho) \frac{K}{\sqrt{K^2 + \kappa_{\rom{1}}^2}}\: e^{-\sqrt{K^2 + \kappa_{\rom{1}}^2}\:\left|z - a\right|}   \notag\\ \notag\\ + \frac{q^2}{\varepsilon_{\rom{1}}}\int_0^\infty dK J_0(K\rho)\frac{\xi\sqrt{K^2 + \kappa_{\rom{1}}^2} - \sqrt{K^2 + \kappa_{\rom{2}}^2}}{\xi\sqrt{K^2 + \kappa_{\rom{1}}^2} +\sqrt{K^2 + \kappa_{\rom{2}}^2}}\frac{K}{\sqrt{K^2 + \kappa_{\rom{1}}^2}} \: e^{-\sqrt{K^2 + \kappa_{\rom{1}}^2}\:(z + a)}
\end{align}

\noindent Given that

\begin{align}
\frac{q^2}{\varepsilon_{\rom{1}}}\int_0^\infty dK J_0(K\rho) \frac{K}{\sqrt{K^2 + \kappa_{\rom{1}}^2}}\: e^{-\sqrt{K^2 + \kappa_{\rom{1}}^2}\:\left|z - a\right|} = \frac{e^{-\kappa_{\rom{1}}\sqrt{(z - \alpha)^2 + \rho^2}}}{\sqrt{(z - \alpha)^2 + \rho^2}} \:,
\end{align}

\noindent equation A.21 becomes

\begin{align}
q\Phi_{\rom{1}}(\rho, z) = \frac{e^{-\kappa_{\rom{1}}\sqrt{(z - \alpha)^2 + \rho^2}}}{\sqrt{(z - \alpha)^2 + \rho^2}} \notag\\ \notag\\ + \frac{q^2}{\varepsilon_{\rom{1}}}\int_0^\infty dK J_0(K\rho)\frac{\xi\sqrt{K^2 + \kappa_{\rom{1}}^2} - \sqrt{K^2 + \kappa_{\rom{2}}^2}}{\xi\sqrt{K^2 + \kappa_{\rom{1}}^2} +\sqrt{K^2 + \kappa_{\rom{2}}^2}}\frac{K}{\sqrt{K^2 + \kappa_{\rom{1}}^2}} \: e^{-\sqrt{K^2 + \kappa_{\rom{1}}^2}\:(z + a)} \:.
\end{align}

\section{Image Potential Energy for Gold and ITO with Hydrocarbon Films}

The use of electrode systems possessing a hydrocarbon surface film is gaining traction due to the idea that nanoparticles functionalised with similar hydrocarbon-based ligands will be able to enjoy a soft landing at the electrode\cite{2016 Edel}. As such, the illustration of the use of the expression for the image potential energy could prove useful in elucidating some qualitative effects of the presence of the film.\par

In figures A.2 and A.3, the effect of the hydrocarbon film on gold and ITO electrode system behaviour may be seen. Under the assumption that the layered electrode material has a dielectric constant of 3 (Usually this value is roughly 2 for hydrocarbons, but we will assume some leakage of water into the material, in addition to structural defects), the plots of the two electrode systems behave in a similar fashion. In deGennes style, the Thomas-Fermi length of the gold system is taken to be the thickness of the film. We will take this value to be 10 nm. In the ITO system, the effective Thomas-Fermi length will be taken to be the sum of the Thomas-Fermi length of ITO and the film thickness, ca. 17 nm altogether. Figure A.4 shows that the subcutaneous electrode material, under the hydrocarbon "skin" does not make itself felt in either system, the plots overlay each other almost perfectly. In effect, both setups behave as if the electrode was made of hydrocarbon. As such, already at film thicknesses of 10 nm, it can be observed that the hydrocarbon film dominates the behaviour of the image potential energy profile. A logical next step would be to see to what extent this is the case for materials with largely differing Thomas-Fermi lengths.




\begin{figure}[h]
\centering
\includegraphics[scale=0.7]{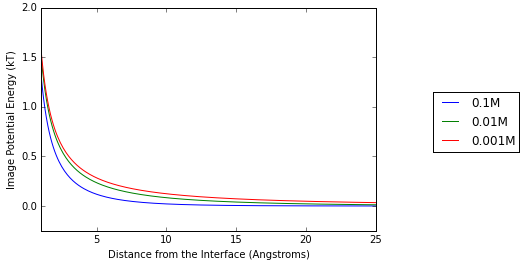}
\caption{Point Particle next to Gold with Film.}
\label{fig:Point Particle next to Gold with Film.}
\end{figure}


\begin{figure}[h]
\centering
\includegraphics[scale=0.7]{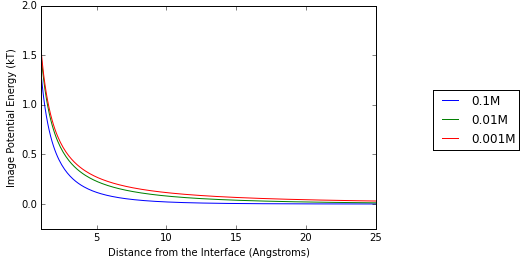}
\caption{Point Particle next to ITO with Film.}
\label{fig:Point Particle next to ITO with Film.}
\end{figure}


\begin{figure}[h]
\centering
\includegraphics[scale=0.7]{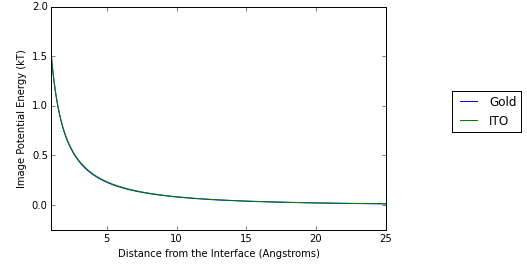}
\caption{Comparison of Gold with Film and ITO with Film.}
\label{fig:Comparison of Gold with Film and ITO with Film.}
\end{figure}

\clearpage

\section{Point Particle Pair Interaction Energy}

Considering the method of images in modelling the interaction between two charges and a metal, the most intuitive way of transforming the charge distribution corresponds to a 'pancake' of charge, with characteristic radius

\begin{align}
\left(\frac{\varepsilon_\rom{1}}{\varepsilon_{\rom{2}}}\right)\kappa_{\rom{2}}^{-1} \:.
\end{align}

\noindent The image problem is illustrated graphically in figures A.5 and A.6. 

\begin{figure}[H]
\centering
\includegraphics[scale=0.3]{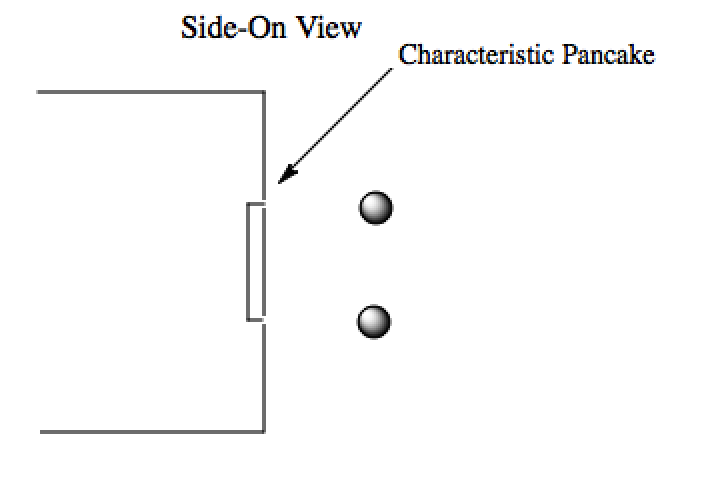}
\caption{Side-On View of the Characteristic Pancake.}
\label{fig:Side-On View of the Characteristic Pancake.}
\end{figure}

\begin{figure}[H]
\centering
\includegraphics[scale=0.3]{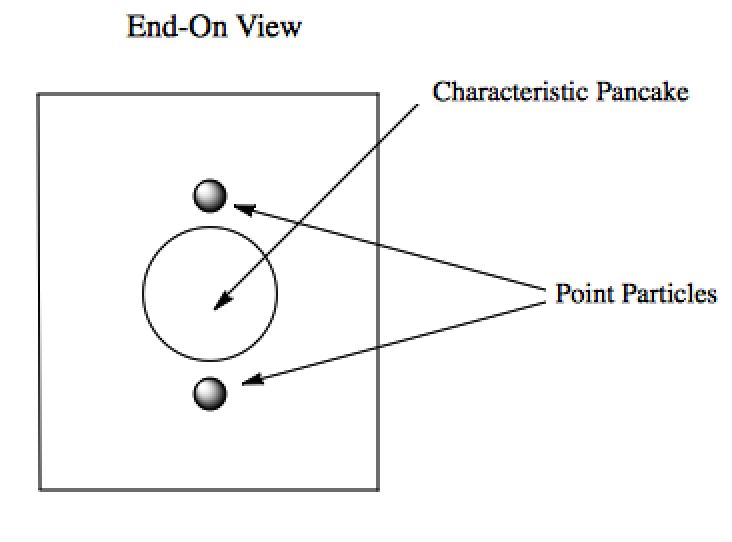}
\caption{End-On View of the Characteristic Pancake.}
\label{fig:End-On View of the Characteristic Pancake.}
\end{figure}

\noindent Again results will be considered here for electrodes possessing a hydrocarbon layer on their surface. As such, in this case, the characteristic radius of the charge distribution becomes

\begin{align}
\left(\frac{\varepsilon_\rom{1}}{\varepsilon_{\rom{2}}}\right)L_{f}^{-1} \:,
\end{align}

\noindent where $L_{f}^{-1}$ is the thickness of the film. The three limiting cases that it will be necessary to examine are\cite{1979 Vorotyntsev}:

\begin{enumerate}
\item The case when the two charges are close to the interface and the distance between them is within the effective radius.
\item The case when the two charges are close to the interface and the distance between them is larger than the effective radius.
\item The case when the two charges are far from the interface.
\end{enumerate}

\noindent The expression for the general situation, in reference to figure A.1, when 

\begin{align}
a \approx \rho \ll \left(\frac{\varepsilon_1}{\varepsilon_{2}}\kappa_{2}^{-1} \:, \kappa_{1}^{-1}\right) \:,
\end{align}

\noindent is

\begin{align}
\frac{1}{\rho} + \left[\frac{(\varepsilon_1 - \varepsilon_2)}{(\varepsilon_1 + \varepsilon_2)}\right]\left(\rho^2 + 4a^2\right)^{-\frac{1}{2}} \:.
\end{align}

\noindent Another limit is that of the point charges being within the effective radius of the pancake charge distribution, but farther away from each other than from the surface,

\begin{align}
\frac{2}{\rho} \frac{\varepsilon_1}{\varepsilon_1 + \varepsilon_2} \:.
\end{align}

\noindent When the point charges are outside the effective radius of the pancake and close to the interface the expression is

\begin{align}
\frac{2 \left(a + \frac{\varepsilon_1}{\varepsilon_2}\kappa_1^{-1}\right)^2}{\rho^3} \:.
\end{align}

\noindent In figures A.7 - A.9, the general form of the expression for the pair interaction energy, the dimensionless analogue of equation A.21, will be plotted in the limiting cases outlined above. In all cases, the plot title shows the conditions of the asymptotic law against which the general expression is plotted. The x-axis will always represent the variation of $\rho$, the distance between point particles. In the asymptotic comparisons, the parameters taken for use in the equations are those corresponding to gold. The dielectric constant of water is taken to be ca. 80 while the dielectric constant of gold is taken to be ca. 6. Taking a Thomas-Fermi length of 0.5\AA, this results in a characteristic radius of size ca. 6.5\AA.

\begin{figure}[H]
\centering
\includegraphics[scale=0.6]{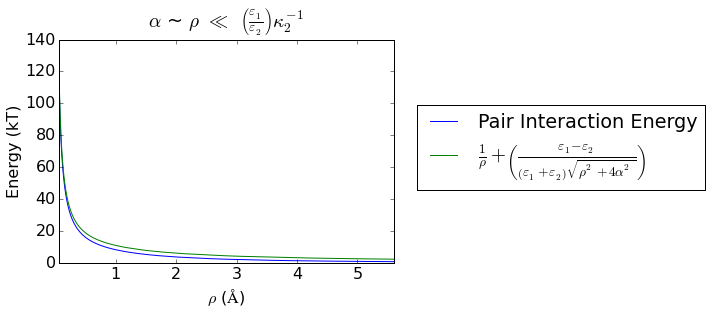}
\caption{General Law over full Range.}
\end{figure}

\noindent In figure A.7 it may be seen that the general expression for the pair interaction potential converges with the general asymptotic law at short interparticle separations. One can see that when the values of $\rho$ approach that of the characteristic radius, the curves begin to deviate.


\begin{figure}[H]
\centering
\includegraphics[scale=0.6]{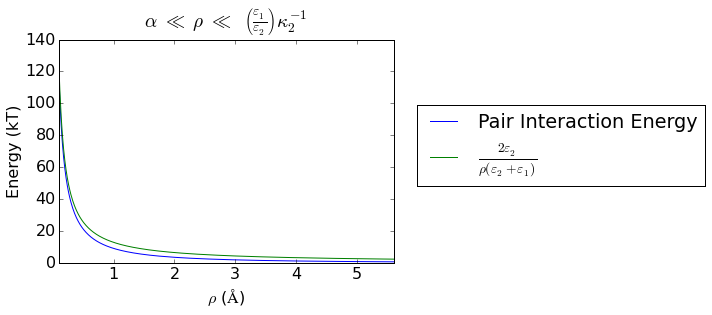}
\caption{Short Distance Law.}
\end{figure}

\noindent Figures A.8 and A.9 depict similar agreement in the associated limits.

\begin{figure}[H]
\centering
\includegraphics[scale=0.6]{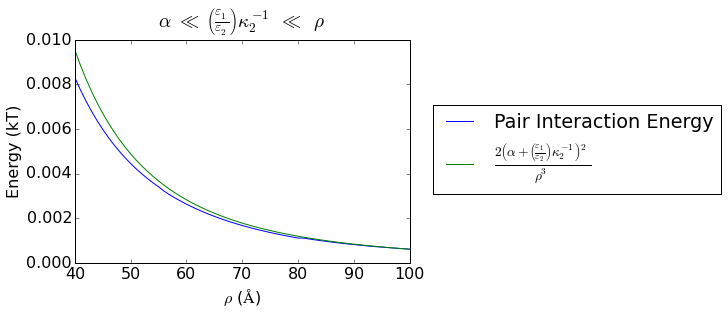}
\caption{Long Distance Law.}
\end{figure}

\noindent In figure A.10, the general expression for the pair interaction energy is plotted against the classical image law describing the effects of attraction to an ideal metal. At large distances from the interface, i.e. large $\alpha$, the results converge in figure A.10, irrespective of the value of $\rho$.

\begin{figure}[H]
\centering
\includegraphics[scale=0.6]{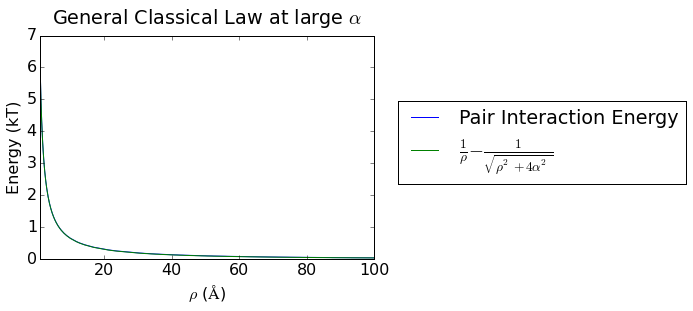}
\caption{General Classical Law.}
\end{figure}

\noindent Figure A.11 shows the agreement with the asymptotic classical image law for large values of $\alpha$ and large values of $\rho$.

\begin{figure}[H]
\centering
\includegraphics[scale=0.6]{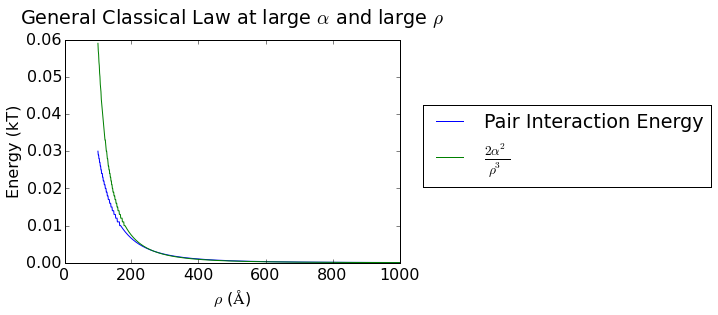}
\caption{Long Distance Classical Law.}
\end{figure}

\noindent  In figures A.12 through A.15, the pair interaction law is shown for four different systems:

\begin{enumerate}
\item Gold
\item ITO
\item Gold with a Hydrocarbon Film
\item ITO with a Hydrocarbon Film
\end{enumerate}

\begin{figure}[H]
    \centering
    \begin{subfigure}[h]{0.485\textwidth}
        \includegraphics[width=\textwidth]{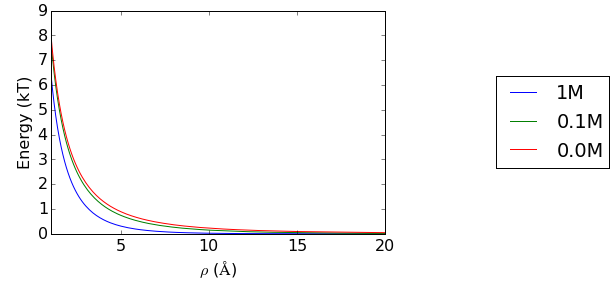}
        \caption{Gold - Closer than the Characteristic Radius.}
        \label{fig: Gold - Closer than the Characteristic Radius.}
    \end{subfigure}
    ~ 
    \begin{subfigure}[h]{0.485\textwidth}
        \includegraphics[width=\textwidth]{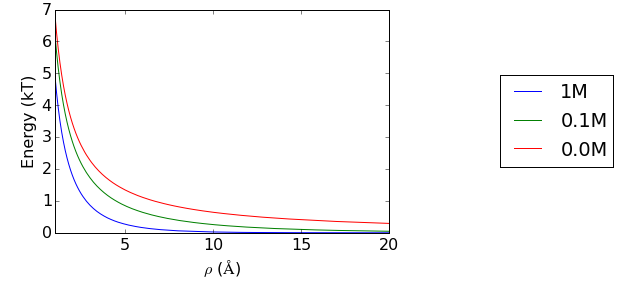}
        \caption{Gold - Farther than the Characteristic Radius.}
        \label{fig: Gold - Farther than the Characteristic Radius.}
    \end{subfigure}
    \caption{Gold.}
    \label{fig: Gold.}
\end{figure}




\begin{figure}[H]
    \centering
    \begin{subfigure}[h]{0.485\textwidth}
        \includegraphics[width=\textwidth]{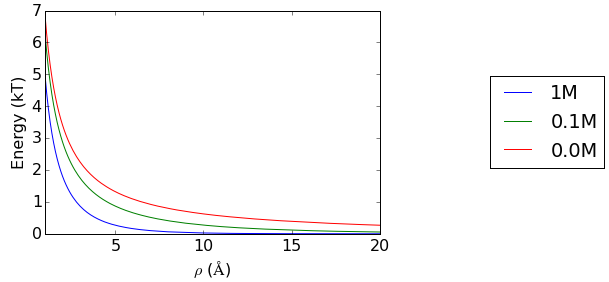}
        \caption{ITO - Closer than the Characteristic Radius.}
        \label{fig: ITO - Closer than the Characteristic Radius.}
    \end{subfigure}
    ~ 
    \begin{subfigure}[h]{0.485\textwidth}
        \includegraphics[width=\textwidth]{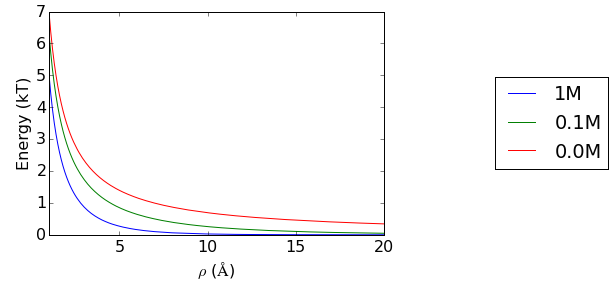}
        \caption{ITO - Farther than the Characteristic Radius.}
        \label{fig: ITO - Farther than the Characteristic Radius.}
    \end{subfigure}
    \caption{ITO.}
    \label{fig: ITO.}
\end{figure}




\begin{figure}[H]
    \centering
    \begin{subfigure}[h]{0.485\textwidth}
        \includegraphics[width=\textwidth]{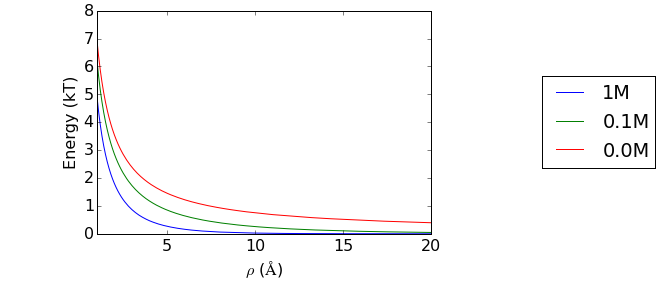}
        \caption{Gold with Film - Closer than the Characteristic Radius.}
        \label{fig: Gold with Film - Closer than the Characteristic Radius.}
    \end{subfigure}
    ~ 
    \begin{subfigure}[h]{0.485\textwidth}
        \includegraphics[width=\textwidth]{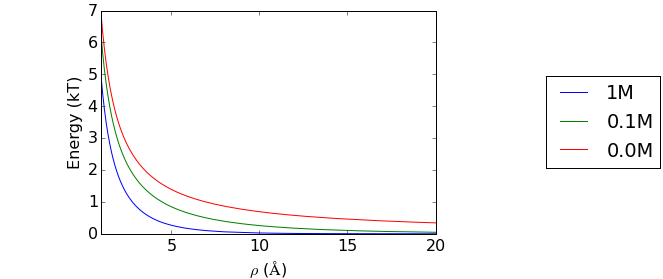}
        \caption{Gold with Film - Farther than the Characteristic Radius.}
        \label{fig: Gold with Film - Farther than the Characteristic Radius.}
    \end{subfigure}
    \caption{Gold with Film.}
    \label{fig: Gold with Film.}
\end{figure}




\begin{figure}[H]
    \centering
    \begin{subfigure}[h]{0.485\textwidth}
        \includegraphics[width=\textwidth]{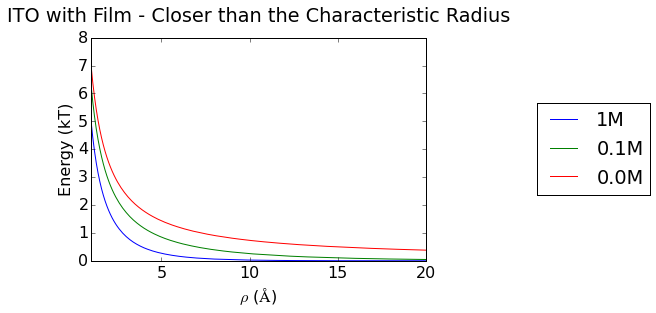}
        \caption{ITO with Film - Closer than the Characteristic Radius.}
        \label{fig: ITO with Film - Closer than the Characteristic Radius.}
    \end{subfigure}
    ~ 
    \begin{subfigure}[h]{0.485\textwidth}
        \includegraphics[width=\textwidth]{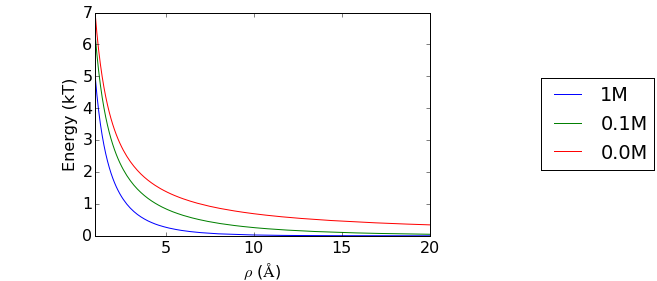}
        \caption{ITO with Film - Farther than the Characteristic Radius.}
        \label{fig: ITO with Film - Farther than the Characteristic Radius.}
    \end{subfigure}
    \caption{ITO with Film.}
    \label{fig: ITO with Film.}
\end{figure}

\noindent The most striking difference in the distance-dependent behaviour of the pair-interaction energy is observed for the purer gold and ITO systems when compared to the film systems. Once again, the plots of the two film systems remain similar, and so this would add further support to the point that a 10 nm hydrocarbon film is capable of stealing the systemic material properties from the underlying electrode material.




\section{Annotated Program for the Computation of the Hamaker Coefficient}

In this section, a commented Python script for the computation of the Hamaker coefficient is provided. The program features the computation of the Hamaker coefficient of a gold-water-gold system using only the parameters of Johnson and Christy for gold\cite{1972 Johnson and Christy}, and those of Parsegian and Weiss\cite{1981 Weiss} for water, for the purposes of illustration.

\begin{lstlisting}

import numpy as np

k = 1.38*(10**-23) # joules per Kelvin.
T = 298.15 # Kelvin.
hbar = 1.0545718*(10**-34) # joule seconds
pi = np.pi

# Matsubara coefficient

Matsubara = (2*pi*k*T)/(hbar)
rad_s = 1.05*(10**11)
eV_conversion = 6.55*(10**-5)

# Conversion from rad s^-1 to electronvolts required.

Matsubara_coeff = lambda n: ((Matsubara/rad_s)*eV_conversion)*n

# Johnson and Christy parameters

wj3 = [0.0,3.0,4.8] # eV
fj3 = [53.0,5.0,104.0] # eV^2
gj3 = [1.8,0.8,4.4] # eV

# Parameters for water found on page 266 of Parsegian's book.
# The damped oscillator model given on this page is used here
# to calculate the Hamaker coefficient.

dj = 74.8 # Dimensionless

tau = 15267.0 # eV^-1

Wwj = [2.07*(10**-2), 6.9*(10**-2), 9.2*(10**-2), 2.0*(10**-1), 
       4.2*(10**-1), 8.25, 
       10.0, 11.4, 13.0, 14.9, 18.5] # eV
       
Wfj = [6.25*(10**-4), 3.5*(10**-3), 1.28*(10**-3), 5.44*(10**-4), 
       1.35*(10**-2), 
       2.68, 5.67, 12.0, 26.3, 33.8, 92.8] # eV^2
       
Wgj = [1.5*(10**-2), 3.8*(10**-2), 2.8*(10**-2), 2.5*(10**-2), 
       5.6*(10**-2),
       0.51, 0.88, 1.54, 2.05, 2.96, 6.26] # eV
       
# Writing dielectric functions (damped oscillator form) for different 
# parameter sets 
# As Python does not support multiline lambda functions, I have given the
# basic terms of the damped oscillator model below

# 2nd term in damped oscillator model.

# ((dj)/(1 + (Matsubara_coeff(n)*tau)))

# 3rd term in damped oscillator model
# 3rd term in the damped oscillator model is a sum
# and so the term given below represents one of the terms in the sum.
# There will be a separate term for each of the 11 parameters
# given in the WWj, Wfj and Wgj lists provided above.

# (Wfj[0])/(Wwj[0])**2 + Matsubara_coeff(n)**2 + (Matsubara_coeff(n)*Wgj[0]) 

# e.g. the next term in the sum will be:

# (Wfj[1])/(Wwj[1])**2 + Matsubara_coeff(n)**2 + (Matsubara_coeff(n)*Wgj[1]) 

# These terms are added after the ... below.

# The dielectric function for water is just a sum of 

Epsilon_Water = lambda n: 1 + ((dj)/(1 + (Matsubara_coeff(n)*tau))) + ...

# Similarly this partial term for the gold dielectric function is
# continued by the terms give above but for the gold 
# parameter values.

Epsilon_Gold_JC = lambda n: 1 + (fj3[0])/((Matsubara_coeff(n))**2 + ...

# Making separate expressions for the zero frequency term.

Epsilon_Gold_JC_Zero_Freq = lambda n: 1 + (fj3[1])/((wj3[1])**2 + ...

# Implementing the power series form given by Parsegian on page 275 
# for the zero-frequency term.
# This power series form can also be used in the computation of the 
# other frequencies, but tends to result in a negligible contribution to 
# the value of the Hamaker coefficient.
# Zero-Frequency term converges to three decimal places when k is set to 10.

k = 1.0

Zero_Freq_Term_JC = 0.0

while k < 5:
    
    Zero_Freq_Term_JC += 0.5*((((Epsilon_Gold_JC_Zero_Freq(0) 
    - Epsilon_Water(0))/(Epsilon_Gold_JC_Zero_Freq(0) 
    + Epsilon_Water(0)))**(2*k))/(k**3))
    k += 1.0

print Zero_Freq_Term_JC 

# The value of the zero-frequency term comes out as 0.54 kT
# Given that the total value of the Hamaker coefficient is 24.7 kT, 
# the  zero-frequency term accounts for 2.2\% of its magnitude. 
# Additionally, this term will be screened in electrolyte, 
# resulting in a further decrease in its magnitude. 
# As such, it has been neglected in the results of this work.
    
Hamaker_JC = 0.0

# Summing over the quantised Matsubara frequencies, designated as n.
# The values for the Hamaker coefficient would seem to converge to the 
# accuracy of two decimal places in the units of kT at n = 10000.
# It is also possible to integrate over the full range of frequencies. 
# The magnitude of the Hamaker coefficient however,
# is largely invariant on the approach taken.

n = 1.0

while n < 10000.0:
    k = 1.0
    while k < 5.0:
    
        Hamaker_JC += (((Epsilon_Gold_JC(n) - 
        Epsilon_Water(n))/(Epsilon_Gold_JC(n) 
        + Epsilon_Water(n)))**(2*k))/(k**3)
        
        k +=1.0
        
    n += 1.0
  
# Summing the n = 1-10,000 terms with the zero-frequency term 
# And multiplying by 3.0/2.0 in order to get the Hamaker coefficient in 
# units of kT.
# It's necessary to be careful in Python to always specify numbers 
# as floats.
# If numbers are not specified as floats this could incur rounding 
# errors - for example
# 3/2 is interpreted by Python as being 1 instead of 1.5.

Hamaker_Johnson = (Hamaker_JC + Zero_Freq_Term_JC)*(3.0/2.0)

print Hamaker_Johnson # value of 24.7 kT obtained.

\end{lstlisting}

\clearpage

\section{Drude-Lorentz Model Parameters for ITO}

The parameter values for the Drude-Lorentz model of ITO,

\begin{align} \label{equation 1}
\varepsilon_{ITO}(\omega) = 1 - \frac{\omega_p^2}{\omega^2 + i\omega_{\tau}\omega} - \sum_{j=1}^{2} \frac{f_j\omega_j^2}{ \omega^2 - \omega_j^2  + i\gamma_j\omega}\:,
\end{align}

\noindent are provided in the following table.

\vspace{10 mm}

\begin{tabular}{| c | c | c | c | c | c | c | c | c | c | }
  \hline
  Parameters & $\omega_p$ & $\omega_{\tau}$ & $\gamma_1$ & $\gamma_2$ & $\omega_1$ & $\omega_2$ & $f_1$ & $f_2$ \\
  \hline 
  Values (eV) & $1.8$ & $0.05$ & $0.2$ & $0.1$ & $4.82$ & $7.1$ & $0.47$ & $2.3$ \\
  \hline
\end{tabular}

\chapter{Epilogue}

Although an appealing prospect, utilising theory to meet the demands for the development of practical applications is not always a realistic goal. As has been featured throughout this report, the treatment of novel physical effects does not always lend itself to pragmatism. Such was the case for the Van der Waals forces of chapter 3, where a size-dependent Hamaker coefficient, a reasonable addition to the theory of the Van der Waals interaction, is not of great interest to the broader scientific community due to its applicability only at atomic length scales, and not the more popular nanoscale dimensions. Similarly, in the case of the image potential results in electrolytic solution, accounting for field penetration into the electrode, the magnitude of the force was deemed inconsequential for the practical consideration of achieving electrovariability. The appeal of theoretical results however, would appear to lie with their permanence. Although it may not be the case that a newly derived theory is immediately applicable to the popular science of the time, it endures regardless. One may take the example of the mathematics of Carl Friedrich Gauss on higher dimensional surfaces. Although the work didn't seem to have any firm purpose at the time, it was used by Einstein in the following century to underpin his general theory of relativity. Occasionally, when working on theory, the allure of permanent results may overpower the need to develop pragmatic models for the sensible applications of the day. It is with this spirit that this epilogue introduces some unfinished, but highly time-intensive work on one such permanent result for the image potential created by a sphere.\par

\section{A Geometrically Accurate Result for a Sphere}

\noindent Starting from the Debye equation, the solutions for the potential in the two media are given by

\begin{align}
\nabla\phi = (K^2 + \kappa_{\rom{1}}^2)\phi - 4\pi\rho(\vec{R}) \:,
\end{align}

\noindent for medium 1 corresponding to electrolyte, where $\rho$ denotes the charge density of the sphere. The potential in the metal (medium 2), is given by the solution of 

\begin{align}
\nabla\phi = (K^2 + \kappa_{\rom{2}}^2)\phi\:.
\end{align}

\noindent Defining the Fourier relations,

\begin{align}
\phi(\vec{R}, z) = \int e^{i\vec{K}\vec{R}}\widetilde{\phi}(\vec{K}, z) d\vec{K}\:,
\end{align}

\noindent and

\begin{align}
\widetilde{\phi}(\vec{K}, z) = \frac{1}{(2\pi)^2} \int e^{-i\vec{K}\vec{R}}\phi(\vec{R}, z)d\vec{R}\:,
\end{align}

\noindent the boundary conditions are

\begin{align}
\phi_1(K, z = 0) = \phi_2(K, z = 0)\:,
\end{align}

\noindent and

\begin{align}
\varepsilon_1 \frac{\partial }{\partial z}\phi_1(K, z)\bigg|_{z = 0} = \varepsilon_2 \frac{\partial}{\partial z}\phi_2(K, z)\bigg|_z = 0\:.
\end{align}

\noindent $\widetilde{\rho}(\vec{K}, z)$, the charge density required to obtain a solution to the potentials in the two media from equations B.1 and B.2, may be obtained as

\begin{align}
\widetilde{\rho}(\vec{K}, z) = \frac{1}{(2\pi)^2} \int e^{-i\vec{K}\vec{R}}\rho(\vec{R}, z) d\vec{R}\:.
\end{align}

\noindent As such, it will be necessary to obtain an expression for $\rho(\vec{R}, z)$. In the cylindrical coordinate system, $\rho$ is the circular coordinate and $z$ the axial coordinate. The expression for $\rho(\vec{R}, z)$ is

\begin{align}
\rho(\vec{R}, z) = \frac{\delta\left(\vec{R} - \sqrt{a^2 - (z - z_0)^2}\right)}{A}\:.
\end{align}

\noindent In order to calculate the normalisation constant $A$ it is necessary to integrate over all space in the radial coordinate, but just from $z_0 - a$ to $z_0 + a$ in the axial coordinate.

\begin{align}
\frac{1}{A}\int_0^\infty dR\:2\pi  R \:  \int_{z_0 - a}^{z_0 + a} dz\: \delta\left(\vec{R} - \sqrt{a^2 - (z - z_0)^2}\right) = 1
\end{align}

\noindent Following some manipulation,

\begin{align}
\frac{1}{A}\int_{z_0 - a}^{z_0 + a} dz \: 2\pi \int_0^\infty dR \: R \: \delta\left(\vec{R} - \sqrt{a^2 - (z - z_0)^2}\right) = 1 \:.
\end{align}

\noindent Using the relation

\begin{align}
\int F(\vec{r}) \: \delta(\vec{r} - \vec{r_0}) d\vec{r} = F(\vec{r_0})\:,
\end{align}

\noindent we get

\begin{align}
\frac{1}{A}\int_{z_0 - a}^{z_0 + a} dz \: 2\pi \sqrt{a^2 - (z - z_0)^2} = 1\:,
\end{align}

\noindent yielding, following a variable substitution $t = z - z_0$

\begin{align}
A = \pi^2a^2\:.
\end{align}

\noindent This variable substitution has been checked in the case of $z_0 = 0$ in page 2 of the notes of the 17th of March. If there is an error, it is likely to be before this point.

\begin{align}
\rho(\vec{R}, z) = \left(\frac{1}{\pi^2a^2}\right) \delta\left(\vec{R} - \sqrt{a^2 - (z - z_0)^2}\right)\:.
\end{align}

From equation B.14 we go to

\begin{align}
\widetilde{\rho}(\vec{K}, z) = \frac{1}{(2\pi)^2} \int e^{-i\vec{K}\vec{R} \: cos\theta}\: d\theta \: \rho(\vec{R}, z) R \: d\vec{R}\:.
\end{align}

\noindent From equation B.15, we go to equation B.16.

\begin{align}
\widetilde{\rho}(\vec{K}, z) = \frac{1}{(2\pi)^2} \int_0^\infty \vec{R} \: d\vec{R} \frac{1}{\pi^2a^2} \: \delta\left(\vec{R} - \sqrt{a^2 - (z - z_0)^2}\right) 2 \int_{-\frac{\pi}{2}}^{+\frac{\pi}{2}} d\theta \: e^{i\vec{K}\vec{R} \: cos\theta}\:.
\end{align}

\noindent Yielding finally

\begin{align}
\widetilde{\rho}(\vec{K}, z) = \frac{1}{(2\pi)^2}\frac{1}{\pi^2a^2}\left[\sqrt{a^2 - (z - z_0)^2}J_0\left(K\sqrt{a^2 - (z - z_0)^2}\right)\:.\right]
\end{align}

\noindent The image potential energy for the system may be obtained by subtracting the potential created by the sphere itself from the full expression for the potential according to

\begin{align}
\frac{q^2}{2}\big[\Phi(z, R) - \Phi_0(z, R)\big]\:,
\end{align}

\noindent where

\begin{align}
\delta\Phi = \big[\Phi(z, R) - \Phi_0(z, R)\big]\:.
\end{align}

\noindent The image potential energy is then given as 

\begin{align}
W_s = \frac{q^2}{2} \int \delta\Phi(\vec{R}, z) \: d\vec{R} \: dz \: \rho(\vec{R}, z) \:.
\end{align}

\noindent In terms of $\vec{K}$,

\begin{align}
W_s = \frac{q^2}{2} \int_{-\infty}^\infty dz \: \int d\vec{K} \delta\Phi(\vec{K}, z) \rho(-\vec{K}, z)\:.
\end{align}

\noindent Hopefully this may give a flavour of what is to come in terms of the latest additions to the theory of the image potential.


\begin{thebibliography}{88}


\bibitem{2003 Kelly}
K. L. Kelly, 
E. Coronado, 
L. L. Zhao, 
G. C. Shatz,
\textit{J. Phys. Chem. B},
2003,
\textbf{107},
668-677.


\bibitem{1998 Feldheim}
D. L. Feldheim,
C. D. Keating,
\textit{Chem. Soc. Rev.},
1998,
\textbf{27},
1-12.


\bibitem{2005 Duan}
H. Duan,
D. Wang,
N. S. Sobal,
M. Giersig,
D. G. Knuth,
H. Mohwald,
\textit{Nano Lett.},
2005
\textbf{5},
949-952.


\bibitem{2010 Crossley}
S. Crossley,
J. Faria,
M. Shen,
D. E. Resasco,
\textit{Science},
2010,
\textbf{327},
68-72.


\bibitem{2009 Velev}
O. D. Velev,
S. Gupta,
\textit{Adv. Mater.},
2009,
\textbf{21},
1897-1905.


\bibitem{2013 Kanekiyo}
M. Kanekiyo,
C. -J. Wei,
H. M. Yassine,
P. M. McTamney,
J. C. Boyington,
J. R. R. Whittle,
S. S. Rao,
W. -P. Kong,
L. Wang,
G. J. Nabel,
\textit{Nature}
2013,
\textbf{499},
102-106.


\bibitem{2016 Yang}
J. Yang, 
M. K. Choi, 
D. -H. Kim, 
T. Hyeon,
\textit{Adv. Mater.}
2016,
\textbf{28},
1176-1207.


\bibitem{2010 Nie}
Z. Nie,
A. Petukhova,
E. Kumacheva,
\textit{Nat. Nanotech.},
2010,
\textbf{5},
15-25.


\bibitem{2010 Liu}
S. Liu,
Z. Tang,
\textit{J. Mater. Chem.},
2010,
\textbf{20},
24-35.


\bibitem{2011 Stockman Today}
M. I. Stockman,
\textit{Phys. Today},
2010,
\textbf{64},
39-44.


\bibitem{2003 Barnes}
W. L. Barnes,
A. Dereux,
T. W. Ebbesen,
\textit{Nature},
2003,
\textbf{424},
824-830.


\bibitem{Maier Book}
S. A. Maier,
\textit{Plasmonics: Fundamentals and Applications},
Springer,
New York,
1st edn.,
2007.


\bibitem{2008 Anker}
J. N. Anker,
W. P. Hall,
O. Lyandres,
N. C. Shah,
J. Zhao,
R. P. Van Duyne,
\textit{Nat. Mater.},
2008,
\textbf{7},
442-453.


\bibitem{2013 Cecchini}
M. P. Cecchini,
V. A. Turek,
J. Paget,
A. A. Kornyshev,
J. B. Edel,
\textit{Nat. Mater.},
2013,
\textbf{12},
165-171.


\bibitem{2014 Heavy Metal}
M. P. Cecchini,
V. A. Turek,
A. Demetriadou,
G. Britovsek,
T. Welton,
A. A. Kornyshev,
J. D. E. T. Wilton-Ely,
J. B. Edel,
\textit{Adv. Opt. Mater.},
2014,
\textbf{2},
966-977.


\bibitem{2012 Nanoplasmonic Metamaterials}
O. Hess,
J. B. Pendry,
S. A. Maier,
R. F. Oulton,
J. M. Hamm,
K. L. Tsakmakidis,
\textit{Nat. Mater.},
2012,
\textbf{11},
573-584.


\bibitem{2011 Metamaterials}
Y. Liu,
X. Zhang,
\textit{Chem. Soc. Rev.},
2011,
\textbf{40},
2494-2507.


\bibitem{2006 Invisible}
J. B. Pendry,
D. Schurig,
D. R. Smith,
\textit{Science},
2006,
\textbf{312},
1780-1782.


\bibitem{2011 Stockman OSA}
M. I. Stockman,
\textit{Opt. Express},
2011,
\textbf{19},
22029-22106.


\bibitem{2012 Hafner}
N. J. Halas,
S. Lal,
S. Link,
W. -S. Chang,
D. Natelson,
J. H. Hafner,
P. Nordlander,
\textit{Adv. Mater.},
2012,
\textbf{24},
4842-4877.


\bibitem{1988 Efrima}
D. Yogev,
S. Efrima,
\textit{J. Phys. Chem.},
1988,
\textbf{92},
5754-5760.


\bibitem{2003 Borra}
H. Yockell-Lelievre,
E. F. Borra,
A. M. Ritcey,
L. Vieira da Silva Jr.,
V. A. Parsegian,
\textit{Appl. Opt.},
2003,
\textbf{42},
1882-1887.


\bibitem{2005 Truong}
L. Truong,
E. F. Borra,
R. Bergamasco,
N. Caron,
P. Laird,
A. Ritcey,
\textit{Appl. Opt.},
2005,
\textbf{44},
1595-1600.


\bibitem{2006 Gingras}
J. Gingras,
J. -P. Dery,
H. Yockell-Lelievre,
E. Borra,
A. M. Ritcey,
\textit{Colloids Surf., A},
2006,
\textbf{279},
79-86.


\bibitem{2010 Flatte}
M. E. Flatte,
A. A. Kornyshev,
M. Urbakh,
\textit{J. Phys. Chem. C},
2010,
\textbf{114},
1735-1747.


\bibitem{2010 Furst}
M. Grzelczak,
J. Vermant,
E. M. Furst,
L. M. Liz-Marzan,
\textit{ACS Nano},
2010,
\textbf{4},
3591-3605.


\bibitem{2016 Edel}
J. B. Edel, 
A. A. Kornyshev, 
A. R. Kucernak, 
M. Urbakh,
\textit{Chem. Soc. Rev.},
2016,
\textbf{45},
1581-1596.


\bibitem{2010 Sperling}
R. A. Sperling,
W. J. Parak,
\textit{Phil. Trans. R. Soc. A},
2010,
\textbf{368},
1333-1383.


\bibitem{2012 Turek}
V. A. Turek,
M. P. Cecchini,
J. Paget,
A. R. Kucernak,
A. A. Kornyshev,
J. B. Edel,
\textit{ACS Nano},
2012,
\textbf{6},
7789-7799.


\bibitem{2004 Girault}
B. Su,
J. -P. Abid,
D. J. Fermin,
H. H. Girault,
H. Hoffmannova,
P. Krtil,
Z. Samek,
\textit{J. Am. Chem. Soc.},
2004,
\textbf{126},
915-919.


\bibitem{2008 Flatte}
M. E. Flatte,
A. A. Kornyshev,
M. Urbakh,
\textit{J. Phys. Condens. Matter},
2008,
\textbf{20},
073102.


\bibitem{2012 Marinescu}
A. A. Kornyshev,
M. Marinescu,
J. Paget,
M. Urbakh,
\textit{Phys. Chem. Chem. Phys.},
2012,
\textbf{14},
1850-1859.


\bibitem{2014 Walpole}
J. Paget,
V. Walpole,
M. B. Jorquera,
J. B. Edel,
M. Urbakh,
A. A. Kornyshev,
A. Demetriadou,
\textit{J. Phys. Chem. C},
2014,
\textbf{118},
23264-23273.


\bibitem{2016 Girault}
E. Smirnov,
P. Peljo,
M. D. Scanlan,
F. Gumy,
H. H. Girault,
\textit{Nanoscale},
2016,
\textbf{8},
7723-7737.


\bibitem{2008 Israelachvili}
Y. Min,
M. Akbulut,
K. Kristiansen,
Y. Golan,
J. Israelachvili,
\textit{Nat. Mater.},
2008,
\textbf{7},
527-538.


\bibitem{2013 Edel}
J. B. Edel,
A. A. Kornyshev,
M. Urbakh,
\textit{ACS Nano},
2013,
\textbf{7},
9526-9532.


\bibitem{Parsegian}
V. A. Parsegian,
\textit{Van der Waals Forces: A Handbook for Biologists, Chemists, Engineers and Physicists},
Cambridge University Press,
New York,
2006.


\bibitem{1937 Hamaker}
H. C. Hamaker,
\textit{Physica},
1937,
\textbf{4},
1058-1072.


\bibitem{1948 Casimir}
H. B. G. Casimir,
D. Polder,
\textit{Phys. Rev.},
1948,
\textbf{73},
360-372.


\bibitem{1956 Lifshitz}
E. M. Lifshitz,
\textit{Soviet Physics},
1956,
\textbf{2},
73-83.


\bibitem{1961 Lifshitz}
E. Dzyaloshinskii,
E. M. Lifshitz,
L. P. Pitaevskii,
\textit{Sov. Phys. Usp.},
1961,
\textbf{4},
153-176.


\bibitem{1956 Derjaguin}
B. V. Derjaguin,
I. I. Abrikosova,
E. M. Lifshitz,
\textit{Q. Rev. Chem. Soc.},
1956,
\textbf{10},
295-329.


\bibitem{Israelachvili}
J. N. Israelachvili,
\textit{Intermolecular and Surface Forces},
Academic Press,
Boston,
3rd edn.,
2011.


\bibitem{Palik Quote}
E. D. Palik,
\textit{Handbook of Optical Constants of Solids \rom{2} },
Academic,
San Diego,
CA,
1991.


\bibitem{1997 Bergstrom}
L. Bergstrom,
\textit{Adv. Colloid Interface Sci.},
1997,
\textbf{70},
125-169.


\bibitem{1963 Enuston}
B. V. Enuston,
J. Turkevich,
\textit{J. Am. Chem. Soc.},
1963,
\textbf{85},
3317-3328.


\bibitem{2015 Wijenayaka}
L. A. Wijenayaka,
M. R. Ivanov,
C. M. Cheatum,
A. J. Haes,
\textit{J. Phys. Chem. C},
2015,
\textbf{119},
10064-10075.


\bibitem{2015 Gambinossi}
F. Gambinossi,
S. E. Mylon,
J. K. Ferri,
\textit{Adv. Colloid Interface Sci.},
2015,
\textbf{222},
332-349.


\bibitem{1961 Vold}
M. J. Vold,
\textit{J. Coll. Sci. Imp. Tok.},
1961,
\textbf{16},
1-12.


\bibitem{1973 Vincent}
B. Vincent,
\textit{J. Colloid Interface Sci.},
1973,
\textbf{42},
270-285.


\bibitem{2012 Pinchuk}
A. O. Pinchuk,
\textit{J. Phys. Chem. C},
2012,
\textbf{116},
20099-20102.


\bibitem{2003 Pinchuk}
A. Pinchuk,
U. Kreibig,
\textit{New J. Phys.},
2003,
\textbf{5},
151.1-151.15.


\bibitem{2012 Schatz}
R. Esquivel-Sirvent,
G. C. Schatz,
\textit{J. Phys. Chem. C},
2012,
\textbf{116},
420-424.


\bibitem{2014 Pendry}
Y. Luo,
R. Zhao,
J. B. Pendry,
\textit{Proc. Natl. Acad. Sci. USA},
2014,
\textbf{111},
18422-18427.


\bibitem{2014 Mortenson}
N. A. Mortenson,
S. Raza,
M. Wubs,
T. Sondergaard,
S. I. Bozhevolnyi,
\textit{Nat. Commun.},
2014,
\textbf{5},
3809.


\bibitem{1975 Kreibig}
L. Genzel,
T. P. Martin,
U. Kreibig,
\textit{Z. Phys. B},
1975,
\textbf{21},
339-346.


\bibitem{Kreibig}
U. Kreibig,
M. Vollmer,
\textit{Optical Properties of Metal Clusters},
Springer,
Berlin,
Germany,
1995.


\bibitem{2010 He}
Y. He,
T. Zeng,
\textit{J. Phys. Chem. C},
2010,
\textbf{114},
18023-18030.


\bibitem{Wijenayaka}
L. A. Wijenayaka,
PhD Thesis,
University of Iowa,
2015.


\bibitem{2005 Kim}
T. Kim,
K. Lee,
M. -S. Gong,
S. -W. Joo,
\textit{Langmuir},
2005,
\textbf{21},
9524-9528.


\bibitem{2008 Lundgren}
A. O. Lundgren,
F. Bjorefors,
L. G. M. Olofsson,
H. Elwing,
\textit{Nano Lett.},
2008,
\textbf{8},
3989-3992.


\bibitem{1981 Weiss}
V. A. Parsegian,
G. H. Weiss,
\textit{J. Colloid Interface Sci.},
1981,
\textbf{81},
285-289.


\bibitem{2012 Olmon}
R. L. Olmon,
B. Slovick,
T. W. Johnson,
D. Shelton,
S. -H. Oh,
G. D. Boreman,
M. B. Raschke,
\textit{J. Phys. Rev. B},
2012,
\textbf{86},
235147.


\bibitem{1972 Johnson and Christy}
P. B. Johnson,
R. W. Christy,
\textit{Phys. Rev. B},
1972,
\textbf{6},
4370-4379.


\bibitem{1974 DESY}
H. -J. Hagemann,
W. Gudat,
C. Kunz,
\textit{Optical Constants from the Far Infrared to the X-Ray Regions: Mg, Al, Ca, Ag, Au, Bi, C and Al$_2$O$_3$},
DESY-Report SR-74/77,
DESY,
Hamburg,
Federal Republic of Germany,
1974.


\bibitem{1971 Irani}
G. B. Irani,
T. Huen,
T. Wooten,
\textit{J. Opt. Soc. Am.},
1971,
\textbf{61},
128-129.


\bibitem{2000 Dagastine}
R. R. Dagastine,
D. C. Prieve,
L. R. White,
\textit{J. Colloid Interface Sci.},
2000,
\textbf{231},
351-358.


\bibitem{1996 Roth}
C. M. Roth,
A. M. Lenhoff,
\textit{J. Colloid Interface Sci.},
1996,
\textbf{179},
637-639.


\bibitem{1974 Heller}
J. M. Heller Jr.,
R. N. Hamm,
R. D. Birkhoff,
L. R. Painter,
\textit{J. Chem. Phys.},
1974,
\textbf{60},
3483-3486.


\bibitem{2002 Rizzoli}
M. Losurdo,
M. Griangregorio,
P. Capezzuto,
G. Bruno,
R. De Rosa,
F. Roca,
C. Summonte,
J. Pla,
R. Rizzoli,
\textit{J. Vac. Sci. Technol. A},
2002,
\textbf{20},
37-42.


\bibitem{2014 Konig}
T. A. F. Konig,
P. A. Ledin,
J. Kerszulius,
M. A. Mahmoud,
M. A. El-Sayed,
J. R. Reynolds,
V. V. Tsukruk,
\textit{ACS Nano},
2014,
\textbf{8},
6182-6192.


\bibitem{1977 Kornyshev}
A. A. Kornyshev,
A. I. Rubinshtein,
M. A. Vorotyntsev,
\textit{Phys. Status Solidi B},
1977,
\textbf{84},
125-132.


\bibitem{Jackson}
J. D. Jackson,
\textit{Classical Electrodynamics},
John Wiley \& Sons, Inc.,
Hoboken,
New Jersey,
3rd edn.,
1999.


\bibitem{2011 Hikita}
Y. Hikita,
M. Kawamura,
C. Bell,
H. Y. Hwang,
\textit{Appl. Phys. Lett.},
2011,
\textbf{98},
192103.


\bibitem{2016 Moerland}
R. J. Moerland,
J. P. Hoogenboom,
\textit{Optica},
2016,
\textbf{3},
112-117.


\bibitem{2007 Neumann}
F. Neumann,
Y. A. Genenko,
C. Melzer,
S. V. Yapolskii,
H. von Seggern,
\textit{Phys. Rev. B},
2007,
\textbf{75},
205322.


\bibitem{2011 Melikyan}
A. Melikyan,
N. Lindenmann,
S. Walheim,
P. M. Leufke,
S. Ulrich,
J. Ye,
P. Vincze,
H. Hahn,
T. Schimmel,
C. Koos,
W. Freude,
J. Leuthold,
\textit{Opt. Express},
2011,
\textbf{19},
8855-8869.


\bibitem{Mahanty}
J. Mahanty,
B. W. Ninham,
\textit{Dispersion Forces},
Academic Press,
London-New York,
1976.


\bibitem{2005 Marcovitch}
M. Marcovitch,
H. Diamant,
\textit{Phys. Rev. Lett.},
2005,
\textbf{95},
223203.


\bibitem{1996 Silbey}
M. Cho,
R. J. Silbey,
\textit{J. Chem. Phys.},
1996,
\textbf{104},
8730-8741.


\bibitem{1979 Vorotyntsev}
M. A. Vorotyntsev,
A. A. Kornyshev,
A. I. Rubinstein,
\textit{Dokl. Akad. Nauk SSSR},
1979,
\textbf{248},
1321-1324.

\bibitem{Ginzburg}
Yu. S. Barash,
V. L. Ginzburg,
\textit{Sov. Phys. Usp.},
1984,
\textbf{27},
467-491.

\bibitem{Bardeen}
J. Bardeen,
\textit{Phys. Rev.},
1940,
\textbf{58},
727-736.

\end{thebibliography}
\end{document}